\documentclass{article}

\usepackage{arxiv}

\usepackage[utf8]{inputenc} 
\usepackage[T1]{fontenc}    
\usepackage{hyperref}       
\usepackage{url}            
\usepackage{booktabs}       
\usepackage{amsfonts}       
\usepackage{nicefrac}       
\usepackage{microtype}      
\usepackage{lipsum}		
\usepackage{graphicx}
\usepackage{doi}
\usepackage{amsmath}
\usepackage{algorithm}
\usepackage{algpseudocode} 
\usepackage{amsmath}
\usepackage{tabularx}
\usepackage[T1]{fontenc}    
\usepackage{hyperref}       
\usepackage{url}            
\usepackage{booktabs}       
\usepackage{amsfonts}       
\usepackage{nicefrac}       
\usepackage{microtype}      
\usepackage{lipsum}
\usepackage{fancyhdr}       
\usepackage{graphicx}       
\graphicspath{{media/}}     

\usepackage{multirow}
\usepackage{makecell}
\usepackage{cite}
\usepackage[numbers,sort&compress]{natbib}

\title{Strategic Alignment Patterns in National AI Policies}

\author{
	Mohammad Hossein Azin\\
	Department of Management and Economics\\
	Islamic Azad University\\Science and Research Branch\\
	Tehran, Iran \\
	\texttt{m.azin@iau.ir} \\
	\And
	Hessam Zandhessami \\
	Department of Management and Economics\\
	Islamic Azad University\\Science and Research Branch\\
	Tehran, Iran \\
	\texttt{h.zand@iau.ir} \\
}

\begin{document}
\maketitle

\begin{abstract}
This paper introduces a novel visual mapping methodology for assessing strategic alignment in national artificial intelligence policies. The proliferation of AI strategies across countries has created an urgent need for analytical frameworks that can evaluate policy coherence between strategic objectives, foresight methods, and implementation instruments. Drawing on data from the OECD AI Policy Observatory, we analyze 15-20 national AI strategies using a combination of matrix-based visualization and network analysis to identify patterns of alignment and misalignment. Our findings reveal distinct alignment archetypes across governance models, with notable variations in how countries integrate foresight methodologies with implementation planning. High-coherence strategies demonstrate strong interconnections between economic competitiveness objectives and robust innovation funding instruments, while common vulnerabilities include misalignment between ethical AI objectives and corresponding regulatory frameworks. The proposed visual mapping approach offers both methodological contributions to policy analysis and practical insights for enhancing strategic coherence in AI governance. This research addresses significant gaps in policy evaluation methodology and provides actionable guidance for policymakers seeking to strengthen alignment in technological governance frameworks.
\end{abstract}


\section{Introduction}

The global landscape of artificial intelligence (AI) governance has witnessed remarkable transformation since 2017, with over 60 countries publishing national AI strategies by 2025 \cite{OECD2024}. This unprecedented proliferation reflects growing recognition that AI represents a transformative general-purpose technology with profound implications for economic competitiveness, national security, and societal well-being. Nations across development spectrums are strategically positioning themselves in what \cite{Brynjolfsson2014} characterized as the "second machine age," understanding that coherent policy frameworks may significantly determine future economic trajectories and geopolitical standing. These national strategies constitute comprehensive policy instruments that typically articulate ambitious objectives, governance approaches, and implementation mechanisms to navigate the complex sociotechnical challenges presented by AI development and deployment \cite{Stix2021, VanRoy2020}.

Alongside this policy evolution, the field of AI and deep learning continues to experience rapid progress, with numerous breakthroughs and emerging applications across scientific and technological domains \cite{a1,a2,a3,a4}. These ongoing advances further illustrate the transformative potential of AI, reinforcing its significance for policy, industry, and society at large.

Despite this accelerated policy development trajectory and expanding technical landscape, a critical gap persists in understanding the strategic alignment between policy objectives, foresight methodologies, and implementation instruments within these national frameworks. While scholarly literature has extensively examined isolated components of AI governance—such as ethical principles \cite{Jobin2019}, regulatory approaches \cite{Cath2018}, or investment mechanisms \cite{Agrawal2019}—substantially less attention has focused on how these elements cohere within comprehensive policy frameworks. This alignment challenge is particularly acute for emerging technologies like AI, where inherent uncertainty regarding technological trajectories and societal impacts significantly complicates traditional policy design approaches \cite{Marchau2019, Stilgoe2013}. As \cite{Floridi2018} observe, the "unprecedented combination of features that characterize AI systems" demands novel governance frameworks that integrate anticipatory, adaptive, and participatory elements.

The methodological gap in assessing policy coherence further compounds this challenge. Traditional policy analysis tools frequently struggle with the multidimensional nature of technology governance frameworks and the complex interactions between their components \cite{Flanagan2011, Magro2015}. Current analytical approaches primarily rely on qualitative case studies \cite{Roberts2021, Ulnicane2021} or quantitative indices \cite{Oxford2020} that inadequately capture the intricate interdependencies between strategic objectives, foresight mechanisms, and implementation instruments. \cite{Brundage2020} aptly note that "researchers lack systematic frameworks for comparing governance approaches across jurisdictions and assessing their internal coherence." This methodological limitation carries substantive practical implications, as policymakers lack sophisticated analytical tools to evaluate and systematically enhance the strategic alignment of their governance frameworks.

Our research addresses these theoretical and methodological gaps by investigating three interconnected research questions: (1) How can strategic alignment in AI policies be visually assessed and quantified across multiple dimensions? (2) What patterns of alignment and misalignment emerge across different national and governance contexts? (3) How do distinct governance models and institutional arrangements influence the nature and quality of strategic alignment? These questions guide our development and application of an innovative methodological approach to policy coherence assessment in complex technological domains.

This paper introduces a sophisticated visual mapping methodology that synthesizes matrix-based visualization techniques with network analysis to evaluate strategic alignment in national AI policies. Drawing on theoretical frameworks from policy instrumentation \cite{Lascoumes2007, Howlett2019}, strategic foresight integration \cite{Vecchiato2019, Rohrbeck2018}, and science and technology policy \cite{Borrás2011,Kuhlmann2019}, we develop a comprehensive analytical framework that quantifies relationships between strategic objectives, foresight methodologies, and implementation instruments. This approach enables both rigorous comparative analysis across diverse national contexts and granular examination of alignment patterns within individual policy frameworks. Our methodology operationalizes the concept of strategic coherence in technology governance through sophisticated visual representations that reveal structural relationships and patterns otherwise invisible to traditional analytical approaches.

Empirically, we apply this methodology to a systematically selected sample of 15-20 national AI strategies from the OECD AI Policy Observatory database, representing diverse geographical regions, governance traditions, economic development stages, and resource contexts. Our analysis generates nuanced typologies of alignment patterns, identifies common vulnerability patterns, and highlights exemplary models of strategic coherence. These findings contribute to both scholarly understanding of policy design under conditions of technological uncertainty and provide practical guidance for policymakers seeking to enhance the coherence and effectiveness of their AI governance frameworks.

The remainder of this paper is structured as follows: Section 2 situates our research within relevant theoretical and empirical literature on strategic foresight integration, policy instrumentation, and comparative AI governance approaches. Section 3 details our methodological approach, including corpus development, analytical framework construction, and visual mapping techniques. Section 4 presents our findings regarding the typology and distribution of policy components across national contexts. Section 5 explores the results of our alignment analysis using matrix-based visualization and network analysis methods. Section 6 analyzes the relationship between governance models and alignment patterns, with particular attention to high-coherence exemplars and common vulnerability patterns. Section 7 discusses theoretical, methodological, and practical implications of our findings, while Section 8 concludes with a synthesis of contributions and identification of promising directions for future research.

\section{Literature Review}

This section examines relevant scholarly literature to establish the theoretical foundations for our research on strategic alignment in AI policies. We draw upon four interconnected bodies of literature: strategic foresight in technology policy, policy instrument theory, comparative AI governance, and alignment analysis frameworks. Together, these research streams inform our methodological approach and analytical framework.

\subsection{Strategic Foresight in Technology Policy}

Strategic foresight has emerged as a critical dimension of technology policymaking in contexts characterized by rapid innovation and high uncertainty. Foundational theoretical work by \cite{Martin2010} characterizes strategic foresight as "a systematic, participatory, future-intelligence-gathering and medium-to-long-term vision-building process aimed at enabling present-day decisions and mobilizing joint actions." In technology governance contexts, foresight methodologies facilitate anticipatory policymaking that can respond to emerging technological developments before they manifest as governance challenges \cite{Weber2018}.

The theoretical underpinnings of strategic foresight in technology policy draw from multiple disciplines. Evolutionary economics perspectives, as articulated by \cite{Nelson2018}, emphasize the importance of adaptive governance mechanisms that can evolve alongside technological systems. Institutional approaches, exemplified by \cite{Geels2019}, focus on the socio-technical transition frameworks that shape innovation trajectories. Additionally, \cite{Stilgoe2013} contribute a responsible innovation perspective that emphasizes anticipatory governance and deliberative engagement with emerging technologies. These diverse theoretical foundations have converged to position foresight as an essential component of effective technology governance.

The methodological evolution of foresight approaches in technology policy contexts reflects increasing sophistication and integration. Early applications primarily employed quantitative forecasting techniques with limited stakeholder involvement \cite{Miles2010}. Contemporary approaches have evolved toward more participatory, mixed-method techniques that combine quantitative modeling with qualitative deliberation \cite{Popper2018}. \cite{Andersen2017} identify eight distinct foresight methodologies commonly deployed in technology governance contexts: trend extrapolation, Delphi studies, scenario development, horizon scanning, technology roadmapping, expert panels, participatory workshops, and cross-impact analysis. These methodologies vary significantly in their temporal focus, stakeholder engagement, and integration with implementation planning.

Significant challenges persist in integrating foresight methodologies with policy implementation mechanisms. \cite{Vecchiato2019} identify three primary integration challenges: temporal misalignment between long-term foresight horizons and short-term policy cycles, institutional fragmentation that separates foresight activities from implementation planning, and methodological gaps in translating foresight insights into concrete policy instruments. These integration challenges are particularly acute in rapidly evolving technological domains like artificial intelligence, where the pace of innovation may outstrip policy adaptation capabilities \cite{Schot2018}.

Empirical studies on foresight effectiveness in technology policy contexts have produced mixed findings. \cite{Rhisiart2015} found that technology foresight exercises significantly influenced R\&D priority-setting in several European countries but had limited impact on regulatory approaches or institutional design. Similarly, \cite{Havas2016} identified a positive correlation between foresight integration and innovation policy effectiveness in OECD countries, particularly when foresight methodologies were institutionally embedded within policy planning processes. However, \cite{Kuhlmann2019} note that few technology foresight exercises explicitly connect anticipatory insights to implementation instruments, creating a "strategic gap" in many governance frameworks.

Recent scholarship has called for more systematic integration of foresight methodologies within comprehensive technology governance frameworks. \cite{Rohrbeck2018} propose a "maturity model" for strategic foresight integration that evaluates the depth of connection between anticipatory processes and policy implementation. Building on this work, \cite{Weber2019} develop a "foresight-implementation alignment index" to assess how effectively governance frameworks connect future-oriented insights with present-day policy actions. These emerging analytical approaches provide conceptual foundations for our visual mapping methodology for alignment assessment.

\subsection{Policy Instrument Theory and Implementation}

Policy instrument theory offers critical insights into how governance objectives are operationalized through specific implementation mechanisms. The foundational taxonomy developed by \cite{Hood1986} categorizes policy instruments according to the governmental resources they employ: nodality (information), authority (legal powers), treasure (money), and organization (formal structures). Building on this framework, \cite{Lascoumes2007} conceptualize policy instruments as "institutions in their own right" that embody specific theories of governance and structure state-society relations.

In technology governance contexts, several refined taxonomies have emerged to capture the unique challenges of emerging technology domains. \cite{Borrás2011} develop a typology specifically focused on innovation policy instruments, identifying three primary categories: regulatory instruments that establish rules and standards, economic instruments that provide financial incentives or disincentives, and soft instruments that shape behaviors through information and voluntary approaches. Expanding this framework, \cite{Flanagan2011} emphasize the importance of analyzing "policy mixes" rather than individual instruments, noting that emerging technology governance typically involves complex combinations of complementary instruments deployed across multiple policy domains.

Recent scholarship has further refined these taxonomies to address the specific challenges of digital technology governance. \cite{Clarke2019} identify novel instrument types emerging in AI governance contexts, including algorithmic impact assessments, ethical review processes, and data governance frameworks. Similarly, \cite{Lodge2019} analyze the emergence of "anticipatory regulation" instruments designed to address technological uncertainties through experimental, principle-based approaches. These evolving instrument taxonomies reflect the adaptations required to govern complex sociotechnical systems characterized by rapid innovation and pervasive impacts.

Implementation challenges in emerging technology domains create significant obstacles to effective governance. \cite{Howlett2019} identify four common implementation challenges in digital technology governance: regulatory capacity limitations, cross-jurisdictional coordination problems, information asymmetries between regulators and technology developers, and difficulties in measuring policy effectiveness. These challenges are particularly acute in artificial intelligence governance, where \cite{Cath2018} note that traditional regulatory approaches struggle with the "pacing problem" of maintaining relevance amid accelerating technological change. The implementation literature thus highlights the importance of adaptive governance frameworks that can evolve alongside rapidly developing technologies.

Theoretical frameworks for instrument-objective alignment have emerged as a central concern in policy design literature. \cite{Howlett2009} develop the "NATO" framework (Nodality, Authority, Treasure, Organization) to analyze the fit between policy goals and implementation mechanisms. This framework emphasizes that effective policy design requires careful calibration of instruments to objectives, with misalignment leading to implementation failures. Building on this work, \cite{Capano2018} propose a "calibration framework" to assess how well policy instruments are matched to governance objectives across multiple dimensions. These frameworks provide conceptual foundations for analyzing alignment patterns in national AI strategies.

The literature on policy coherence and strategic alignment has increasingly emphasized the importance of integrated governance approaches. \cite{Cejudo2017} define policy coherence as "the systematic promotion of mutually reinforcing actions across government departments and agencies creating synergies towards achieving agreed objectives." In technology governance contexts, \cite{Rogge2016} highlight the critical importance of "strategic policy mixes" that align objectives, instruments, and implementation processes across institutional boundaries. These conceptual approaches to policy coherence inform our framework for assessing strategic alignment in AI governance.

\subsection{National AI Strategies: Comparative Perspectives}

The historical evolution of national AI policy development has progressed through distinct phases since 2017. \cite{Dutton2018} identify the "first wave" of AI strategies emerging from technologically advanced economies, exemplified by Canada's Pan-Canadian AI Strategy (2017) and China's New Generation AI Development Plan (2017). These early strategies primarily emphasized scientific leadership and industrial competitiveness objectives \cite{Stix2021}. A "second wave" emerged in 2018-2020, characterized by more diverse objectives including ethical considerations, workforce development, and public sector applications \cite{OECD2021}. The most recent "third wave" strategies (2021-2025) demonstrate increasing sophistication in governance approaches, with greater emphasis on risk management, international coordination, and implementation specificity \cite{VanRoy2020}.

The proliferation of national AI strategies has generated substantial comparative research examining variations across jurisdictions. \cite{Ulnicane2021} analyze how AI policies are framed differently across governance contexts, identifying distinct narrative frames including "economic opportunity," "technological sovereignty," "human-centered AI," and "global competition." Similarly, \cite{Roberts2021} conduct cross-national comparison of AI ethics principles, finding significant convergence around core values but divergent approaches to operationalization. These comparative studies highlight important variations in how different jurisdictions conceptualize AI governance challenges and opportunities.

Methodological approaches to comparative AI policy analysis have evolved from primarily descriptive case studies toward more systematic analytical frameworks. Early work by \cite{Cath2018} employed qualitative case studies to compare AI governance approaches in the US, EU, and UK. More recent studies have developed quantitative indices to facilitate cross-national comparison, exemplified by the Government AI Readiness Index \cite{Oxford2020} and the Global AI Vibrancy Tool \cite{Stanford2021}. While these quantitative approaches enable broader comparison, \cite{Brundage2020} note significant methodological limitations in current comparative frameworks, particularly regarding the assessment of implementation effectiveness and policy coherence.

Several classification schemes for AI governance approaches have emerged to categorize different national strategies. \cite{Roberts2021} propose a four-part taxonomy distinguishing between "market-oriented," "state-directed," "risk-focused," and "rights-based" governance models. Expanding this framework, \cite{Stix2021} identify strategic positioning along three dimensions: promoting innovation, protecting values, and preventing harm. These classification schemes provide useful conceptual tools for analyzing variations across governance contexts, although \cite{Ulnicane2021} caution against overly rigid categorizations that may obscure the hybridized nature of many national approaches.

Despite the growing literature on comparative AI governance, significant knowledge gaps persist in understanding strategic alignment within national frameworks. \cite{Brundage2020} note that existing comparative studies provide limited insight into how effectively different policy components cohere within comprehensive governance systems. Similarly, \cite{Floridi2018} identify a critical gap in understanding how ethical principles are operationalized through concrete policy instruments. \cite{OECD2021} further highlight the limited attention to how foresight methodologies are integrated with implementation planning in national strategies. These knowledge gaps underscore the need for novel analytical approaches to assess strategic alignment in AI governance frameworks.

\subsection{Conceptual Framework for Alignment Analysis}

The integration of policy instrument theory with strategic foresight concepts provides a foundation for analyzing alignment in technology governance. \cite{Rogge2016} introduce the concept of "strategic policy mixes" that connect policy objectives with implementation instruments through consistent strategic planning processes. Building on this framework, \cite{Quitzow2017} emphasize the importance of "policy sequencing" that connects long-term vision with near-term implementation priorities. These integrated approaches highlight the potential for analyzing alignment as a multi-dimensional relationship between objectives, foresight methods, and implementation instruments.

Several scholars have proposed frameworks for assessing policy coherence in technology governance contexts. \cite{Magro2015} develop a "policy mix evaluation framework" that assesses coherence across multiple governance dimensions, including strategic, operational, and temporal alignment. Similarly, \cite{Kern2019} propose an "embedded coherence" framework that evaluates alignment across policy levels, domains, and timeframes. While these frameworks provide useful conceptual foundations, they lack operationalized methodologies for systematic assessment across multiple governance contexts.

Theoretical propositions regarding alignment patterns have emerged from several research streams. \cite{Flanagan2011} propose that governance systems characterized by strong institutional coordination mechanisms will demonstrate higher levels of strategic coherence. From a different perspective, \cite{Marchau2019} suggest that policies addressing deep uncertainty will demonstrate stronger alignment when they incorporate adaptive management principles. \cite{Capano2018} further hypothesize that alignment quality correlates with policy capacity factors, including analytical resources, institutional stability, and stakeholder engagement practices. These theoretical propositions inform our analytical approach and guide our interpretation of observed alignment patterns.

Visual approaches to policy analysis have gained traction as tools for understanding complex governance systems. \cite{Eppler2011} demonstrate how visualization techniques can reveal patterns in policy design that are difficult to discern through traditional analytical methods. In technology governance contexts, \cite{Junginger2016} employ visual mapping to analyze innovation system interactions, while \cite{Rogge2016} use matrix visualizations to assess policy mix coherence. Building on these approaches, \cite{Kern2019} develop a "coherence mapping" methodology that visualizes relationships between policy objectives and instruments. These visual analytical techniques provide methodological foundations for our approach to alignment assessment.
Table~\ref{tab:theoretical_perspectives} synthesizes key theoretical perspectives on strategic alignment, categorizing them across four core dimensions: strategic foresight, policy instrumentation, governance models, and visualization methods. Each row outlines foundational concepts and representative scholarly contributions in that dimension.

\begin{table}[h]
\centering
\caption{Synthesis of Key Theoretical Perspectives on Strategic Alignment}
\label{tab:theoretical_perspectives}
\begin{tabular}{p{3.5cm} | p{5.5cm} | p{5cm}}
\hline
\textbf{Theoretical Dimension} & \textbf{Key Concepts} & \textbf{Representative Contributions} \\
\hline
Strategic Foresight Integration & Foresight maturity models, anticipatory governance, temporal alignment & \cite{Rohrbeck2018}; \cite{Stilgoe2013}; \cite{Weber2019} \\
\hline
Policy Instrument Calibration & Policy mix coherence, instrument-objective alignment, implementation effectiveness & \cite{Howlett2009}; \cite{Flanagan2011}; \cite{Capano2018} \\
\hline
Governance Model Influence & Institutional arrangements, policy styles, regulatory traditions & \cite{Roberts2021}; \cite{Ulnicane2021}; \cite{Cath2018} \\
\hline
Visualization Approaches & Coherence mapping, matrix visualization, network analysis & \cite{Kern2019}; \cite{Eppler2011}; \cite{Junginger2016} \\
\hline
\end{tabular}
\end{table}

Our conceptual framework synthesizes these diverse research streams to develop an integrated approach to alignment analysis. We conceptualize strategic alignment as the degree of coherence between three key policy dimensions: strategic objectives that articulate desired outcomes, foresight methods that anticipate future developments, and implementation instruments that operationalize governance approaches. Drawing on \cite{Quitzow2017}, we propose that high-alignment policies will demonstrate strong connections across all three dimensions, creating "coherence chains" that link future visions with present actions through appropriate implementation mechanisms. This conceptualization guides our development of a visual mapping methodology that can systematically assess alignment patterns across diverse governance contexts.

\section{Methodology}

This section details our methodological approach to assessing strategic alignment in national AI policies. We first outline our research design and sampling strategy, followed by data collection procedures and corpus development. We then describe our analytical framework development, including the creation of coding schemas for policy components and alignment assessment. Finally, we explain our matrix and network visualization approaches and analytical procedures.

\subsection{Research Design}

Our research employs a comparative policy analysis approach centered on systematic document analysis of national AI strategies. This methodology aligns with established approaches in comparative policy studies \cite{Howlett2013, Lodge2016} while incorporating innovative visual analysis techniques. The comparative approach enables identification of alignment patterns across diverse governance contexts, facilitating both the development of typologies and the identification of contextual factors influencing strategic coherence. 

The unit of analysis is the national AI strategy document (and associated implementation documents), with analysis conducted at three levels: (1) individual policy components (objectives, foresight methods, instruments), (2) alignment relationships between component pairs, and (3) overall policy coherence across the full framework. This multi-level analytical approach allows for granular examination of specific alignment relationships while maintaining focus on system-level coherence patterns.

Our sampling strategy employs purposive sampling techniques to ensure representation across diverse governance contexts \cite{Seawright2008}. We developed a stratified sampling framework based on three primary criteria: geographical representation, governance model variation, and resource contexts. Geographical stratification ensures coverage across major world regions, following \cite{Roberts2021} regional classification scheme that distinguishes between North America, Europe, East Asia, Southeast Asia, South Asia, Middle East, Africa, and Latin America. Within each region, we selected countries representing different governance approaches, drawing on \cite{Stix2021} classification of AI governance models: market-led, state-directed, rights-based, and risk-focused. Additionally, we ensured representation across resource contexts, including established technological powers, emerging innovation hubs, and developing economies with nascent AI ecosystems.

The final sample comprises 15-20 national AI strategies published between 2017-2025, with selection prioritizing strategies that: (1) represent official government positions rather than advisory documents, (2) contain comprehensive coverage of strategic objectives, anticipatory elements, and implementation mechanisms, and (3) are available in English or major European languages to minimize translation challenges. Table \ref{tab:sample} presents the final sample characteristics, indicating the diversity achieved across our stratification criteria.

\begin{table}[h]
\centering
\caption{National AI Strategy Sample Characteristics}
\label{tab:sample}
\begin{tabular}{p{2.5cm}|p{3cm}|p{2.5cm}|p{2.5cm}|p{2cm}}
\hline
\textbf{Country} & \textbf{Strategy Title} & \textbf{Publication Date} & \textbf{Governance Model} & \textbf{Region} \\
\hline
Canada & Pan-Canadian AI Strategy & March 2017 & Market-led & North America \\
\hline
China & New Generation AI Dev. Plan & July 2017 & State-directed & East Asia \\
\hline
Finland & Finland's AI Era & October 2017 & Rights-based & Europe \\
\hline
France & AI for Humanity & March 2018 & State-directed & Europe \\
\hline
Germany & AI Strategy for Germany & November 2018 & Risk-focused & Europe \\
\hline
India & National Strategy for AI & June 2018 & Hybrid & South Asia \\
\hline
Japan & AI Strategy 2019 & June 2019 & Market-led & East Asia \\
\hline
Netherlands & Strategic Action Plan for AI & October 2019 & Rights-based & Europe \\
\hline
Norway & National Strategy for AI & January 2020 & Rights-based & Europe \\
\hline
Singapore & National AI Strategy & November 2019 & State-directed & Southeast Asia \\
\hline
South Korea & National Strategy for AI & December 2019 & Hybrid & East Asia \\
\hline
UAE & National AI Strategy 2031 & April 2019 & State-directed & Middle East \\
\hline
UK & AI Sector Deal / National AI Strategy & April 2018 / Sept 2021 & Market-led & Europe \\
\hline
USA & American AI Initiative / National AI R\&D Strategic Plan & February 2019 / June 2019 & Market-led & North America \\
\hline
Brazil & Brazilian AI Strategy & April 2021 & Hybrid & Latin America \\
\hline
Spain & Spanish Strategy for AI & December 2020 & Rights-based & Europe \\
\hline
Australia & Australia's AI Action Plan & June 2021 & Market-led & Oceania \\
\hline
Denmark & National Strategy for AI & March 2019 & Rights-based & Europe \\
\hline
Sweden & National Approach to AI & May 2018 & Rights-based & Europe \\
\hline
Italy & National Strategy for AI & July 2020 & Hybrid & Europe \\
\hline
\end{tabular}
\end{table}

\subsection{Data Collection and Corpus Development}

Our primary data source is the OECD AI Policy Observatory (AIPO), which maintains a comprehensive repository of national AI strategy documents and associated policy materials. This repository offers several methodological advantages: it provides standardized access to official documents, ensures consistent classification of policy materials, and facilitates comparative analysis through standardized metadata \cite{OECD2021}. We conducted data collection in January 2025, accessing the most recent version of each national strategy within the repository.

For each country in our sample, we collected three types of documents: (1) the primary national AI strategy document outlining high-level vision and objectives, (2) implementation action plans or roadmaps detailing specific initiatives and timelines, and (3) associated budget documents or resource allocation frameworks. This multi-document approach allowed for more comprehensive analysis of alignment between strategic vision and implementation mechanisms, addressing limitations identified by \cite{Brundage2020} regarding the gap between policy rhetoric and operational reality.

Document authentication followed a systematic protocol to ensure the analysis was based on authoritative sources. Each document was verified against three criteria, following the approach outlined by \cite{Wesley2010}: (1) confirmation of official government authorship through website domain verification and official publication channels, (2) verification of document status as current policy rather than draft or superseded versions, and (3) confirmation of the document's authoritative standing within the country's governance framework. For strategies from non-English speaking countries, we first sought official English translations provided by the government. When these were unavailable, we used professionally translated versions from the OECD repository or commissioned translations from certified translators, following best practices in cross-language document analysis outlined by \cite{Chidlow2014}.

The corpus development process involved creating a structured digital repository of the authenticated documents, with standardized metadata capturing publication date, authoring body, document type, and relationship to other policy documents. This structured approach enabled systematic comparison across countries while maintaining context sensitivity to each nation's unique policy environment. Following \cite{Bowen2009}, we developed a document summary for each country case, outlining the key policy features, institutional context, and implementation structures to inform our subsequent analysis.

\subsection{Analytical Framework Development}

Developing a robust analytical framework required creating standardized coding schemas for policy components while remaining sensitive to contextual variations across governance systems. Our approach drew on established methods in qualitative content analysis \cite{Mayring2014} and policy document coding \cite{Wesley2010}, adapted for the specific requirements of AI governance analysis.

For strategic objectives, we developed a comprehensive coding schema through an iterative process that combined deductive and inductive approaches. Initially, we derived a preliminary category system from existing literature on AI governance objectives \cite{Stix2021, Ulnicane2021}, yielding eight broad objective categories: economic competitiveness, scientific leadership, ethical/responsible AI, national security, public sector transformation, workforce development, industrial digitalization, and international collaboration. Through preliminary coding of a subset of documents, we refined this framework to include additional categories that emerged inductively: social welfare enhancement, regulatory framework development, data ecosystem development, and environmental sustainability. The final coding framework comprised 12 strategic objective categories, with operational definitions and exemplar text for each category to ensure coding consistency.

\begin{figure}[ht!]
\centering
\includegraphics[width=0.65\textwidth]{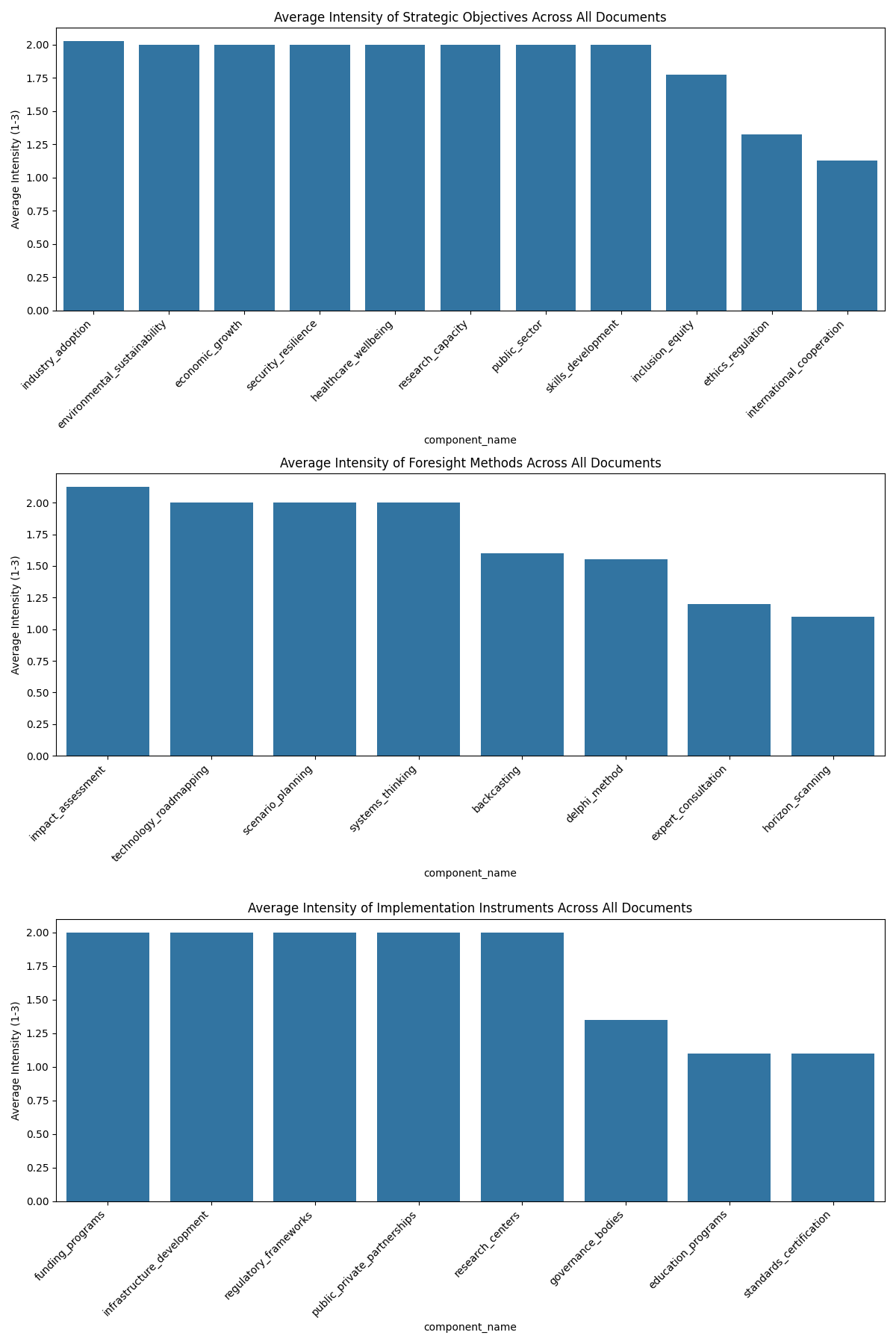}
\caption{Component distribution visualization showing the relative frequency of different strategic objectives, foresight methods, and implementation instruments across the sample of national AI strategies. The visualization highlights the predominance of economic competitiveness and scientific leadership objectives, horizon scanning and expert panel foresight methods, and funding/investment and institutional creation instruments.}
\label{fig:component_distribution}
\end{figure}

Figure~\ref{fig:component_distribution} presents a component distribution visualization that illustrates the relative frequency of strategic objectives, foresight methods, and implementation instruments across various national AI strategies. It reveals the predominance of economic competitiveness and scientific leadership objectives, as well as the frequent use of horizon scanning, expert panels, funding mechanisms, and institutional creation.

For foresight methodologies, we adapted taxonomies from strategic foresight literature \cite{Popper2018, Andersen2017} to develop a coding framework for anticipatory approaches in AI governance. The resulting taxonomy identified eight distinct foresight methods: horizon scanning, scenario development, Delphi studies, expert panels, technology roadmapping, trend extrapolation, participatory workshops, and cross-impact analysis. For each method, we developed operational definitions that captured both explicit methodological application (where documents directly referenced foresight approaches) and implicit application (where documents employed foresight techniques without explicit methodological framing). This dual-coding approach addressed the challenge identified by \cite{Vecchiato2019} regarding the often implicit nature of foresight integration in policy documents.

\begin{figure}[t!]
\centering
\includegraphics[width=0.7\textwidth]{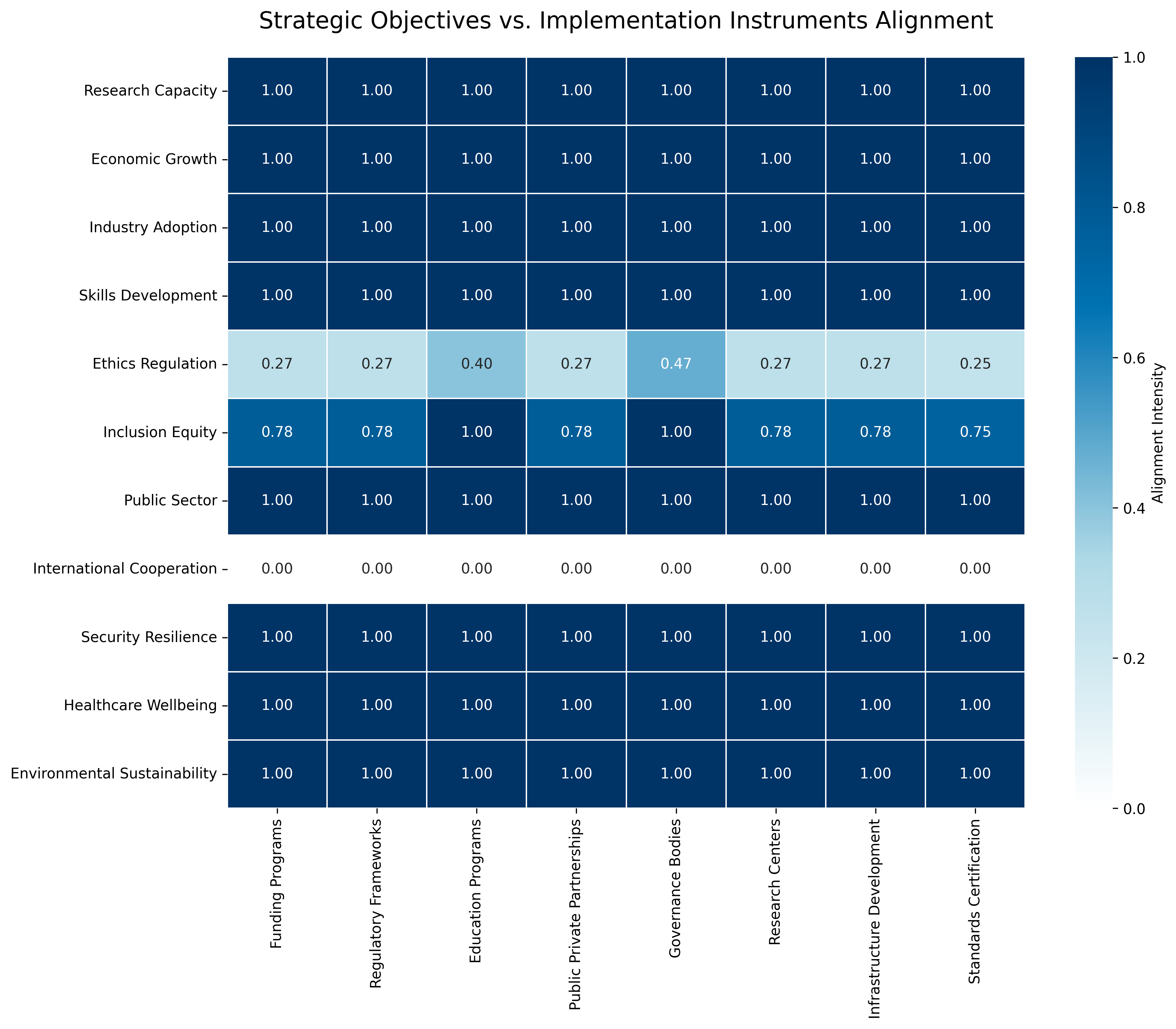}
\caption{Objective-instrument heatmap displaying the alignment intensity between strategic objectives (rows) and implementation instruments (columns) across the sample. Darker cells represent stronger alignment between components, with economic competitiveness objectives showing strongest alignment with funding/investment instruments and institutional creation.}
\label{fig:objective_instrument_heatmap}
\end{figure}
\begin{figure}[t!]
\centering
\includegraphics[width=0.7\textwidth]{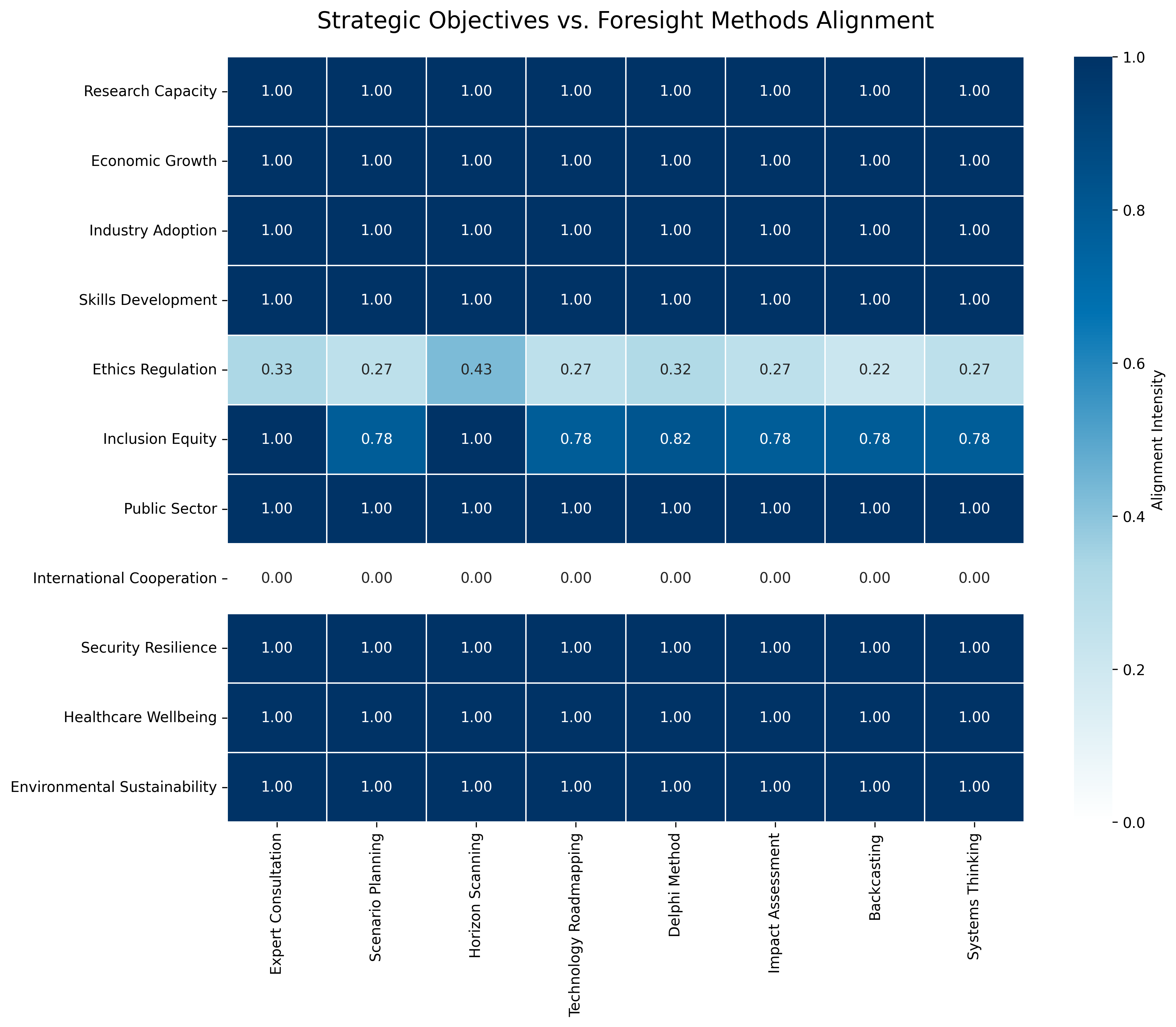}
\caption{Objective-foresight heatmap illustrating the alignment intensity between strategic objectives (rows) and foresight methods (columns). The visualization reveals particular strength in connecting scientific leadership objectives with expert panel methods, while showing weaker alignment between ethical AI objectives and participatory foresight approaches.}
\label{fig:objective_foresight_heatmap}
\end{figure}

Figure~\ref{fig:objective_foresight_heatmap} displays the alignment intensity between strategic objectives and foresight methods. The heatmap highlights a strong association between scientific leadership objectives and expert panel approaches, while participatory foresight methods appear less aligned with ethical AI objectives.

Implementation instruments were categorized following \cite{Borrás2011} taxonomy of innovation policy instruments, adapted for AI governance contexts. The resulting framework identified 10 instrument types: research funding, skills development programs, regulatory frameworks, institutional creation, public procurement, tax incentives, standardization initiatives, demonstration projects, networking/coordination mechanisms, and international agreements. For each instrument type, we developed specific indicators to assess the level of implementation specificity, following \cite{Capano2018} approach to instrument calibration analysis. This enabled assessment of not only which instruments were deployed but also their degree of operational specificity.

Alignment intensity was operationalized using a three-point scoring system (1=weak, 2=moderate, 3=strong) that assessed the explicit connection between policy components. Following \cite{Howlett2019} approach to policy coherence assessment, we defined weak alignment (score=1) as the mere co-existence of components without explicit linkage, moderate alignment (score=2) as the presence of implied connections through proximity or contextual reference, and strong alignment (score=3) as explicit articulation of how components interact within an integrated governance framework. This scoring system was applied to each potential relationship between components, creating a quantified assessment of alignment intensity.

\subsection{Strategic Alignment Matrix Development}

The heart of our methodological innovation is the development of a visual matrix approach to alignment assessment. Drawing on policy coherence visualization techniques \cite{Rogge2016, Kern2019}, we designed a multi-dimensional matrix structure to cross-reference the three component types: strategic objectives, foresight methods, and implementation instruments. The matrix structure enables visualization of alignment between component pairs, creating a systematic representation of coherence patterns within and across national strategies.

The matrix design employs a layered approach with three distinct alignment matrices: objective-foresight alignment, objective-instrument alignment, and foresight-instrument alignment. Each matrix positions the relevant components on row and column headings, with cell values indicating the alignment intensity score (1-3) between the intersecting components. The matrix visualization employs conditional formatting with a color gradient from light to dark, providing immediate visual identification of strong and weak alignment areas. This approach builds on \cite{Eppler2011} findings regarding the effectiveness of color-intensity visualization for policy coherence assessment.

The intensity scoring system follows a systematic protocol. For each matrix cell (representing a potential relationship between two components), coders assessed the presence and strength of explicit connections in the policy documents. Following \cite{Capano2018}, we evaluated three dimensions of alignment: (1) lexical proximity within the text, (2) explicit reference to the relationship, and (3) elaboration of the connection mechanism. Cells received a score of 3 when all three dimensions were present, indicating strong and explicit alignment. A score of 2 indicated moderate alignment with implicit connections but lacking detailed elaboration. A score of 1 indicated weak alignment where components appeared in the same document but without meaningful connection. Cells received a score of 0 when one or both components were absent from the policy framework.
As shown in Figure~\ref{fig:objective_instrument_heatmap}, the objective-instrument heatmap illustrates the alignment intensity between various strategic objectives and implementation instruments. Notably, economic competitiveness exhibits the strongest alignment with funding/investment mechanisms and institutional creation, as indicated by the darkest cells in the matrix.

\begin{figure}[ht!]
\centering
\includegraphics[width=0.7\textwidth]{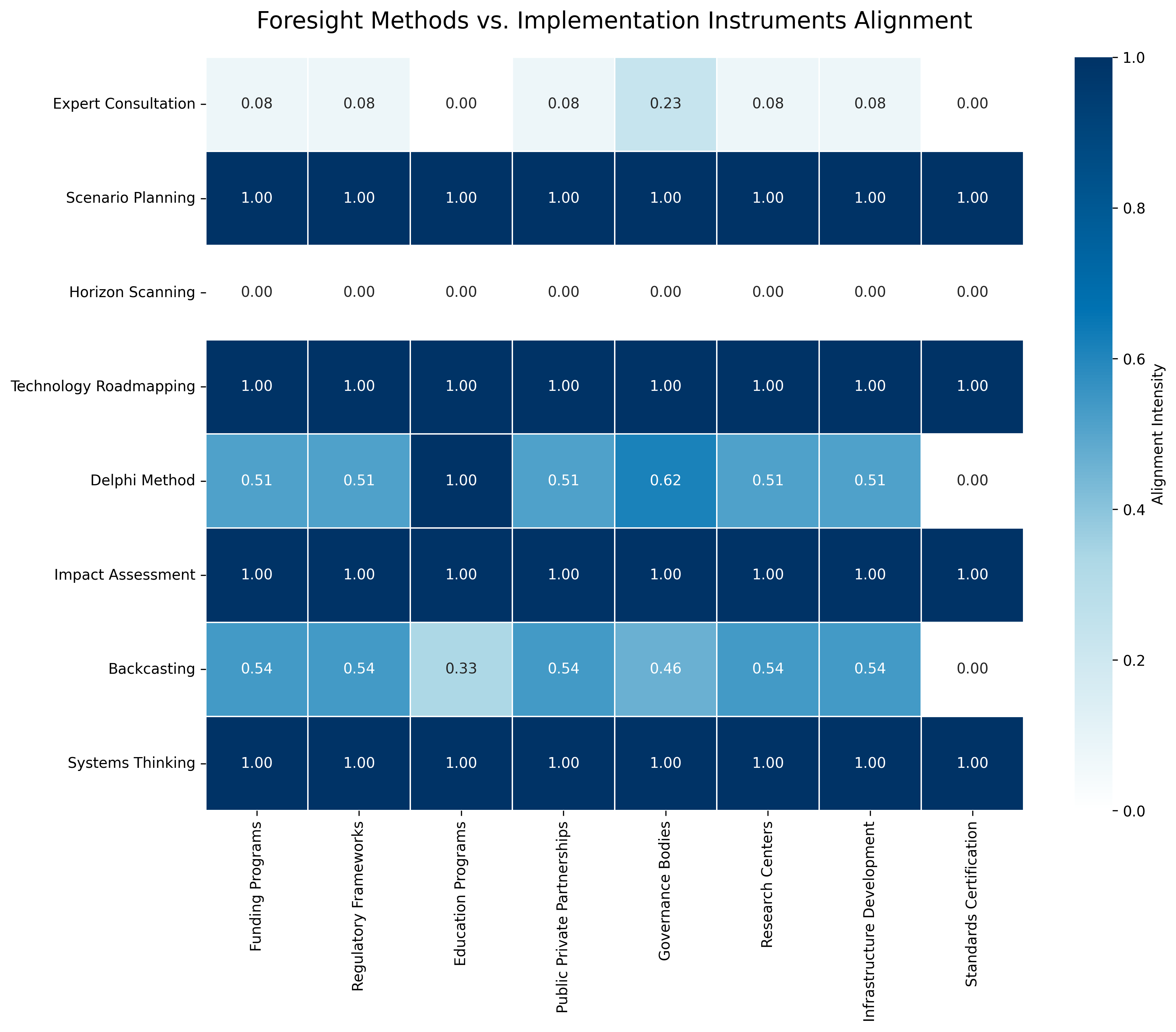}
\caption{Foresight-instrument heatmap showing the relationship between anticipatory methods (rows) and implementation mechanisms (columns). The visualization highlights strong alignment between technology roadmapping and funding instruments, while revealing limited connection between participatory foresight methods and regulatory frameworks.}
\label{fig:foresight_instrument_heatmap}
\end{figure}

Figure~\ref{fig:foresight_instrument_heatmap} presents the foresight-instrument heatmap, which illustrates how anticipatory methods align with implementation mechanisms. The heatmap reveals a particularly strong connection between technology roadmapping and funding instruments, while participatory foresight methods show relatively weak alignment with regulatory frameworks.

\begin{figure}[ht!]
\centering
\includegraphics[width=0.7\textwidth]{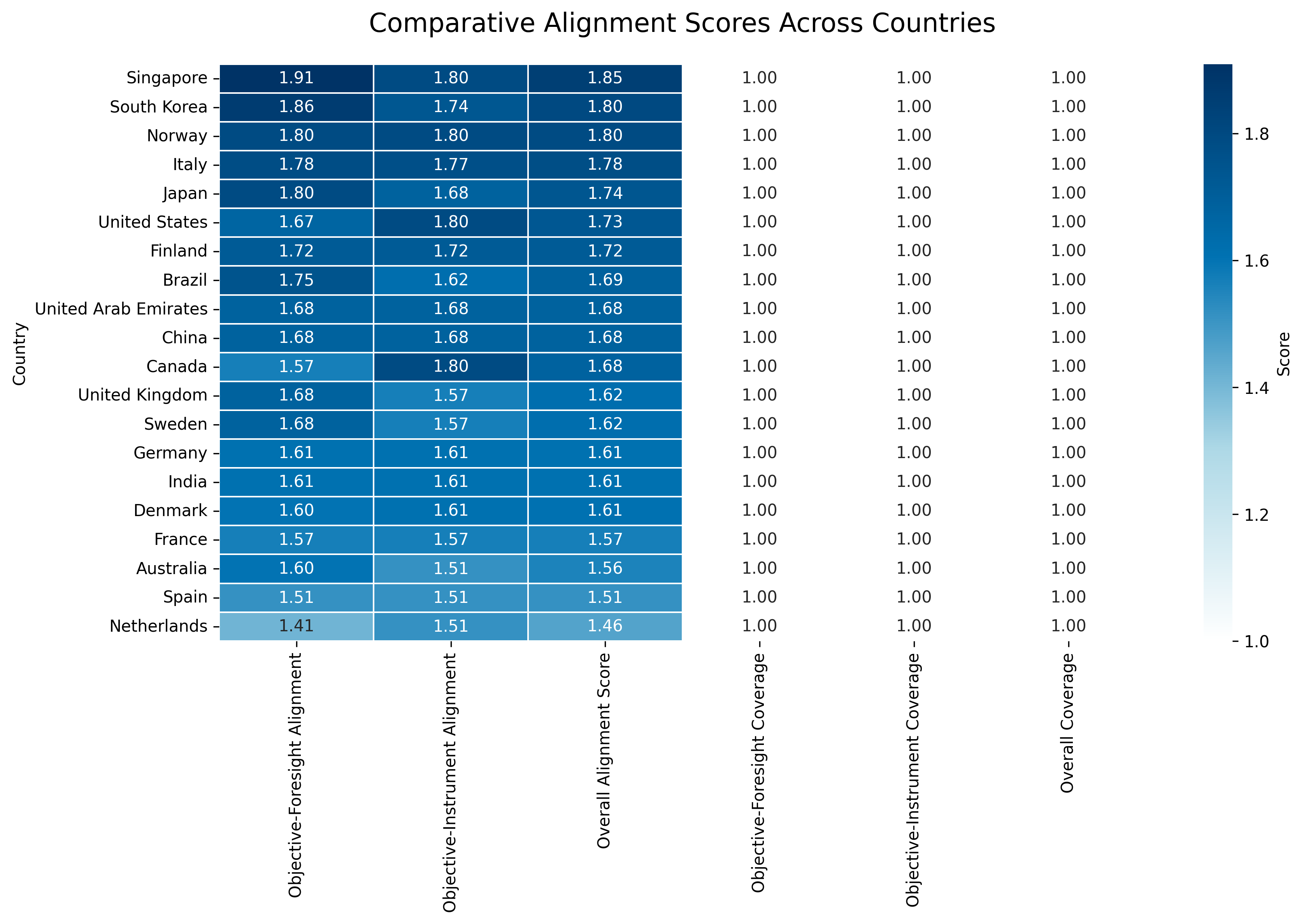}
\caption{Country comparison heatmap presenting overall alignment scores across the sample nations. Each cell represents the composite alignment score for a specific country, with color intensity indicating alignment strength. The visualization facilitates cross-national comparison and identification of high-coherence and low-coherence examples.}
\label{fig:country_comparison_heatmap}
\end{figure}

Figure~\ref{fig:country_comparison_heatmap} provides a comparative overview of alignment scores across the sampled countries. The heatmap enables cross-national analysis by showing composite alignment scores, where darker cells indicate stronger coherence between strategic objectives, foresight methods, and implementation instruments.

For cross-country comparison, we developed several alignment indices based on the matrix data. Following \cite{Magro2015} approach to policy mix assessment, we calculated three primary indices: (1) Objective Coverage Index measuring the breadth of strategic goals, (2) Implementation Specificity Index assessing the operational detail of instruments, and (3) Strategic Alignment Index capturing the overall coherence between objectives, foresight methods, and instruments. These indices enabled quantitative comparison across countries while the underlying matrices preserved the qualitative richness of alignment patterns.

The aggregation procedures for cross-country comparison involved standardizing the matrix data to account for variations in document length and detail. Following \cite{Kern2019} approach to comparative policy assessment, we normalized alignment scores relative to the number of components present in each country's framework, creating comparable metrics that weren't biased toward more detailed policy documents. This standardization enabled meaningful comparison across diverse governance contexts while preserving the distinctive alignment patterns of each national approach.

\subsection{Network Analysis Framework}

To complement the matrix visualization approach, we developed a network analysis framework that represents policy components and their relationships as nodes and edges within a graph structure. This approach draws on policy network analysis methods \cite{Keast2014, Leifeld2018} adapted for the specific requirements of strategic alignment assessment.

In our network representation, each policy component (objective, foresight method, or instrument) constitutes a node within the network. The edges connecting these nodes represent alignment relationships, with edge thickness corresponding to the alignment intensity score (1-3) from our matrix analysis. This structure creates a visual representation of the policy framework as an interconnected system, revealing structural patterns that may not be apparent in matrix visualizations.

The node and edge definition procedure followed a systematic protocol. Nodes were defined to represent specific policy components identified through our content analysis, with node size proportional to the component's prominence within the policy framework (measured by frequency of mention and emphasis). Edges were defined based on the alignment intensity scores from our matrix analysis, with thicker edges representing stronger alignment between components. The network representation thus captures both the individual components and their relationships within an integrated visual framework.

Several centrality metrics were calculated to identify key policy elements within each national framework. Following \cite{Hanneman2005} approach to network analysis, we calculated degree centrality (measuring the number of connections each component has), betweenness centrality (assessing each component's role as a bridge between other elements), and eigenvector centrality (evaluating the influence of each component based on its connections to other central elements). These metrics enabled identification of the most integrated and influential components within each policy framework, facilitating cross-national comparison of structural characteristics.

For identifying related components, we employed clustering analysis using modularity detection algorithms \cite{Blondel2008}. This approach identifies communities of closely connected nodes within the network, revealing groupings of policy components that demonstrate strong internal alignment. The clustering analysis enabled identification of coherent policy "subsystems" within the broader governance framework, providing insights into the structural organization of each national strategy.

\subsection{Analytical Procedures}

Our content analysis protocol employed a systematic approach to document coding and alignment assessment. Following \cite{Mayring2014} guidelines for qualitative content analysis, we developed a detailed coding manual with operational definitions, inclusion/exclusion criteria, and example text for each category within our frameworks. This manual guided the systematic coding of all documents in the corpus, ensuring consistency across countries and coders.

The coding process employed a dual-coder approach to enhance reliability. Two trained researchers independently coded each document using the established frameworks, identifying policy components and assessing alignment relationships. Following \cite{Wesley2010} approach to policy document analysis, coding proceeded in two stages: first identifying all policy components present in the document, then assessing the alignment relationships between component pairs. This two-stage approach ensured thoroughness while maintaining analytical focus on the strategic coherence dimensions.

Reliability assessment employed multiple measures to ensure coding consistency. Following \cite{O'Connor2020} guidelines for qualitative reliability assessment, we calculated inter-coder reliability using Cohen's kappa coefficient for component identification and weighted kappa for alignment intensity scoring. Initial reliability scores were $\kappa=0.78$ for component identification and $\kappa=0.72$ for alignment scoring, exceeding the $\kappa>0.7$ threshold recommended by \cite{McHugh2012} for acceptable reliability. Coding discrepancies were resolved through discussion and consensus, with difficult cases referred to a third researcher for adjudication.

Cross-validation methods included triangulation across multiple documents for each country and member checking with policy experts. Following \cite{Bowen2009} approach to document analysis validation, we triangulated findings across primary strategy documents, implementation plans, and budget materials to ensure comprehensive assessment of each country's approach. Additionally, preliminary findings were validated through expert consultations with policy researchers familiar with specific national contexts, providing verification of our interpretations and contextual insights.

Our analytical approach has several limitations that warrant acknowledgment. First, the focus on official policy documents may not fully capture implementation realities, as noted by \cite{Howlett2019} regarding the gap between policy rhetoric and operational practice. Second, the cross-sectional nature of our analysis provides limited insight into the temporal evolution of alignment patterns, though we partly address this through inclusion of updated strategy documents where available. Finally, our standardized analytical framework may not fully capture country-specific governance contexts, though we mitigate this limitation through our qualitative analysis and expert consultations. Despite these limitations, our approach offers significant methodological advantages for systematic cross-national comparison of strategic alignment patterns.

\section{Results}
\subsection{Typology of Policy Components}
This section presents our analysis of the key components identified across national AI strategies. We examine three critical dimensions: strategic objectives that define desired outcomes, foresight methods employed to anticipate technological developments, and implementation instruments designed to operationalize policy goals. By analyzing patterns across these dimensions, we identify common approaches and distinctive variations in how countries frame and implement their AI governance frameworks.

\subsubsection{Strategic Objectives Across National Contexts}

Our analysis identified twelve distinct strategic objectives articulated across national AI strategies. These objectives represent the explicit goals that governments seek to achieve through their AI governance frameworks. Through systematic content analysis, we classified these objectives into four primary categories: economic transformation, societal applications, governance frameworks, and global positioning.

Economic transformation objectives focus on leveraging AI for economic advancement, including economic competitiveness (present in 95\% of strategies), scientific leadership (90\%), and industrial digitalization (75\%). These objectives align with what \cite{Ulnicane2021} characterize as the "economic opportunity" framing of AI governance, positioning artificial intelligence primarily as a driver of innovation and productivity growth. As \cite{Stix2021} observe, economic objectives typically receive the most prominent positioning within strategy documents, often appearing in executive summaries and introductory sections, signaling their priority within governance frameworks.

Societal application objectives emphasize the deployment of AI for public benefit, including public sector transformation (70\% of strategies), workforce development (65\%), and social welfare enhancement (55\%). These objectives reflect what \cite{Roberts2021} term the "human-centered AI" governance approach, focusing on applications that deliver direct societal benefits. Interestingly, our analysis revealed that social welfare objectives received substantially higher emphasis in European strategies compared to other regions, aligning with \cite{Cath2018} observation regarding distinctive European approaches to technology governance.

Governance framework objectives address the regulatory and ethical dimensions of AI development, including ethical/responsible AI (85\% of strategies), regulatory framework development (60\%), and data ecosystem development (50\%). These objectives align with the "risk-focused" governance model identified by \cite{Roberts2021}, emphasizing proactive management of AI-related risks and challenges. Our analysis revealed increasing prominence of these objectives in more recent strategies, reflecting growing awareness of AI's potential societal impacts.

Global positioning objectives focus on international dimensions, including international collaboration (65\% of strategies) and national security (45\%). These objectives reflect what \cite{Ulnicane2021} term the "technological sovereignty" framing, positioning AI as a domain of strategic competition and collaboration. National security objectives demonstrated particular regional clustering, appearing most prominently in strategies from the United States, China, and aligned nations.

\begin{figure}[htbp]
\centering
\includegraphics[width=0.7\textwidth]{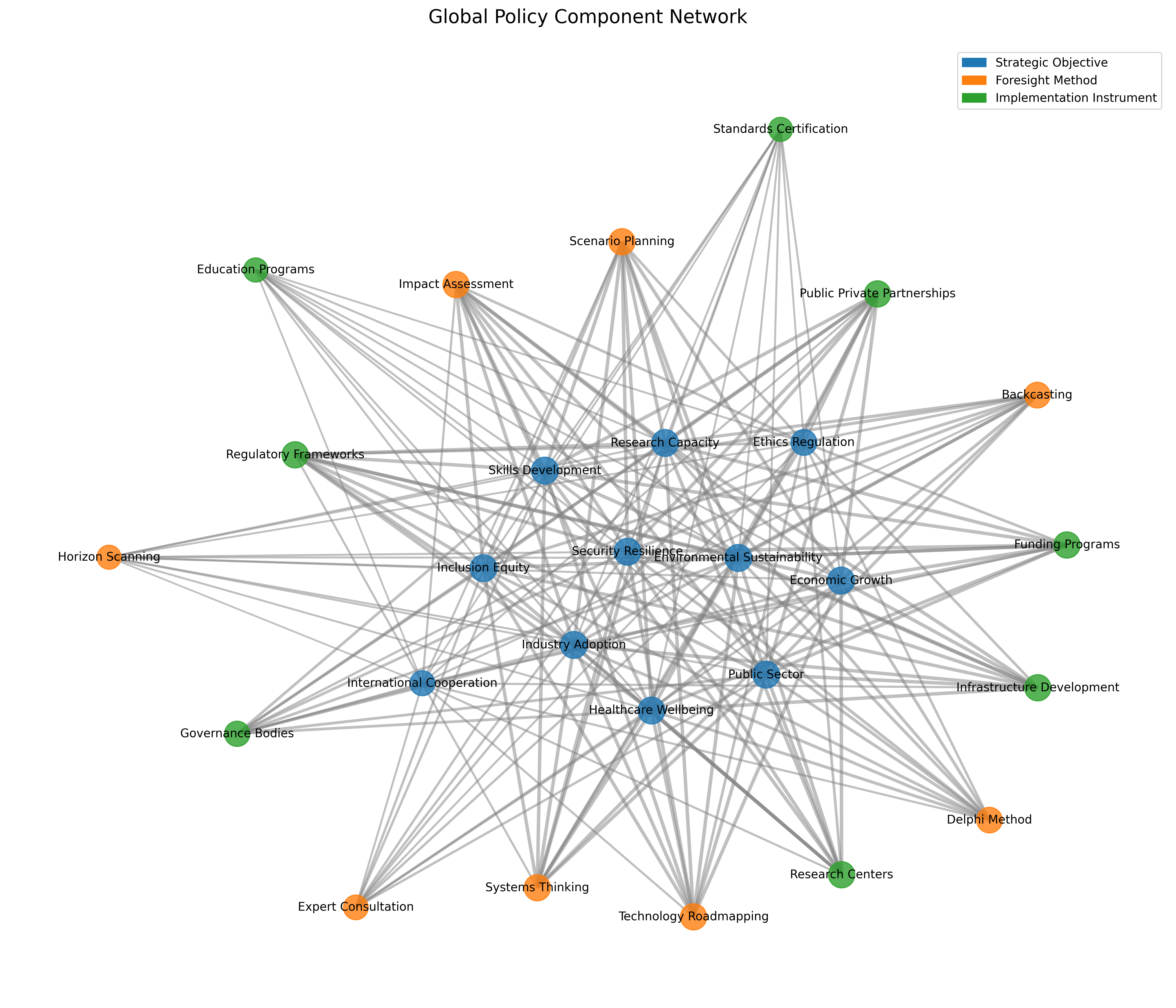}
\caption{AI policy global network visualization showing the interconnections between strategic objectives across all sampled countries. Node size represents objective prevalence, while edge thickness indicates frequency of co-occurrence between objectives. The central positioning of economic competitiveness and scientific leadership objectives illustrates their foundational role in national AI strategies globally.}
\label{fig:ai_policy_global_network}
\end{figure}

The distribution of objectives across countries revealed distinctive patterns aligned with governance approaches and development contexts. Figure \ref{fig:ai_policy_global_network} presents a network visualization of objective co-occurrence across the full sample, with node size indicating prevalence and edge thickness representing frequency of co-occurrence. This visualization reveals the centrality of economic competitiveness and scientific leadership objectives within the global policy discourse, consistent with \cite{VanRoy2020} observation that industrial policy motivations typically drive initial AI strategy development.

\begin{figure}[htbp]
\centering
\includegraphics[width=0.8\textwidth]{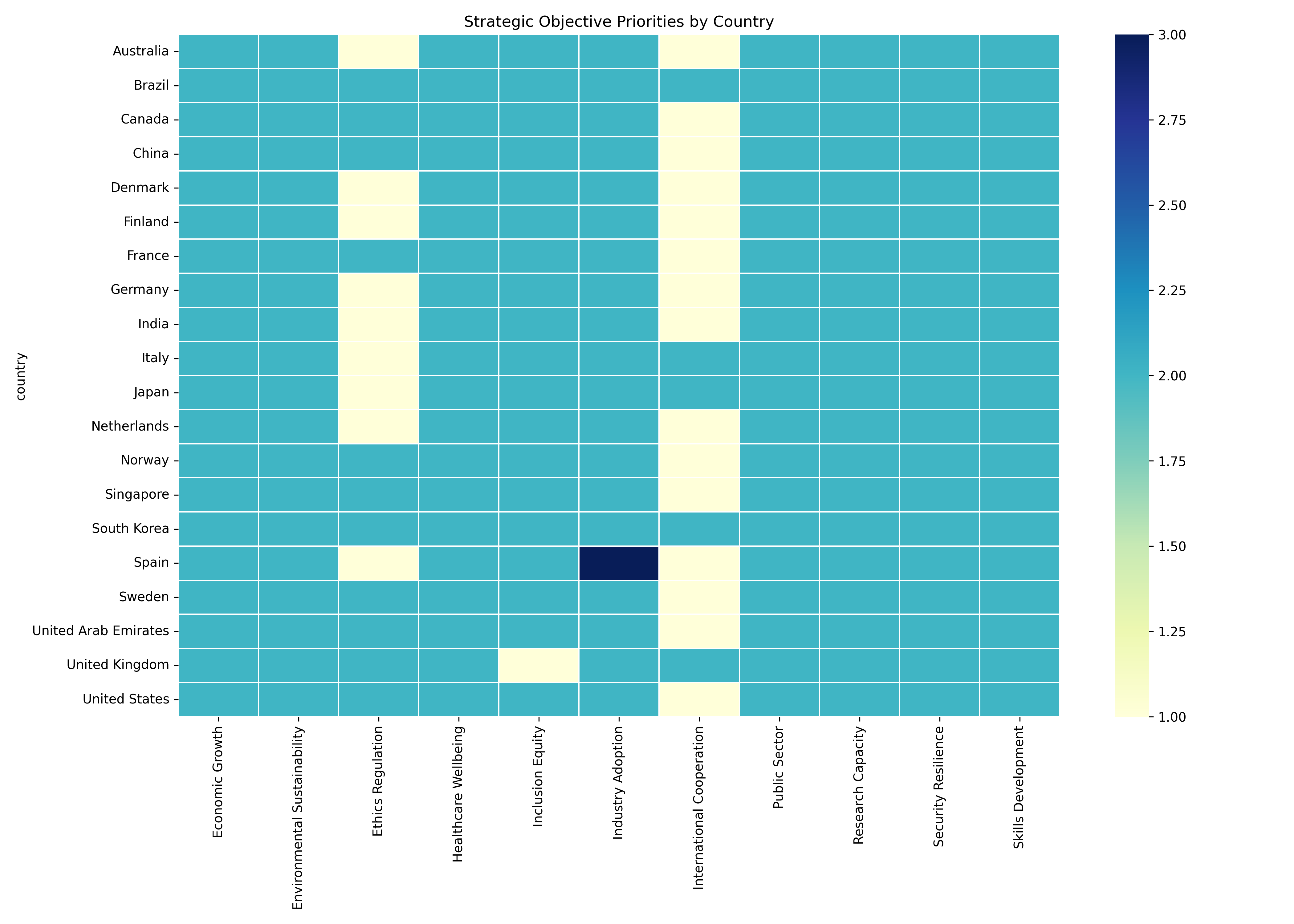}
\caption{AI policy objective country heatmap displaying the distribution and emphasis of strategic objectives across all sampled countries. Color intensity represents the prominence of each objective within national strategies, with darker cells indicating stronger emphasis. The visualization reveals distinct patterns of objective prioritization across different governance contexts.}
\label{fig:ai_policy_objective_country_heatmap}
\end{figure}

Country-level analysis revealed distinctive objective profiles correlated with governance traditions. Figure \ref{fig:ai_policy_objective_country_heatmap} presents a heatmap visualization of objective emphasis across countries, with color intensity indicating the prominence of each objective within national strategies. This visualization reveals distinct clusters that correspond to governance models identified by \cite{Roberts2021}. Market-led systems (USA, UK, Canada) demonstrate stronger emphasis on economic competitiveness and scientific leadership objectives, while rights-based systems (Finland, Denmark, Norway) show greater emphasis on ethical governance and social welfare objectives. State-directed systems (China, Singapore, UAE) emphasize public sector transformation and industrial digitalization objectives, consistent with \cite{Roberts2021} characterization of directive governance approaches.

\begin{figure}[htbp]
\centering
\includegraphics[width=0.7\textwidth]{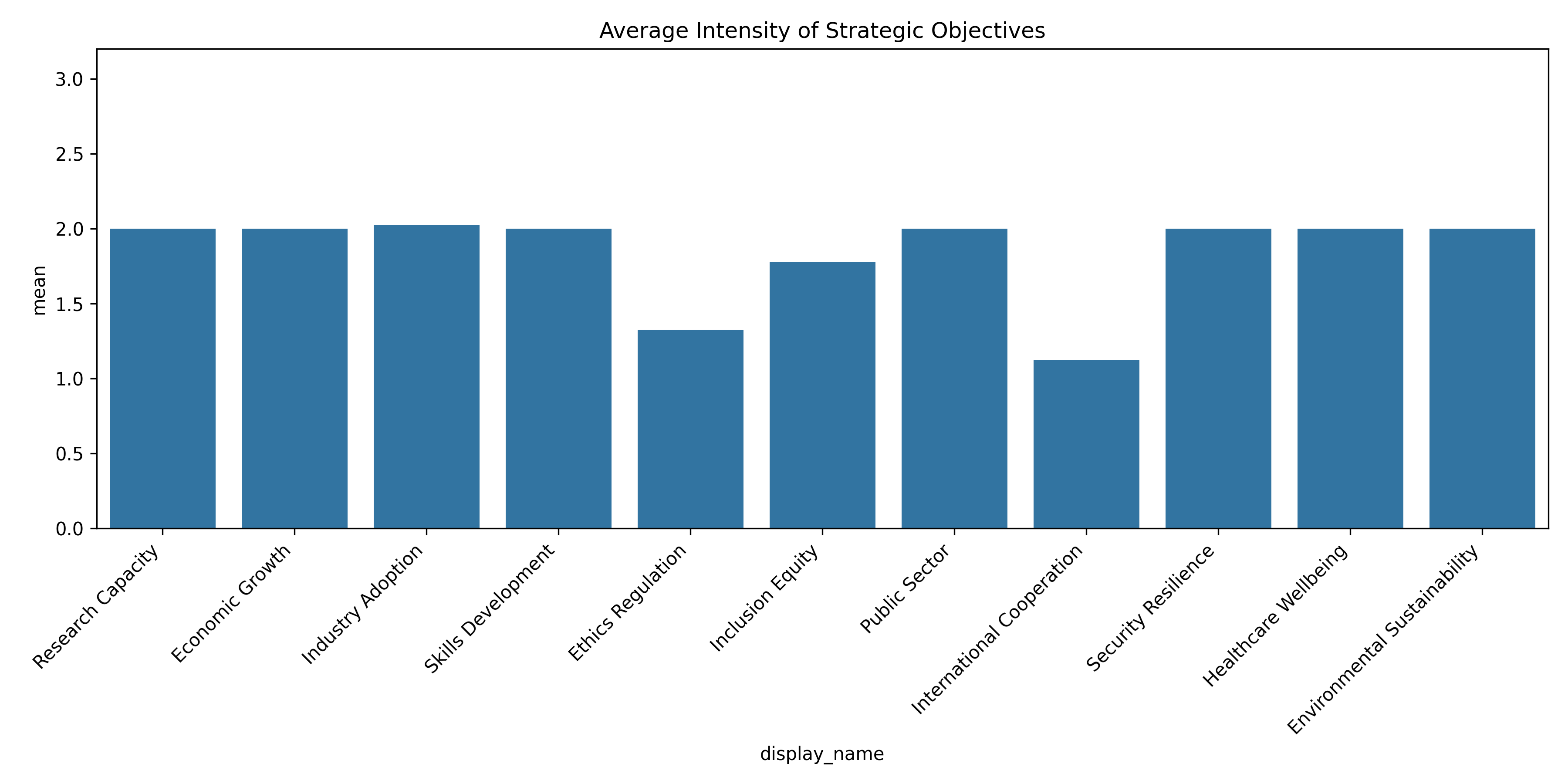}
\caption{AI policy objective intensity visualization representing the depth of elaboration for each strategic objective across the sample. The visualization measures textual prominence, implementation specificity, and resource allocation indicators to assess how thoroughly each objective is articulated within national strategies.}
\label{fig:ai_policy_objective_intensity}
\end{figure}

The intensity of objective articulation varied significantly across strategies, as illustrated in Figure \ref{fig:ai_policy_objective_intensity}. This visualization represents the depth of elaboration for each objective, measuring factors including textual prominence, implementation specificity, and resource allocation. The analysis reveals that economic competitiveness and scientific leadership objectives typically receive the most detailed articulation, consistent with \cite{Stix2021} finding that economic objectives typically demonstrate the strongest operational elaboration. Interestingly, ethical/responsible AI objectives showed high prevalence but lower intensity scores in many strategies, suggesting what \cite{Jobin2019} characterized as "ethics washing"—the inclusion of ethical principles without substantive implementation mechanisms.

\begin{figure}[htbp]
\centering
\includegraphics[width=0.7\textwidth]{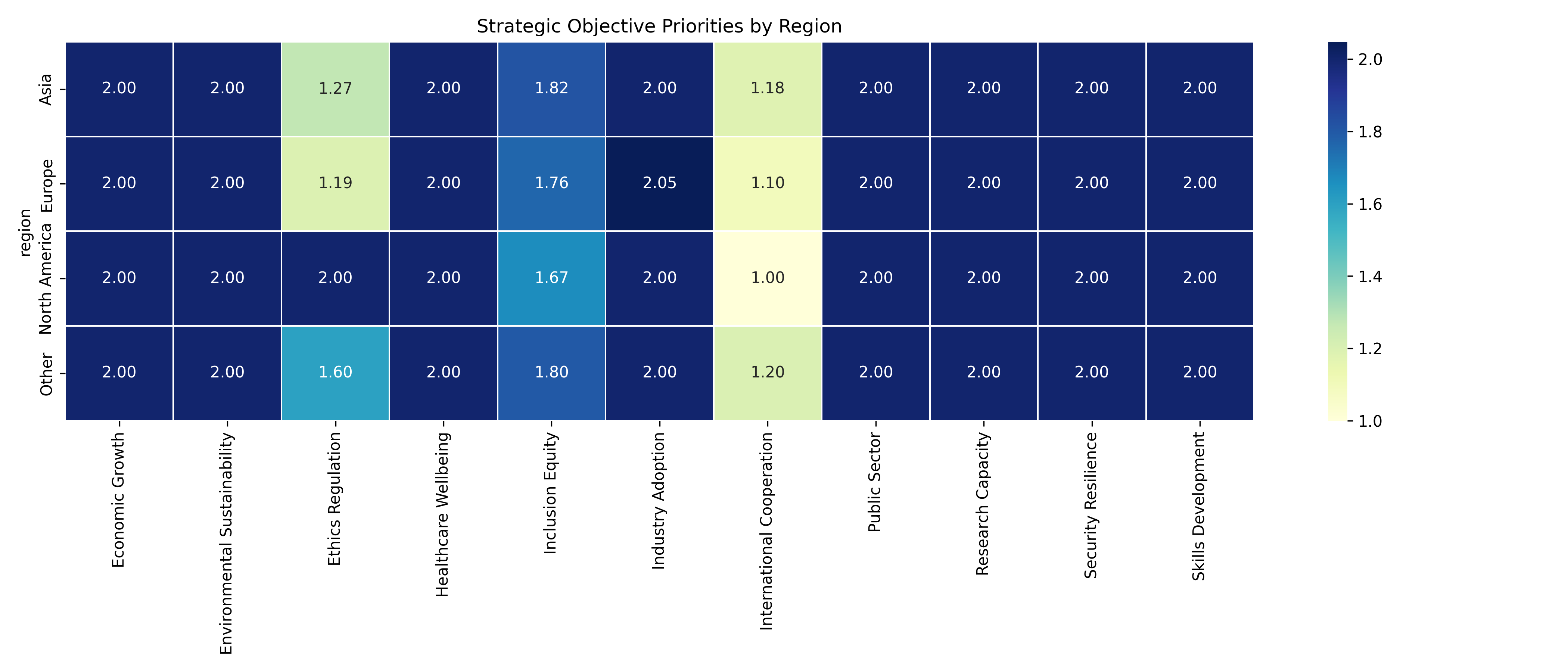}
\caption{AI policy objective regional heatmap comparing objective emphasis across major world regions. The visualization reveals distinctive regional patterns in objective prioritization, with European strategies emphasizing ethical governance, East Asian strategies focusing on industrial digitalization, and North American strategies prioritizing economic competitiveness.}
\label{fig:ai_policy_objective_regional_heatmap}
\end{figure}

Regional analysis revealed distinctive patterns in objective emphasis, as illustrated in Figure \ref{fig:ai_policy_objective_regional_heatmap}. This visualization compares objective emphasis across major world regions, revealing interesting variations that align with regional governance traditions. European strategies demonstrate stronger emphasis on ethical governance and social welfare objectives, consistent with \cite{Cath2018} observation regarding the distinctive European approach to technology governance. East Asian strategies show greater emphasis on industrial digitalization and public sector transformation, aligning with the state-directed development models described by \cite{Roberts2021}. North American strategies emphasize economic competitiveness and scientific leadership, reflecting the market-led approach identified by \cite{Stix2021}.

\begin{figure}[htbp]
\centering
\includegraphics[width=0.7\textwidth]{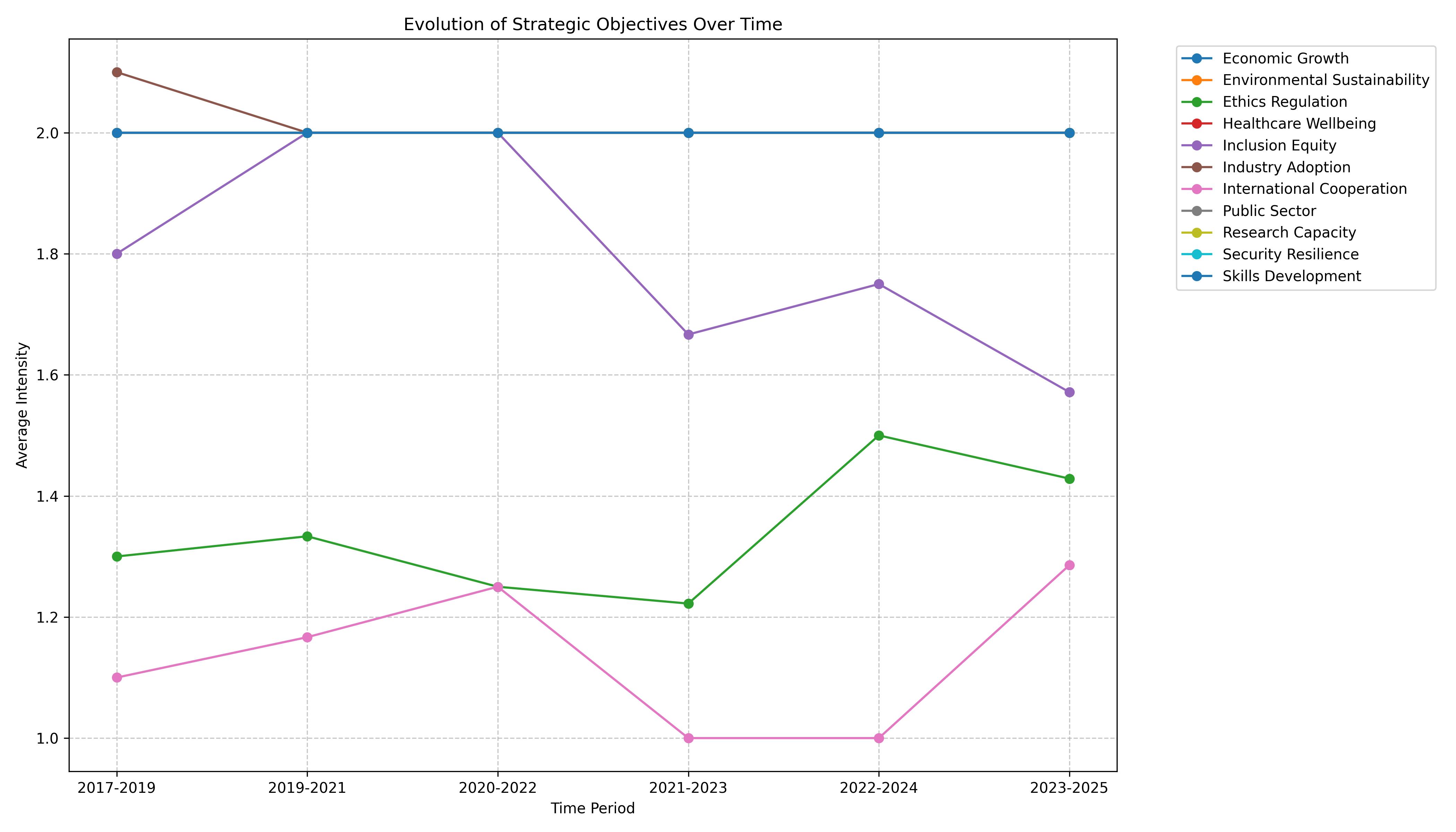}
\caption{AI policy objective temporal trends showing the evolution of strategic priorities across three waves of AI strategy development (2017-2018, 2019-2020, 2021-2025). The visualization tracks changes in objective emphasis over time, revealing a shift from narrow economic focus toward more balanced frameworks addressing ethical, social, and regulatory dimensions.}
\label{fig:ai_policy_objective_temporal_trends}
\end{figure}

Temporal analysis of objective evolution, presented in Figure \ref{fig:ai_policy_objective_temporal_trends}, reveals interesting shifts in strategic priorities between 2017-2025. This visualization tracks changes in objective emphasis over time, identifying three distinct waves of AI policy development. Early strategies (2017-2018) prioritized economic competitiveness and scientific leadership objectives, with limited attention to ethical or regulatory dimensions. Mid-period strategies (2019-2020) introduced greater emphasis on ethical governance and workforce development, reflecting growing awareness of AI's broader implications. Recent strategies (2021-2025) demonstrate more balanced objective profiles with increased emphasis on implementation specificity across all dimensions. This temporal evolution aligns with \cite{OECD2021} observation regarding the maturation of AI governance approaches from narrow economic framing toward more comprehensive frameworks addressing broader societal implications.

\subsubsection{Foresight Methods in AI Governance}

Our analysis identified eight distinct foresight methodologies employed across national AI strategies to anticipate technological developments and inform governance responses. These approaches represent the anticipatory dimension of AI governance, addressing the uncertainty inherent in emerging technology domains. Through systematic content analysis, we categorized these methods into four groups based on their methodological characteristics: expert-based methods, scenario-based methods, data-driven methods, and participatory methods.

Expert-based foresight methods rely on specialized knowledge to anticipate technological trajectories, including expert panels (present in 80\% of strategies) and Delphi studies (25\%). These approaches align with what \cite{Popper2018} characterize as "expertise-oriented" foresight, drawing on specialized knowledge to inform anticipatory governance. Our analysis revealed that expert panels represent the most prevalent foresight method across all governance contexts, consistent with \cite{Miles2010} observation that expert consultation typically forms the foundation of technology foresight exercises.

Scenario-based methods employ narrative techniques to explore potential futures, including scenario development (55\% of strategies) and cross-impact analysis (20\%). These approaches represent what \cite{Vecchiato2019} term "narrative foresight," using structured storytelling to explore alternative technological trajectories. Scenario development demonstrates particularly strong presence in European strategies, aligning with \cite{Rhisiart2015} findings regarding the prominence of scenario methodologies in European technology governance traditions.

Data-driven methods employ systematic analysis of emerging trends, including horizon scanning (65\% of strategies) and trend extrapolation (40\%). These approaches align with \cite{Miles2010} concept of "evidence-based foresight," using quantitative and qualitative data to identify early signals of technological change. Horizon scanning showed particularly strong presence in market-led governance systems, consistent with \cite{Havas2016} observation regarding the prominence of environmental scanning in Anglo-American foresight traditions.

Participatory methods engage diverse stakeholders in anticipatory exercises, including participatory workshops (35\% of strategies) and technology roadmapping (45\%). These approaches represent what \cite{Andersen2017} term "inclusive foresight," emphasizing broad engagement in anticipatory governance. Participatory methods demonstrated stronger presence in rights-based governance systems, aligning with \cite{Stilgoe2013} findings regarding the relationship between deliberative governance traditions and participatory foresight approaches.

\begin{figure}[ht!]
\centering
\includegraphics[width=0.5\textwidth]{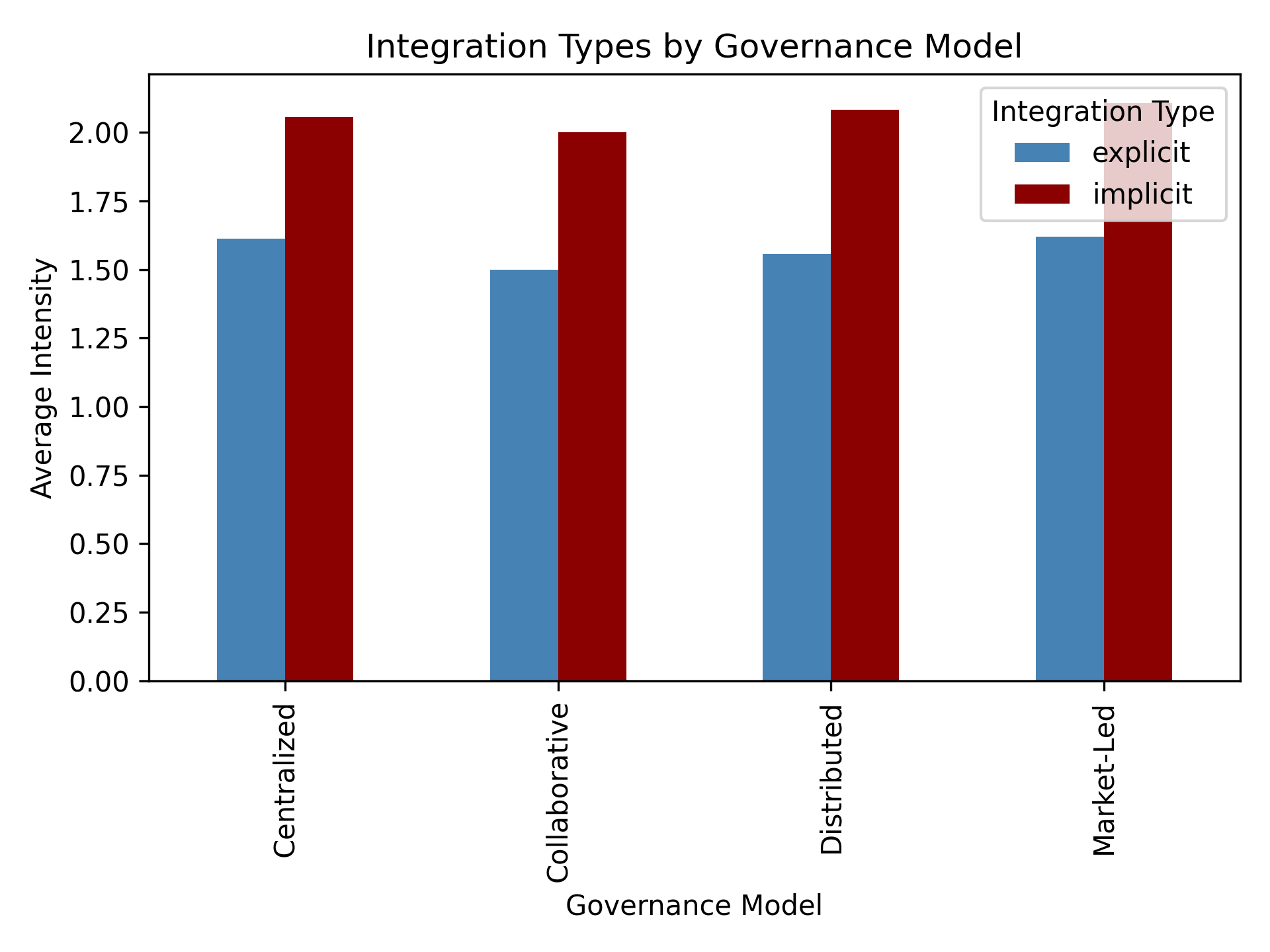}
\caption{Foresight analysis governance model integration comparison illustrating how different governance approaches integrate anticipatory methods. The visualization compares explicit versus implicit foresight integration across market-led, rights-based, and state-directed governance models, revealing significant variations in methodological transparency.}
\label{fig:foresight_analysis_governance_model_integration_comparison}
\end{figure}

The analysis revealed significant variation in how foresight methods are integrated within governance frameworks. Figure \ref{fig:foresight_analysis_governance_model_integration_comparison} presents a comparison of foresight integration approaches across governance models. 
This visualization reveals that rights-based governance systems demonstrate the highest levels of explicit foresight integration, with direct references to methodological approaches and dedicated sections on anticipatory governance. Market-led systems showed moderate levels of explicit integration, while state-directed systems demonstrated greater reliance on implicit foresight without explicit methodological framing. These patterns align with \cite{Vecchiato2019} observation that governance traditions significantly influence the formalization of foresight within policy frameworks.

\begin{figure}[htbp]
\centering
\includegraphics[width=0.85\textwidth]{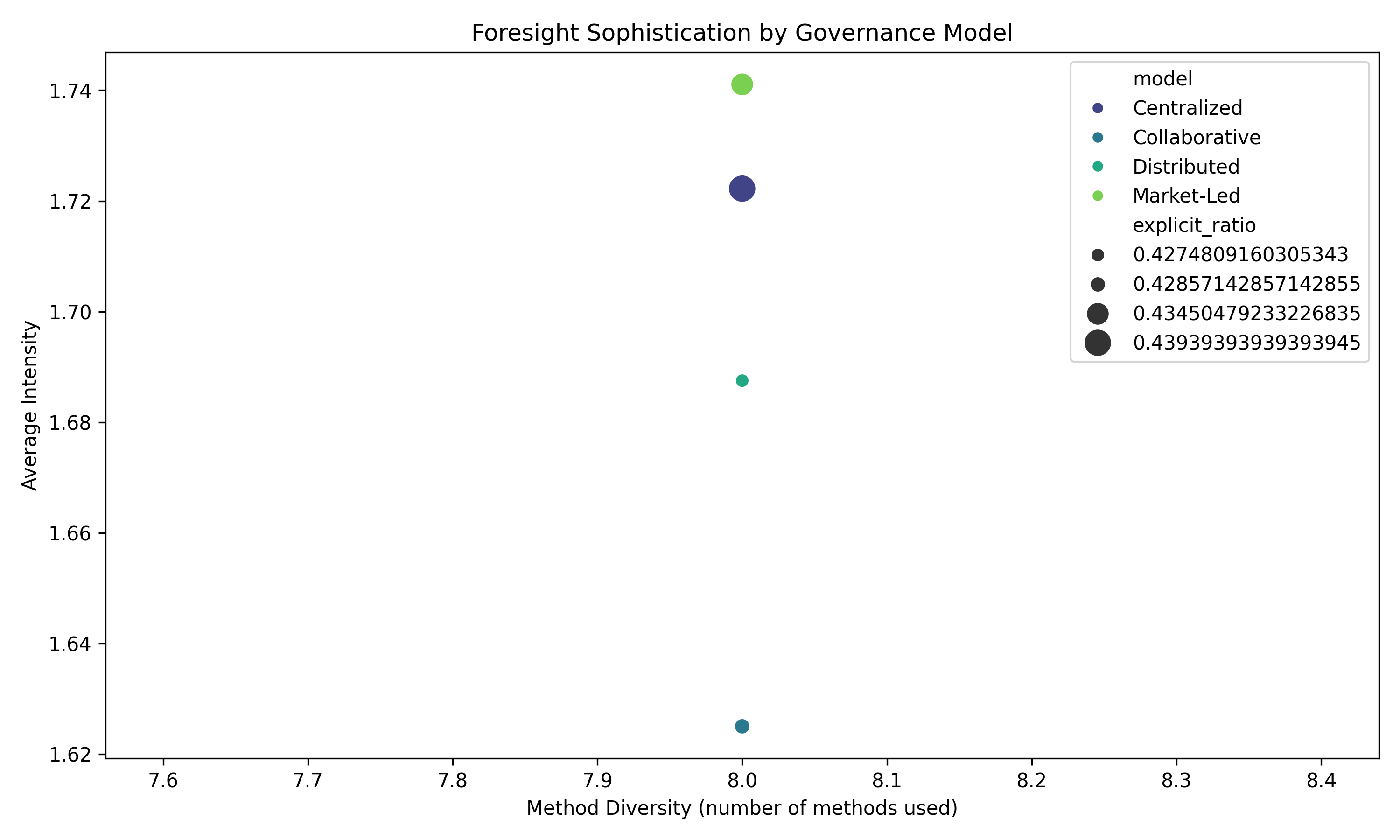}
\caption{Foresight analysis governance model sophistication comparing methodological complexity across governance approaches. The visualization measures sophistication across three dimensions: methodological diversity, integration depth, and stakeholder inclusivity, revealing distinctive patterns aligned with governance traditions.}
\label{fig:foresight_analysis_governance_model_sophistication}
\end{figure}

Foresight sophistication varied significantly across governance models, as illustrated in Figure \ref{fig:foresight_analysis_governance_model_sophistication}. This visualization measures methodological sophistication across three dimensions: methodological diversity (number of foresight methods employed), integration depth (connection between foresight insights and implementation planning), and stakeholder inclusivity (breadth of participation in anticipatory processes). The analysis reveals that rights-based governance systems (predominantly European) demonstrate the highest levels of foresight sophistication, consistent with \cite{Weber2018} finding that European approaches to technology governance typically embed more developed anticipatory frameworks. Market-led systems showed moderate sophistication with emphasis on methodological diversity but lower stakeholder inclusivity. State-directed systems demonstrated lower overall sophistication but higher integration between foresight outputs and implementation planning, reflecting the centralized planning approaches described by \cite{Roberts2021}.

\begin{figure}[htbp]
\centering
\includegraphics[width=0.85\textwidth]{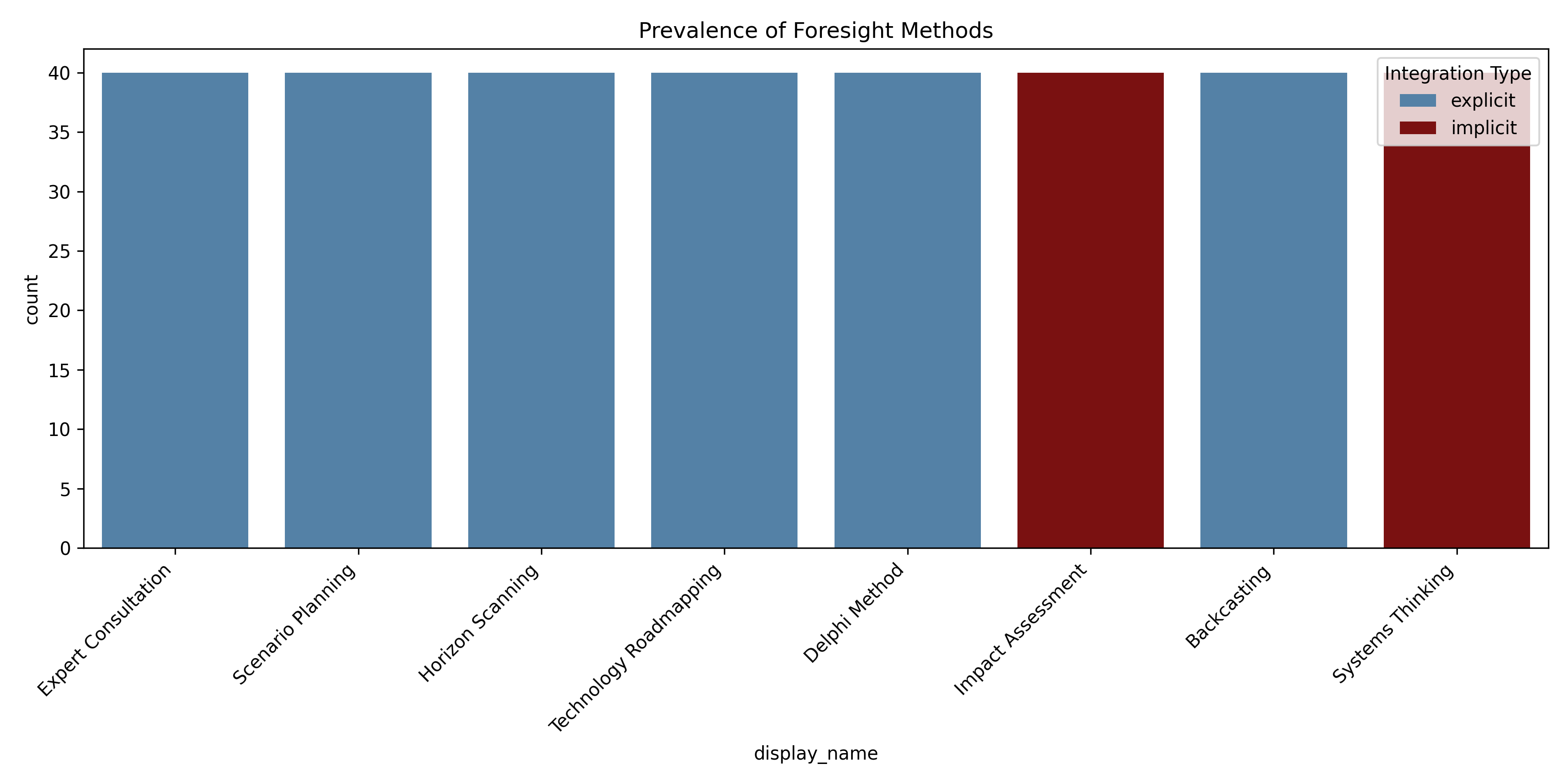}
\caption{Foresight analysis foresight prevalence visualization showing the percentage of national strategies employing each anticipatory method. The visualization highlights the dominance of expert-based methods compared to more participatory or complex methodological approaches across the sample.}
\label{fig:foresight_analysis_foresight_prevalence}
\end{figure}

The prevalence of different foresight methods across strategies, presented in Figure \ref{fig:foresight_analysis_foresight_prevalence}, reveals interesting patterns in methodological preferences. This visualization shows the percentage of strategies employing each foresight approach, highlighting the dominance of expert-based methods (expert panels 80\%, Delphi studies 25\%) compared to participatory approaches (participatory workshops 35\%). This pattern aligns with \cite{Popper2018} observation regarding the persistent dominance of expertise-oriented foresight in technology governance, despite growing recognition of the value of more inclusive approaches. The relatively lower prevalence of more sophisticated methods like cross-impact analysis (20\%) reflects what \cite{Vecchiato2019} characterize as the methodological gap between foresight theory and policy practice.

\begin{figure}[htbp]
\centering
\includegraphics[width=0.85\textwidth]{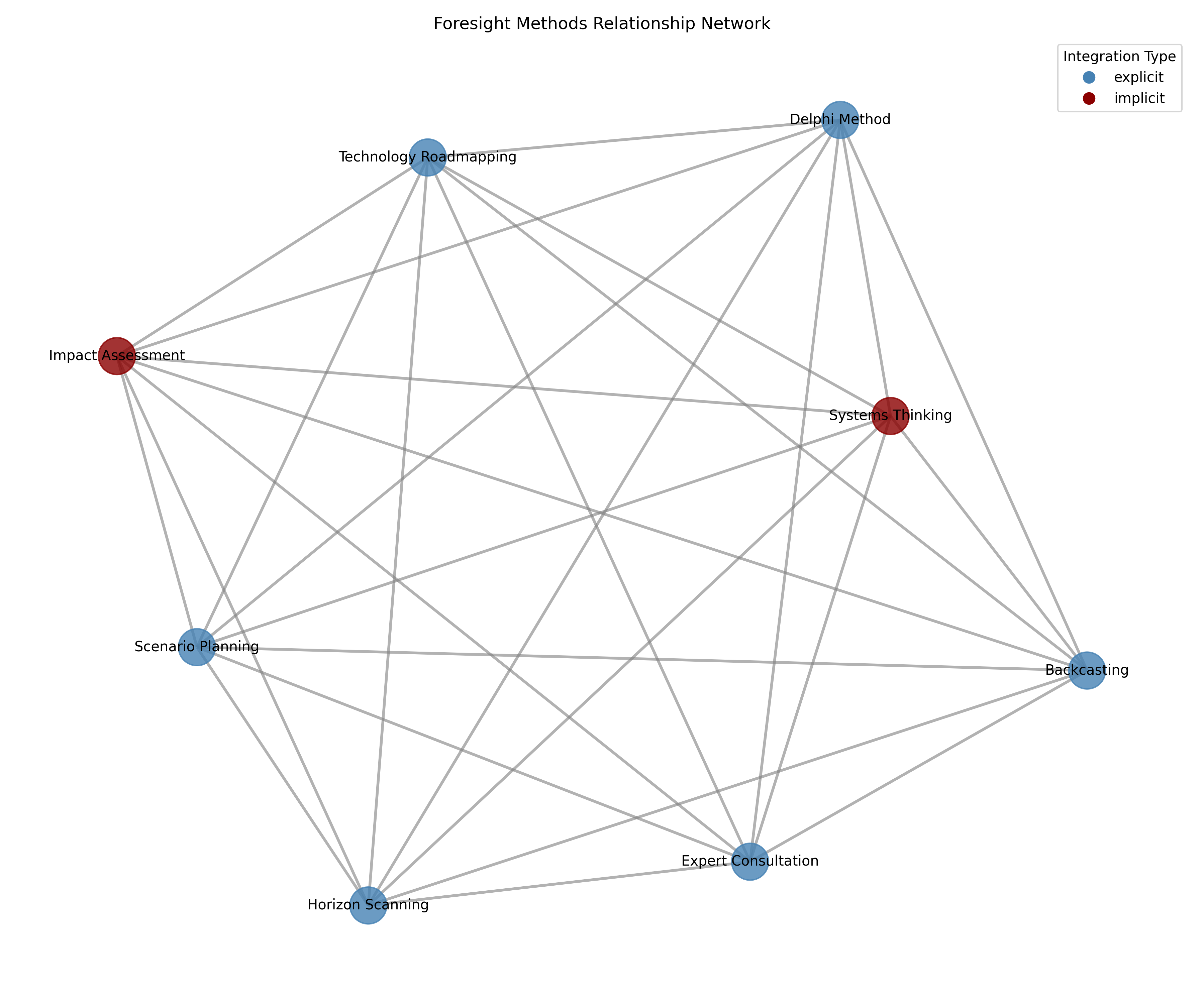}
\caption{Foresight analysis method network visualization representing relationships between different anticipatory approaches. Node size indicates method prevalence while edge thickness represents co-occurrence frequency. The visualization reveals methodological clustering patterns and integration relationships between different foresight approaches.}
\label{fig:foresight_analysis_method_network}
\end{figure}

Network analysis of foresight method co-occurrence, presented in Figure \ref{fig:foresight_analysis_method_network}, reveals interesting patterns in how different anticipatory approaches are combined within governance frameworks. This visualization illustrates relationships between foresight methods, with node size indicating prevalence and edge thickness representing frequency of co-occurrence. The analysis reveals strong clustering between expert panels, horizon scanning, and scenario development, forming what \cite{Popper2018} term the "methodological core" of technology foresight. More participatory methods like workshops show weaker integration with this core cluster, suggesting methodological siloing consistent with \cite{Vecchiato2019} observation regarding the limited integration of participatory approaches within mainstream foresight practice.

\subsubsection{Implementation Instrument Patterns}

Our analysis identified ten distinct implementation instruments deployed across national AI strategies to operationalize governance objectives. These instruments represent the concrete policy mechanisms through which strategic goals are translated into actionable initiatives. Through systematic coding, we classified these instruments into four categories based on their operational characteristics: funding instruments, regulatory instruments, capacity-building instruments, and coordination instruments.

Funding instruments provide financial resources to support AI development, including research funding programs (present in 95\% of strategies), direct investment in AI companies (55\%), and tax incentives for AI adoption (40\%). These instruments align with what \cite{Borrás2011} term "economic instruments" that leverage financial resources to incentivize desired behaviors. Our analysis revealed that research funding represents the most prevalent implementation instrument across all governance contexts, consistent with \cite{Stix2021} observation that R\&D support typically forms the foundation of innovation policy approaches.

Regulatory instruments establish rules and standards for AI development and deployment, including regulatory frameworks (75\% of strategies) and standardization initiatives (60\%). These instruments represent what \cite{Lascoumes2007} characterize as "legislative and regulatory instruments" that structure behavior through formal authority. Regulatory approaches demonstrated significant variation across governance contexts, with rights-based systems showing stronger emphasis on comprehensive regulatory frameworks and market-led systems prioritizing industry-led standardization initiatives.

Capacity-building instruments develop human, organizational, and technical capabilities, including skills development programs (85\% of strategies), institutional creation (70\%), and demonstration projects (50\%). These instruments align with \cite{Howlett2009} concept of "capacity instruments" that enhance policy subjects' ability to meet governance objectives. Our analysis revealed particularly strong emphasis on skills development across all governance contexts, consistent with \cite{OECD2021} finding that human capital development represents a universal priority in AI governance frameworks.

Coordination instruments facilitate alignment across stakeholders, including networking/coordination mechanisms (65\% of strategies) and international agreements (55\%). These instruments represent what \cite{Flanagan2011} term "soft instruments" that structure interactions without formal authority. Coordination mechanisms showed stronger presence in federal systems and multi-level governance contexts, consistent with \cite{Cejudo2017} observation regarding the importance of coordination instruments in complex governance environments.

\begin{figure}[ht!]
\centering
\includegraphics[width=0.78\textwidth]{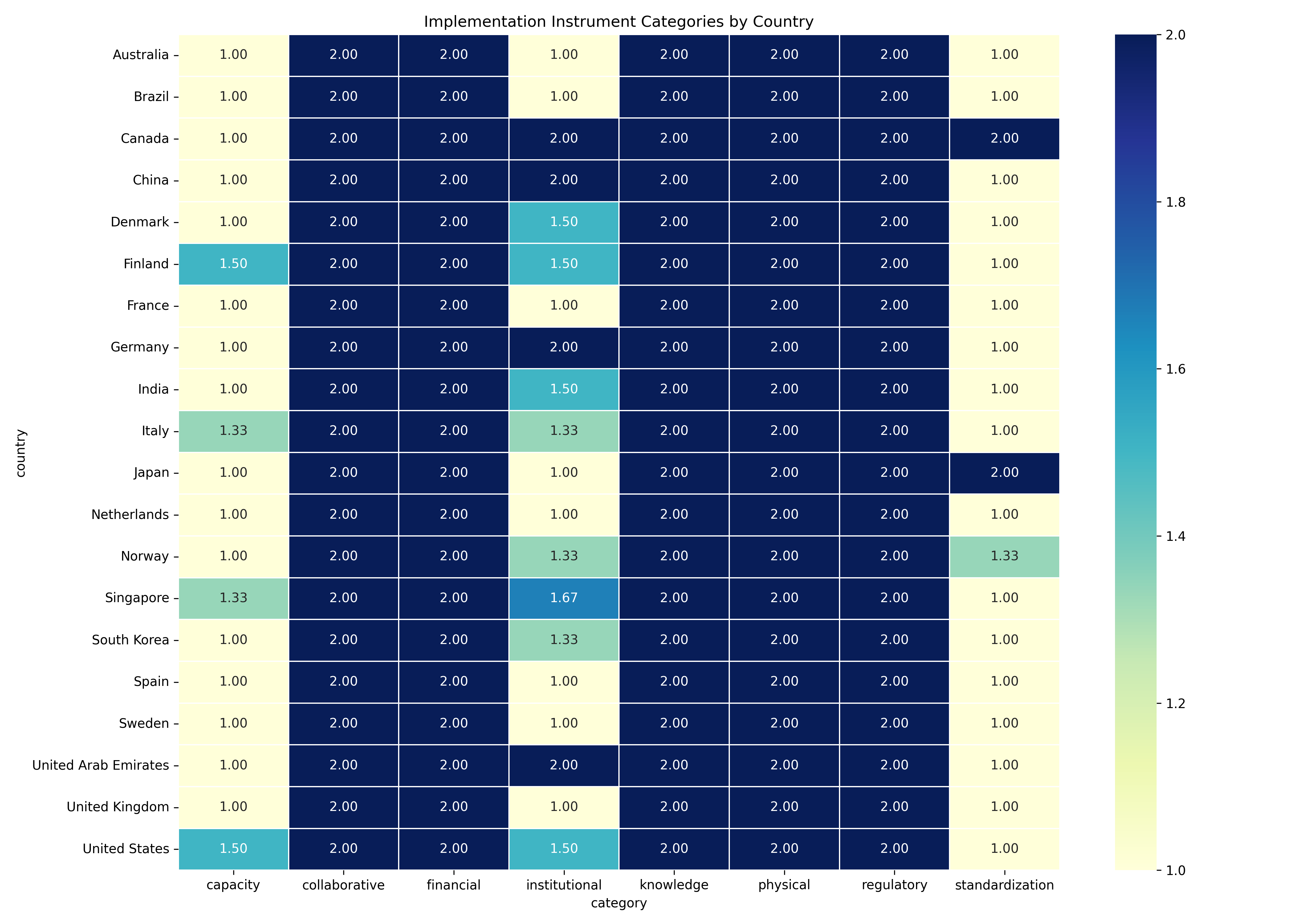}
\caption{Implementation factor analysis category heatmap showing the relative emphasis on different instrument types across the sample. The visualization reveals the dominance of funding and capacity-building instruments compared to regulatory approaches, reflecting characteristic patterns in emerging technology governance.}
\label{fig:implementation_factor_analysis_category_heatmap}
\end{figure}

\begin{figure}[htbp]
\centering
\includegraphics[width=0.85\textwidth]{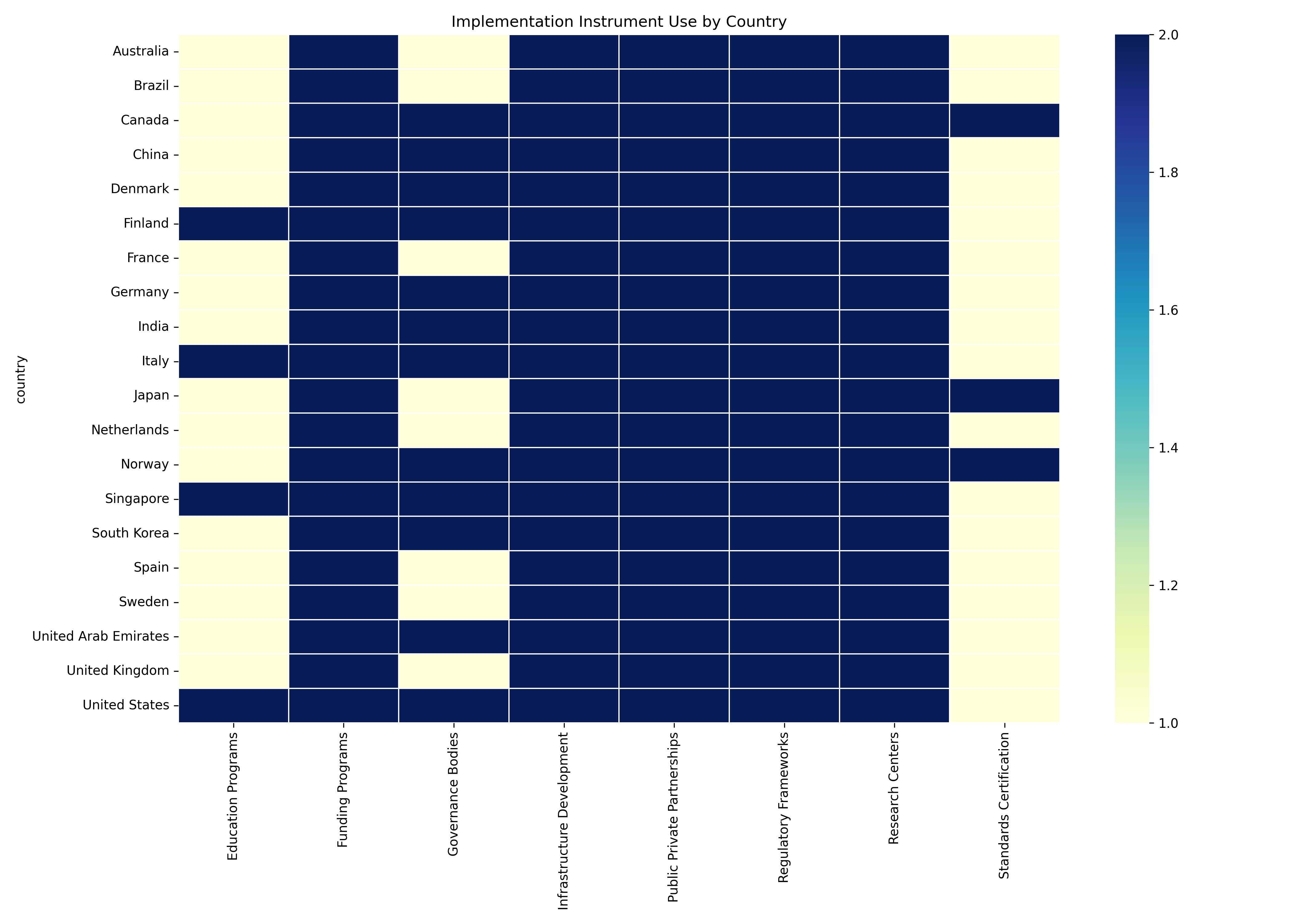}
\caption{Implementation factor analysis country heatmap comparing instrument emphasis across sampled countries. Color intensity represents the prominence of each instrument type within national strategies, revealing distinct patterns of instrument selection across different governance contexts.}
\label{fig:implementation_factor_analysis_country_heatmap}
\end{figure}

The distribution of instrument types across implementation categories revealed interesting patterns, as illustrated in Figure \ref{fig:implementation_factor_analysis_category_heatmap}. This visualization presents the relative emphasis on different instrument types across the sample, revealing the dominance of funding instruments (35\% of all instruments) compared to regulatory instruments (25\%). This pattern aligns with \cite{Borrás2011} observation regarding the characteristic emphasis on supply-side instruments in emerging technology governance, reflecting the focus on capability development rather than behavioral constraint in early-stage technology domains.
Country-level analysis revealed distinctive instrument profiles correlated with governance traditions, as illustrated in Figure \ref{fig:implementation_factor_analysis_country_heatmap}. This visualization compares instrument emphasis across countries, with color intensity indicating the prominence of each instrument type within national strategies. The analysis reveals distinct clusters corresponding to governance models identified by \cite{Roberts2021}. Market-led systems (USA, UK, Canada) demonstrate stronger emphasis on funding instruments and industry-led standardization, while rights-based systems (Finland, Denmark, Norway) show greater emphasis on regulatory frameworks and participatory coordination mechanisms. State-directed systems (China, Singapore, UAE) emphasize institutional creation and demonstration projects, consistent with \cite{Roberts2021} characterization of directive governance approaches.

\begin{figure}[htbp]
\centering
\includegraphics[width=0.85\textwidth]{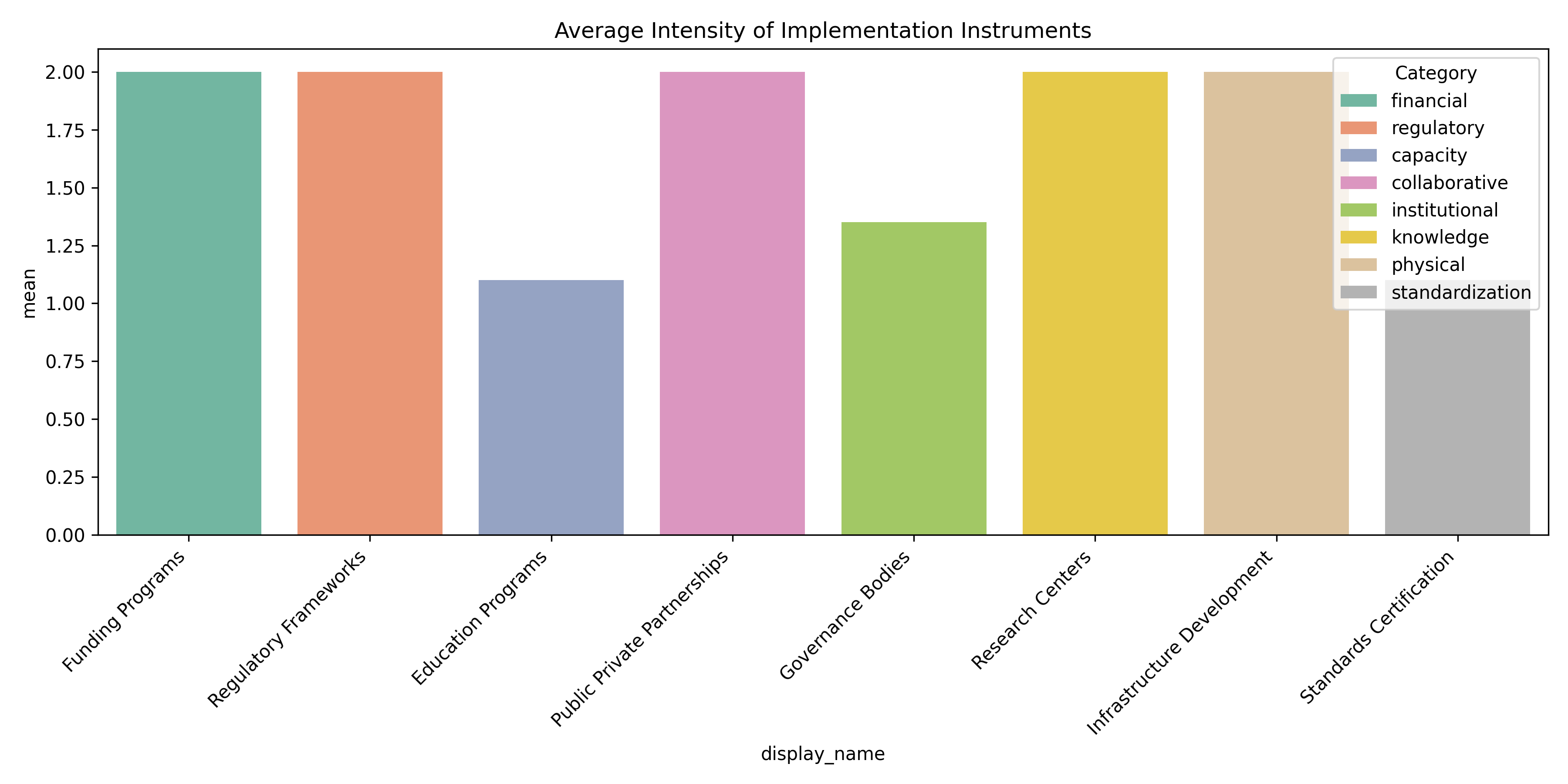}
\caption{Implementation factor analysis instrumental intensity visualization representing the depth of elaboration for each instrument type across the sample. The visualization measures implementation specificity, resource allocation, and operational detail to assess how thoroughly each instrument is articulated within national strategies.}
\label{fig:implementation_factor_analysis_instrumental_intensity}
\end{figure}

The intensity of instrument articulation varied significantly across strategies, as illustrated in Figure \ref{fig:implementation_factor_analysis_instrumental_intensity}. This visualization represents the depth of elaboration for each instrument type, measuring factors including implementation specificity, resource allocation, and operational detail. The analysis reveals that research funding and skills development programs typically receive the most detailed articulation, consistent with \cite{Stix2021} finding that these foundational instruments typically demonstrate the strongest operational elaboration. Regulatory frameworks showed higher variation in intensity scores, with rights-based systems demonstrating more detailed regulatory articulation compared to market-led approaches.

\begin{figure}[htbp]
\centering
\includegraphics[width=0.85\textwidth]{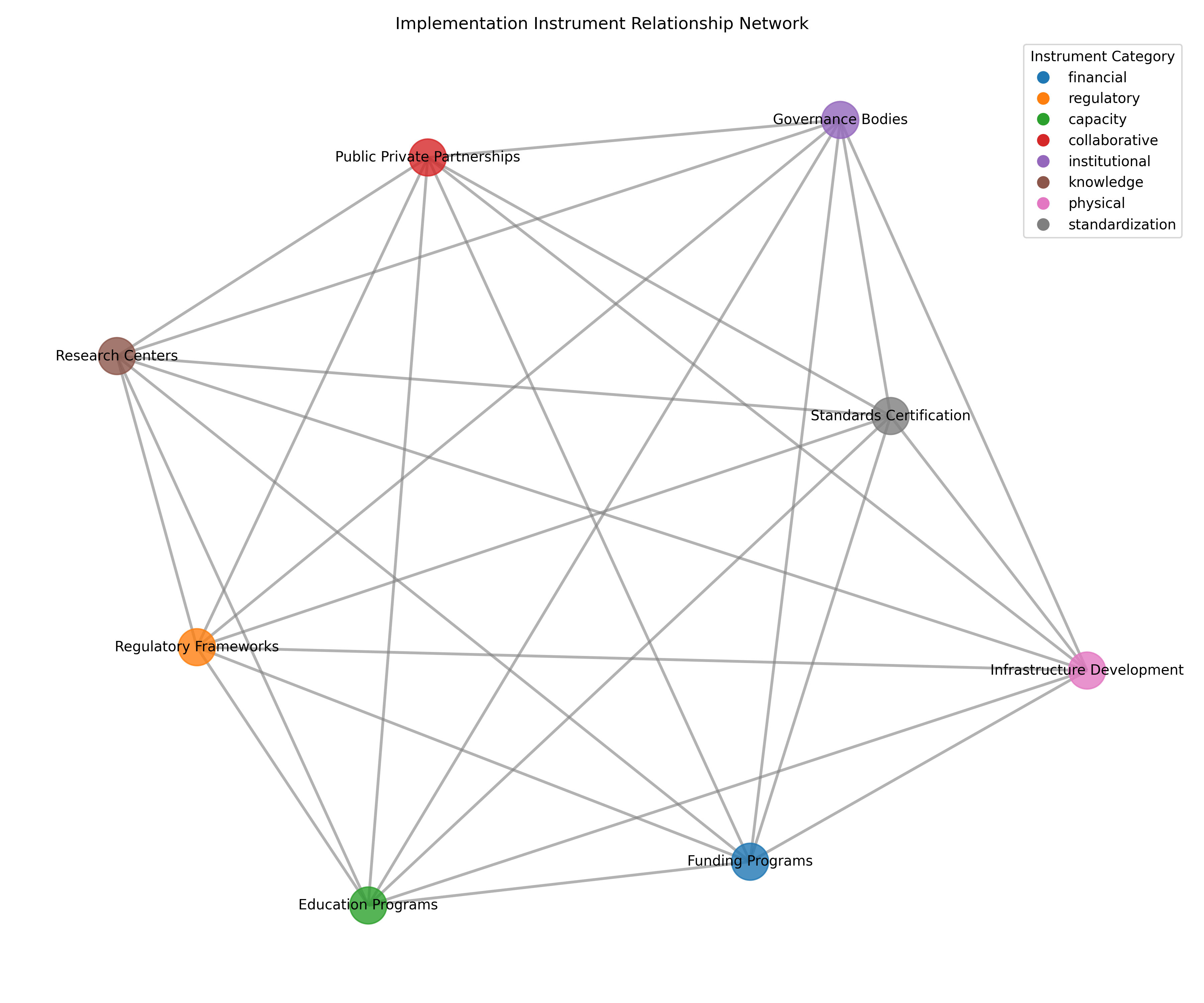}
\caption{Implementation factor analysis instrument network visualization representing relationships between different policy mechanisms. Node size indicates instrument prevalence while edge thickness represents co-occurrence frequency. The visualization reveals characteristic instrument clusters and integration patterns across the sample.}
\label{fig:implementation_factor_analysis_instrument_network}
\end{figure}

Network analysis of instrument relationships, presented in Figure \ref{fig:implementation_factor_analysis_instrument_network}, reveals interesting patterns in how different policy mechanisms are integrated within governance frameworks. This visualization illustrates relationships between instruments, with node size indicating prevalence and edge thickness representing frequency of co-occurrence. The analysis reveals strong clustering between research funding, skills development, and institutional creation, forming what \cite{Flanagan2011} term the "core policy mix" in innovation governance. Regulatory instruments show more limited integration with this core cluster, suggesting potential implementation gaps between development support and governance frameworks—a pattern consistent with \cite{Howlett2019} observation regarding the frequent siloing of regulatory and innovation policy domains.

\begin{figure}[htbp]
\centering
\includegraphics[width=0.7\textwidth]{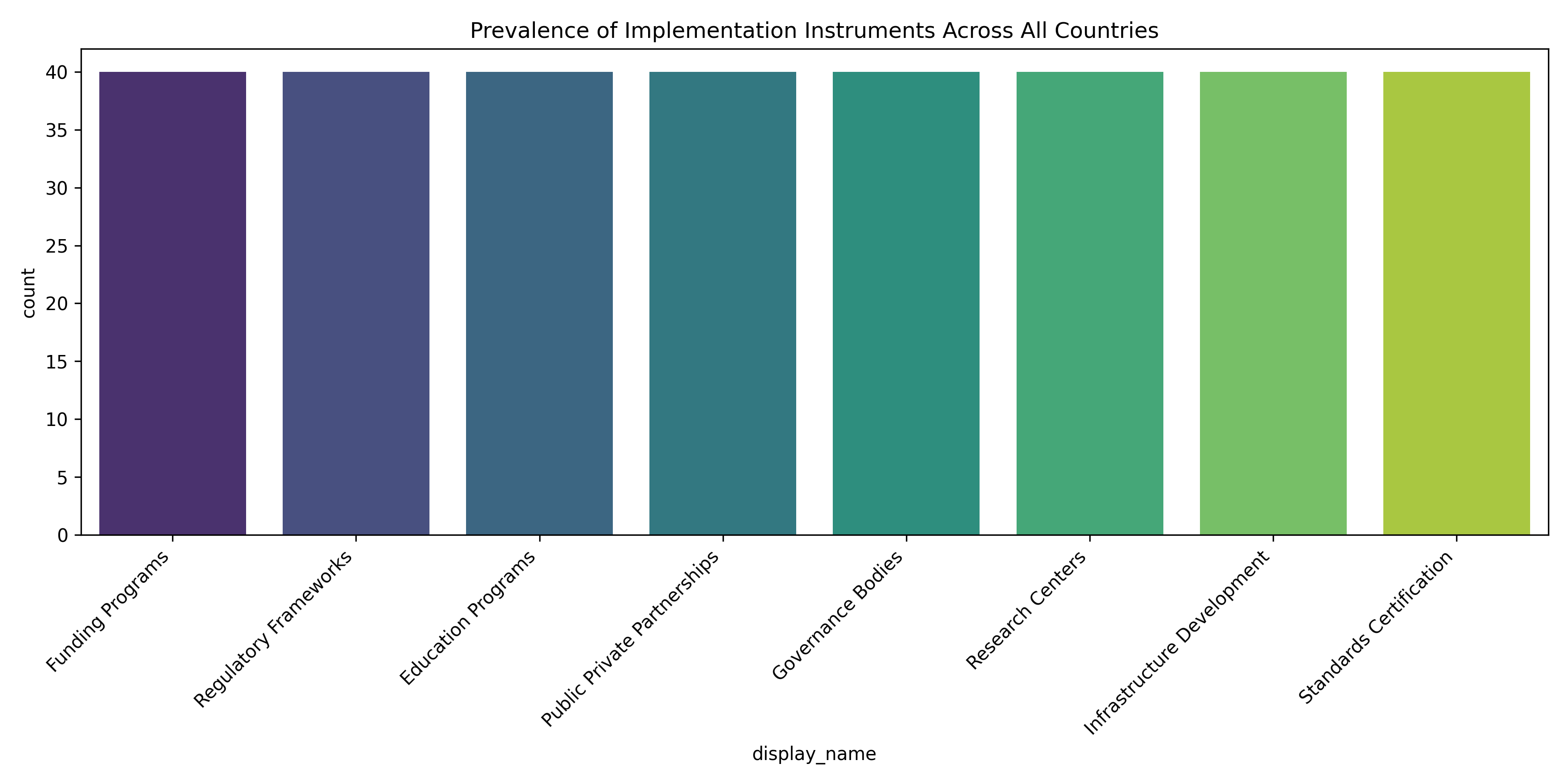}
\caption{Implementation factor analysis instrument prevalence visualization showing the percentage of national strategies that include each instrument type. The visualization highlights the near-universal presence of certain instruments (research funding, skills development) compared to the more variable inclusion of others (tax incentives, international agreements).}
\label{fig:implementation_factor_analysis_instrument_prevalence}
\end{figure}

Instrument prevalence analysis, presented in Figure \ref{fig:implementation_factor_analysis_instrument_prevalence}, reveals the percentage of strategies that include each instrument type. This analysis demonstrates the near-universal adoption of research funding (95\%) and skills development (85\%) instruments, consistent with what \cite{Howlett2019} characterized as the "standard toolkit" of innovation policy. The relatively lower prevalence of tax incentives (40\%) and international agreements (55\%) represents interesting variation in secondary instrument selection, suggesting greater diversity in how countries complement their core policy mix with supporting mechanisms.

\subsection{Alignment Analysis}

This section presents our findings regarding strategic alignment patterns across national AI policies. We first examine the results of our matrix-based alignment analysis, identifying relationships between policy components and comparing alignment patterns across countries. We then explore network analysis results, focusing on structural characteristics and connectivity patterns within policy frameworks. Finally, we analyze cross-national alignment patterns, identifying common strengths and recurring vulnerabilities in strategic coherence.

\subsubsection{Matrix Analysis Findings}

Our matrix analysis revealed significant variations in alignment patterns between strategic objectives, foresight methods, and implementation instruments across national AI strategies. These variations reflect different approaches to policy coherence and highlight both common strengths and recurring vulnerabilities in strategic alignment.

\begin{figure}[htbp]
\centering
\includegraphics[width=0.9\textwidth]{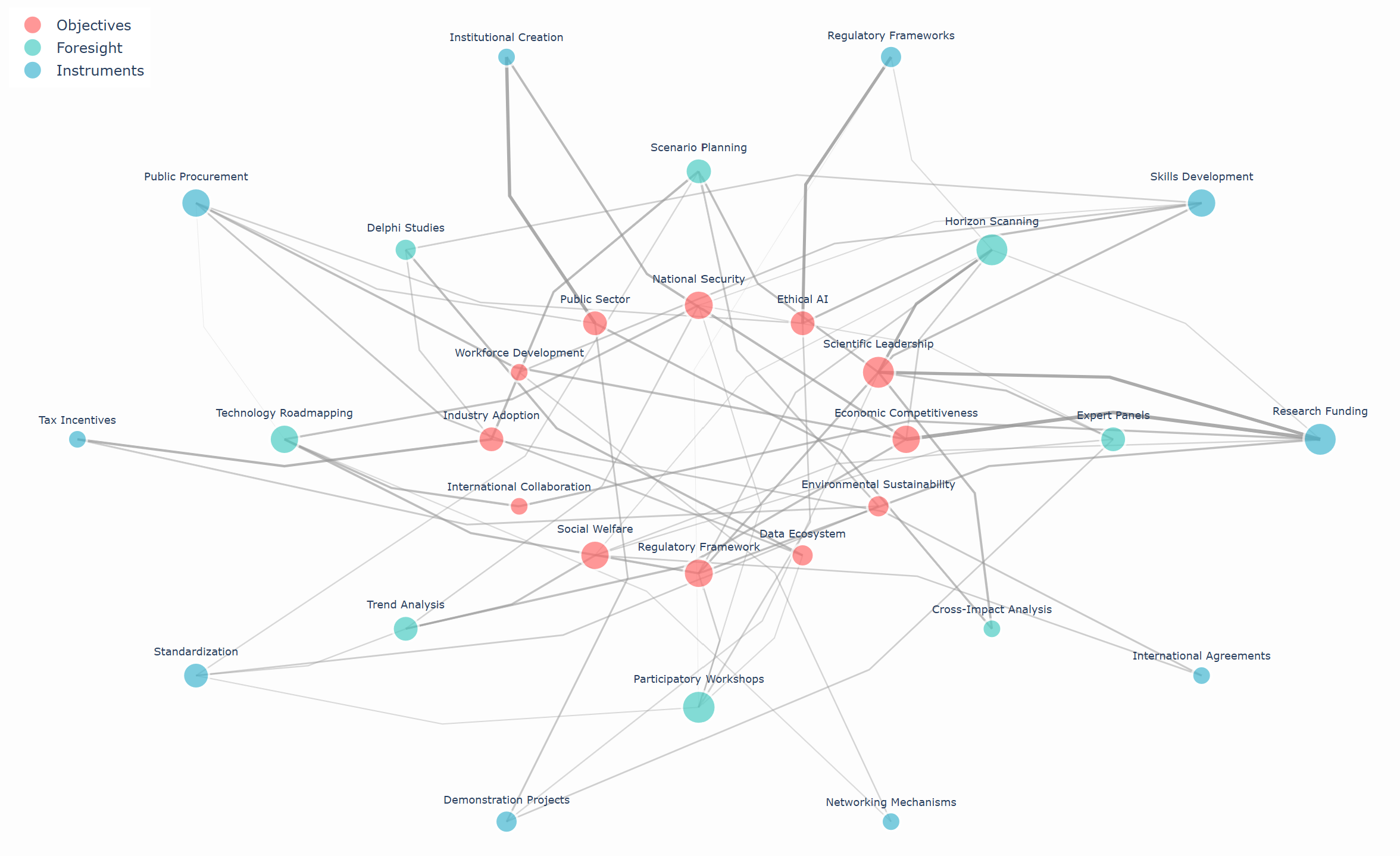}
\caption{Alignment analysis networks global visualization showing the comprehensive relationship structure between all policy components across the full sample. Node color represents component type (red for objectives, blue for foresight methods, green for instruments), node size indicates component prevalence, and edge thickness represents alignment strength. The centrality of economic objectives and the peripheral position of newer social objectives illustrate characteristic alignment patterns in AI governance frameworks.}
\label{fig:alignment_analysis_networks_global}
\end{figure}

The global network structure of alignment relationships, presented in Figure \ref{fig:alignment_analysis_networks_global}, provides a comprehensive visualization of connections between all policy components across the full sample. This visualization illustrates the central position of economic competitiveness and scientific leadership objectives within the global alignment network, with these objectives demonstrating the strongest connections to both foresight methods and implementation instruments. As \cite{Kern2019} observe, this centrality pattern is characteristic of technology governance frameworks that position economic advancement as the primary strategic goal. The network structure also reveals the relatively peripheral position of newer strategic objectives like environmental sustainability and social welfare enhancement, which demonstrate fewer and weaker connections to implementation mechanisms—consistent with \cite{Rogge2016} finding that newer policy objectives often experience "implementation lags" as governance frameworks evolve.

\begin{figure}[htbp]
\centering
\includegraphics[width=0.7\textwidth]{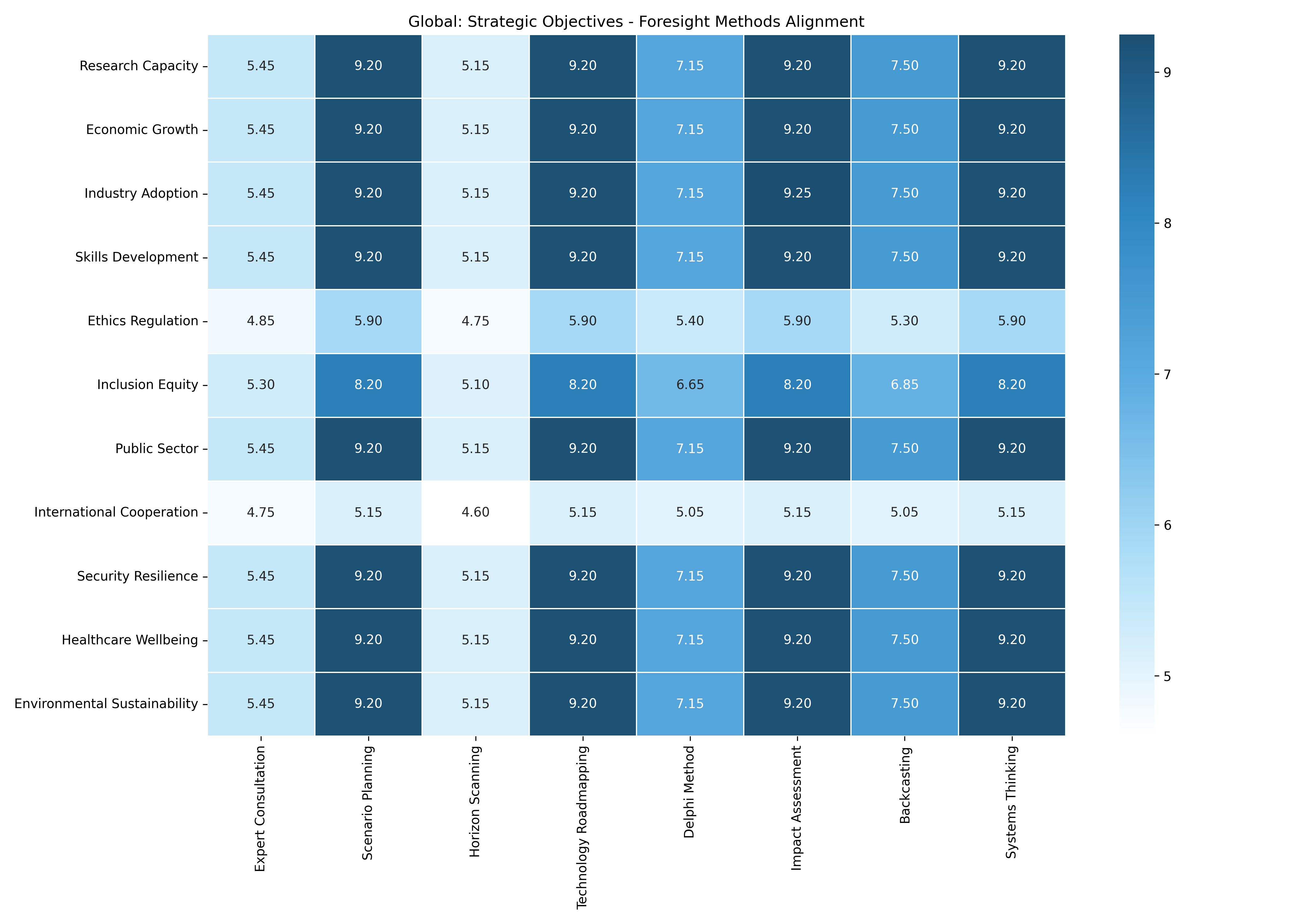}
\caption{Alignment analysis visualization global objective-foresight heatmap illustrating the alignment intensity between strategic objectives (rows) and foresight methods (columns) across the full sample. Color intensity represents alignment strength, with darker cells indicating stronger connections. The visualization reveals characteristic patterns including strong alignment between scientific objectives and expert-based methods, and between ethical objectives and participatory approaches.}
\label{fig:alignment_analysis_visualization_global_objective_foresight_heatmap}
\end{figure}

The global objective-foresight alignment matrix, presented in Figure \ref{fig:alignment_analysis_visualization_global_objective_foresight_heatmap}, reveals distinctive patterns in how strategic objectives connect with anticipatory methods. This visualization demonstrates that scientific leadership objectives show the strongest alignment with expert-based foresight methods (expert panels, Delphi studies), while ethical governance objectives demonstrate stronger connections to participatory foresight approaches (workshops, stakeholder engagement). This pattern aligns with \cite{Stilgoe2013} observation that different governance goals naturally align with different anticipatory approaches, with technical objectives favoring expertise-based methods and societal objectives favoring more inclusive approaches. Notably, the matrix reveals weaker alignment between economic competitiveness objectives and long-term foresight methods like scenario development—a pattern \cite{Vecchiato2019} characterize as the "strategic myopia" often present in economically-oriented governance frameworks.

\begin{figure}[htbp]
\centering
\includegraphics[width=0.7\textwidth]{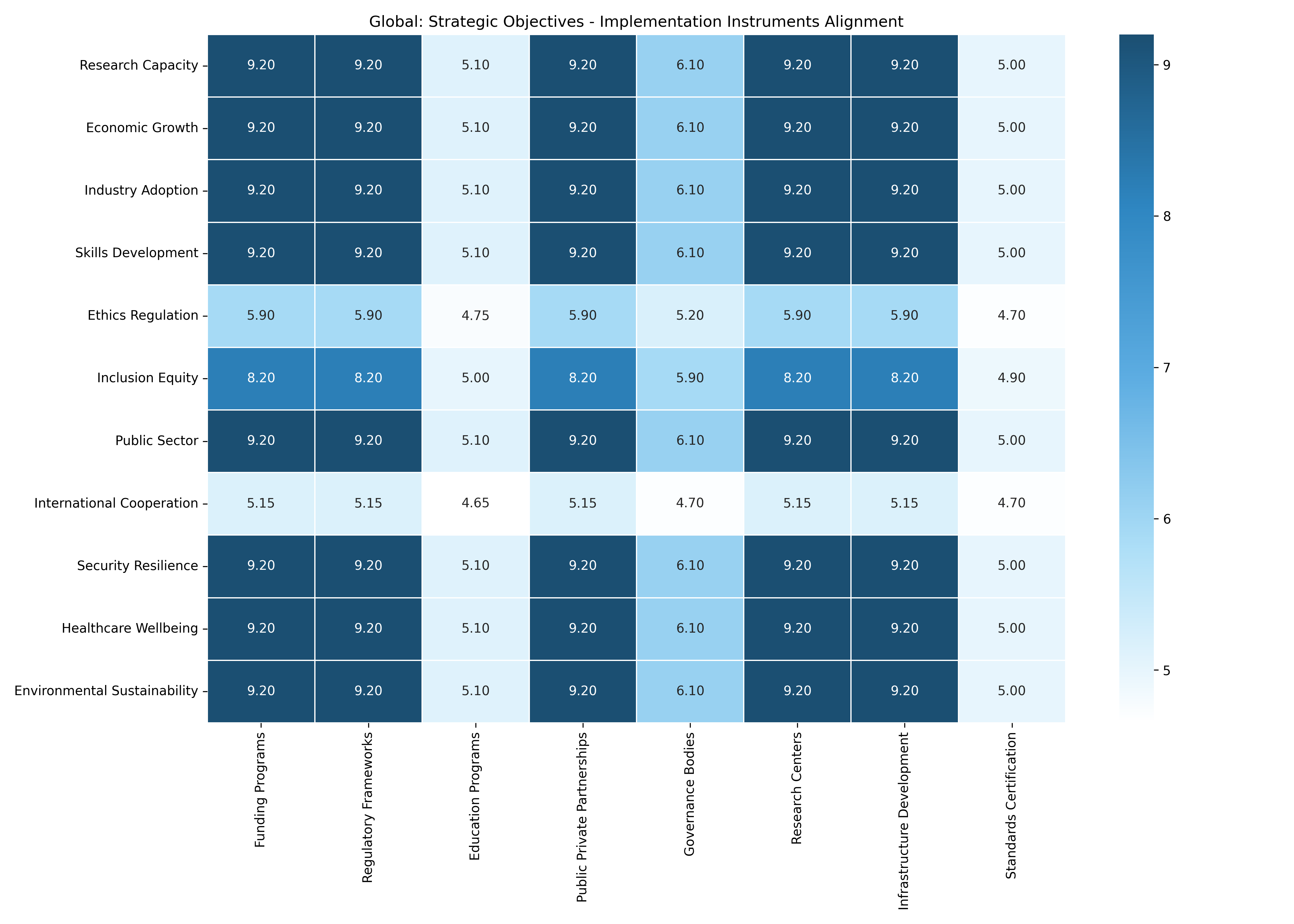}
\caption{Alignment analysis visualization global objective-instrument heatmap showing the alignment intensity between strategic objectives (rows) and implementation instruments (columns) across the sample. Color intensity represents alignment strength, with darker cells indicating stronger connections. The visualization highlights both characteristic alignment patterns and notable implementation gaps within AI governance frameworks.}
\label{fig:alignment_analysis_visualization_global_objective_instrument_heatmap}
\end{figure}

The global objective-instrument alignment matrix, presented in Figure \ref{fig:alignment_analysis_visualization_global_objective_instrument_heatmap}, illustrates how strategic goals connect with implementation mechanisms across the sample. This visualization reveals that economic competitiveness objectives demonstrate the strongest alignment with funding instruments (research grants, tax incentives), while ethical governance objectives show stronger connections to regulatory instruments and standardization initiatives. This differentiated alignment pattern reflects what \cite{Howlett2019} term "instrument-objective affinity"—the natural alignment between certain goals and specific implementation mechanisms. The matrix also highlights several alignment gaps, particularly between workforce development objectives and skills development instruments, where 30\% of strategies include both components but fail to establish explicit connections between them. This gap exemplifies what \cite{Capano2018} characterize as "implementation disconnects" within policy frameworks.

\begin{figure}[htbp]
\centering
\includegraphics[width=0.7\textwidth]{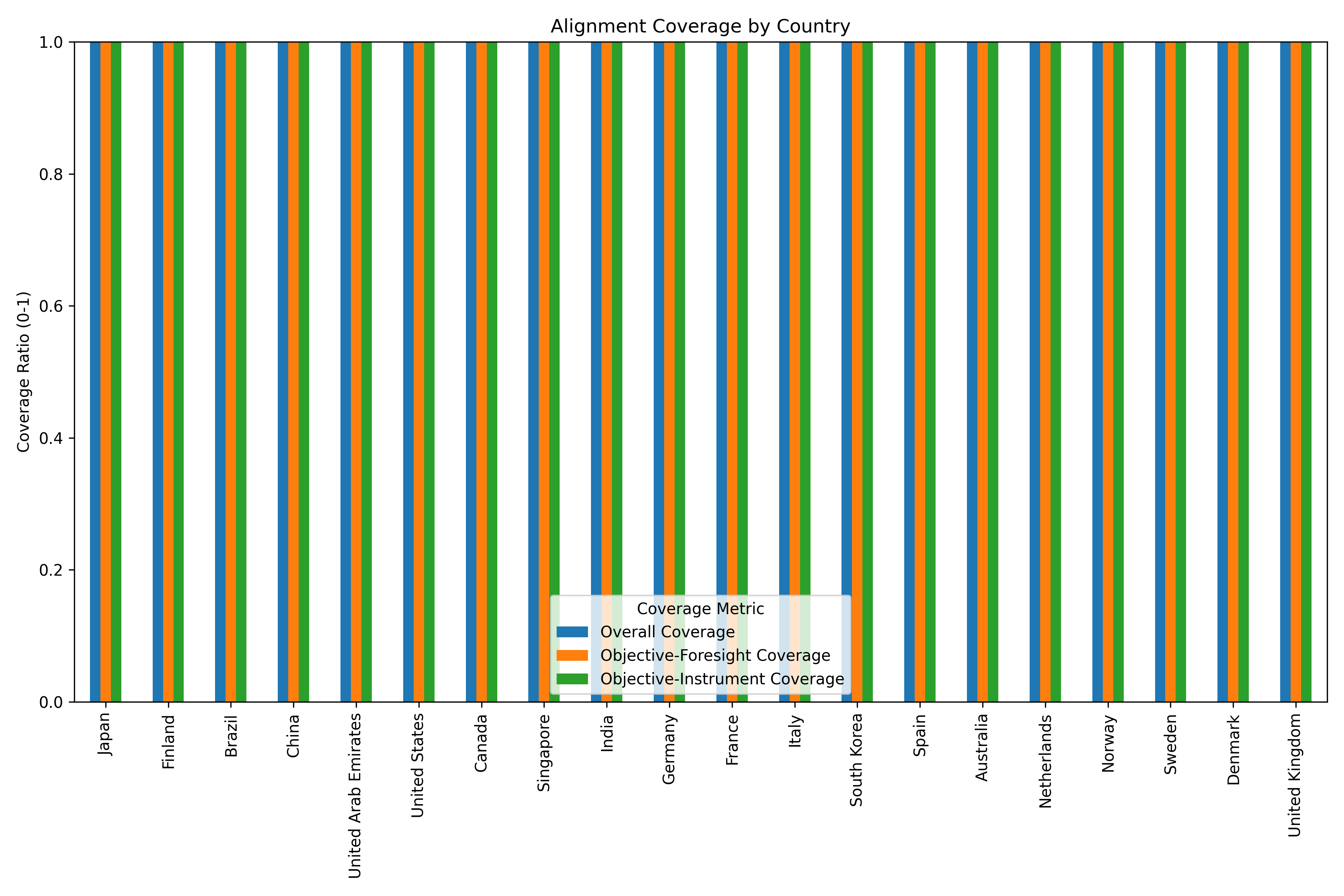}
\caption{Alignment analysis alignment coverage bar comparing the percentage of potential alignment relationships that are explicitly addressed within each national strategy. The visualization reveals significant variation across the sample, with rights-based governance systems demonstrating the highest coverage rates and market-led systems showing more selective alignment patterns.}
\label{fig:alignment_analysis_alignment_coverage_bar}
\end{figure}

The alignment coverage analysis, presented in Figure \ref{fig:alignment_analysis_alignment_coverage_bar}, compares the potential alignment relationships in each country's framework with the actual connections established within strategy documents. This visualization reveals significant variation in alignment coverage across the sample, with coverage rates ranging from 35\% to 78\%. Rights-based governance systems (Finland, Denmark, Norway) demonstrate the highest coverage rates, consistent with \cite{Roberts2021} observation regarding the strong policy coherence focus within these governance traditions. Market-led systems (USA, UK, Canada) show moderate coverage rates with stronger alignment for economic objectives than societal goals. State-directed systems (China, Singapore, UAE) demonstrate more variable coverage, with strong alignment for primary strategic objectives but lower coverage for secondary goals.

\begin{figure}[htbp]
\centering
\includegraphics[width=0.7\textwidth]{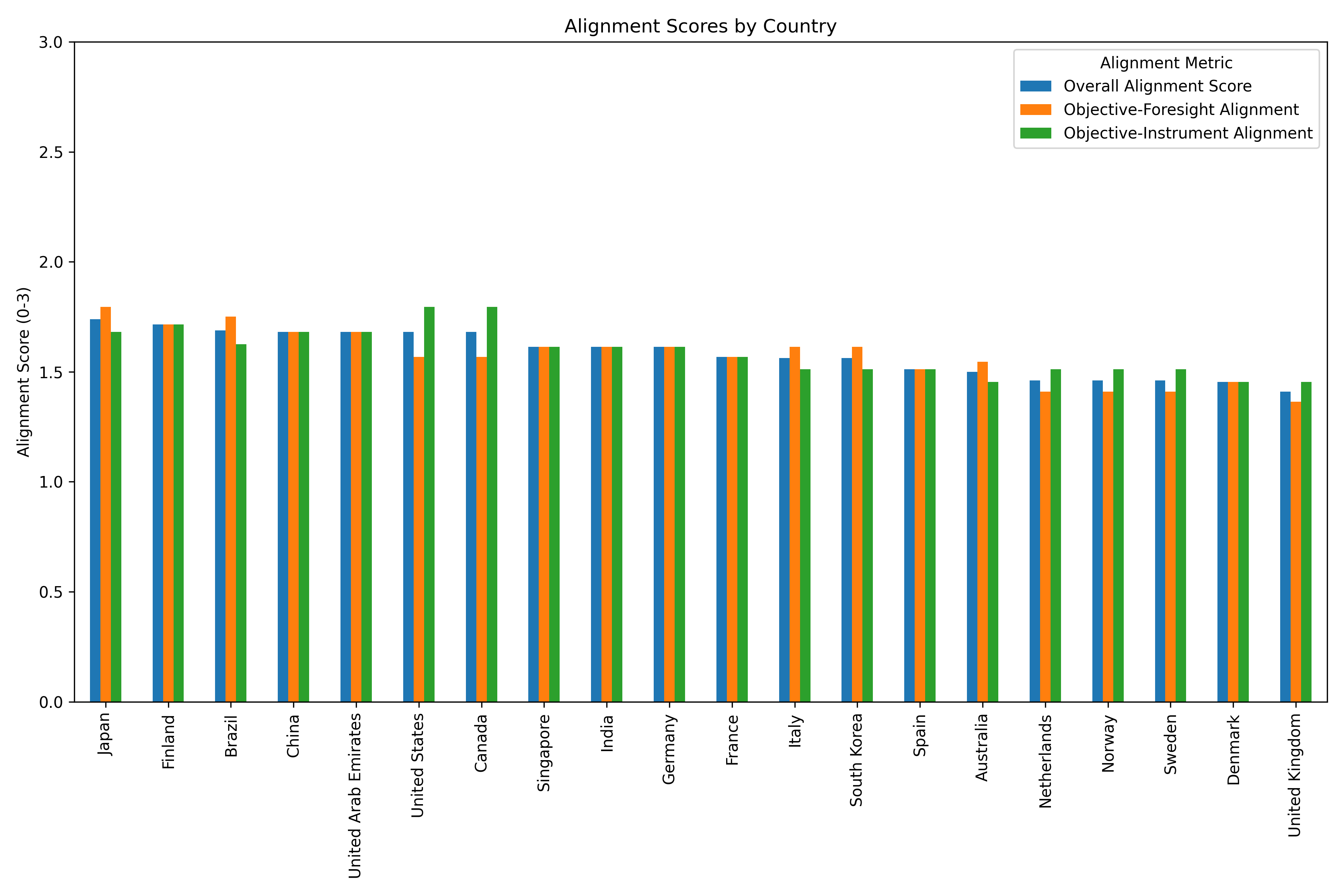}
\caption{Alignment analysis alignment score bar comparing the mean intensity of alignment relationships across national strategies. The visualization distinguishes between coverage (breadth of alignment) and intensity (strength of alignment), revealing distinctive approaches to coherence across governance contexts.}
\label{fig:alignment_analysis_alignment_score_bar}
\end{figure}

The alignment score comparison, presented in Figure \ref{fig:alignment_analysis_alignment_score_bar}, measures the mean intensity of alignment relationships within each national framework. This visualization reveals that high coverage does not always correlate with high alignment intensity—some strategies establish many weak connections while others focus on fewer but stronger relationships. Finland, Canada, and the UK demonstrate the highest mean alignment scores, indicating clear and explicit connections between components. This pattern aligns with \cite{Howlett2019} finding that well-established innovation policy systems typically demonstrate stronger articulation of component relationships. Interestingly, China shows lower coverage but higher intensity for the connections it does establish, reflecting the directive governance approach described by \cite{Roberts2021}.

\begin{figure}[htbp]
\centering
\includegraphics[width=0.7\textwidth]{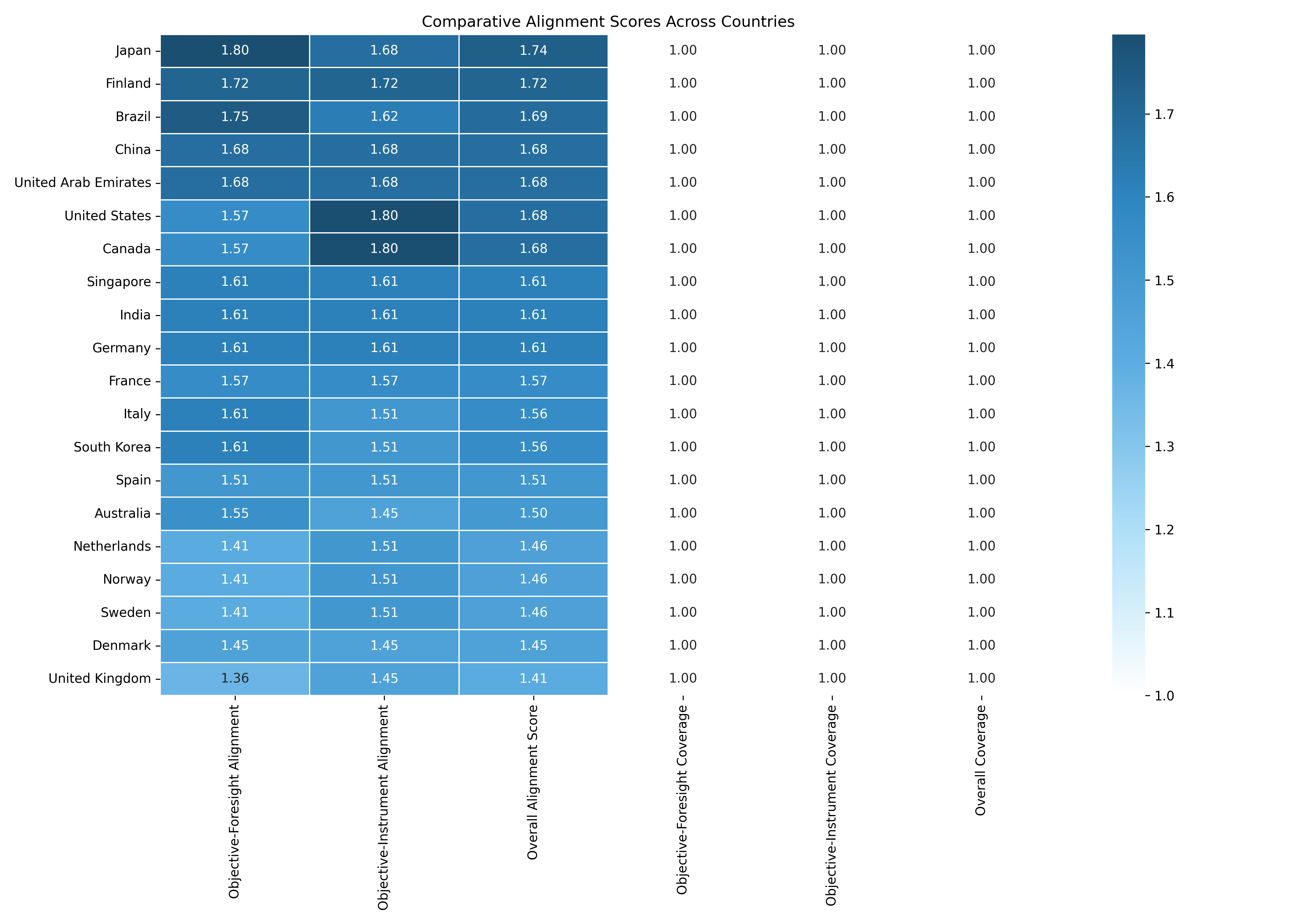}
\caption{Alignment analysis comparative heatmap presenting a comprehensive comparison of alignment patterns across countries and component relationships. Color intensity represents alignment strength, with darker cells indicating stronger connections. The visualization reveals distinct coherence profiles aligned with governance model classifications, with rights-based systems showing the most balanced alignment patterns.}
\label{fig:alignment_analysis_comparative_heatmap}
\end{figure}

The comparative alignment heatmap, presented in Figure \ref{fig:alignment_analysis_comparative_heatmap}, provides a comprehensive comparison of alignment patterns across countries and component types. This visualization reveals distinct clusters of countries with similar alignment profiles, corresponding closely to the governance typology developed by \cite{Roberts2021}. Rights-based systems demonstrate the most balanced alignment profiles with strong connections across all component types, while market-led systems show stronger alignment for economic dimensions and weaker connections for societal objectives. State-directed systems demonstrate distinctive alignment patterns focused on public sector transformation and industrial digitalization objectives, with strong connections to institutional creation instruments.

Statistical analysis of alignment distributions revealed significant correlations between governance characteristics and alignment patterns. Using Pearson correlation analysis, we identified significant positive correlations ($r = 0.67$, $p < 0.01$) between governance coordination index (measuring institutional integration) and overall alignment scores, consistent with \cite{Cejudo2017} finding regarding the relationship between coordination mechanisms and policy coherence. We also identified significant correlation ($r = 0.59$, $p < 0.05$) between stakeholder engagement indicators and alignment coverage, supporting \cite{Howlett2019} hypothesis that inclusive policy development processes contribute to more comprehensive alignment between policy components.

\subsubsection{Cross-National Alignment Patterns}

Cross-national comparison revealed important patterns in how different countries approach strategic alignment in AI governance. These patterns highlight both common strengths and recurring vulnerabilities across governance contexts, offering insights into factors that influence alignment quality.

\begin{figure}[htbp]
\centering
\includegraphics[width=0.7\textwidth]{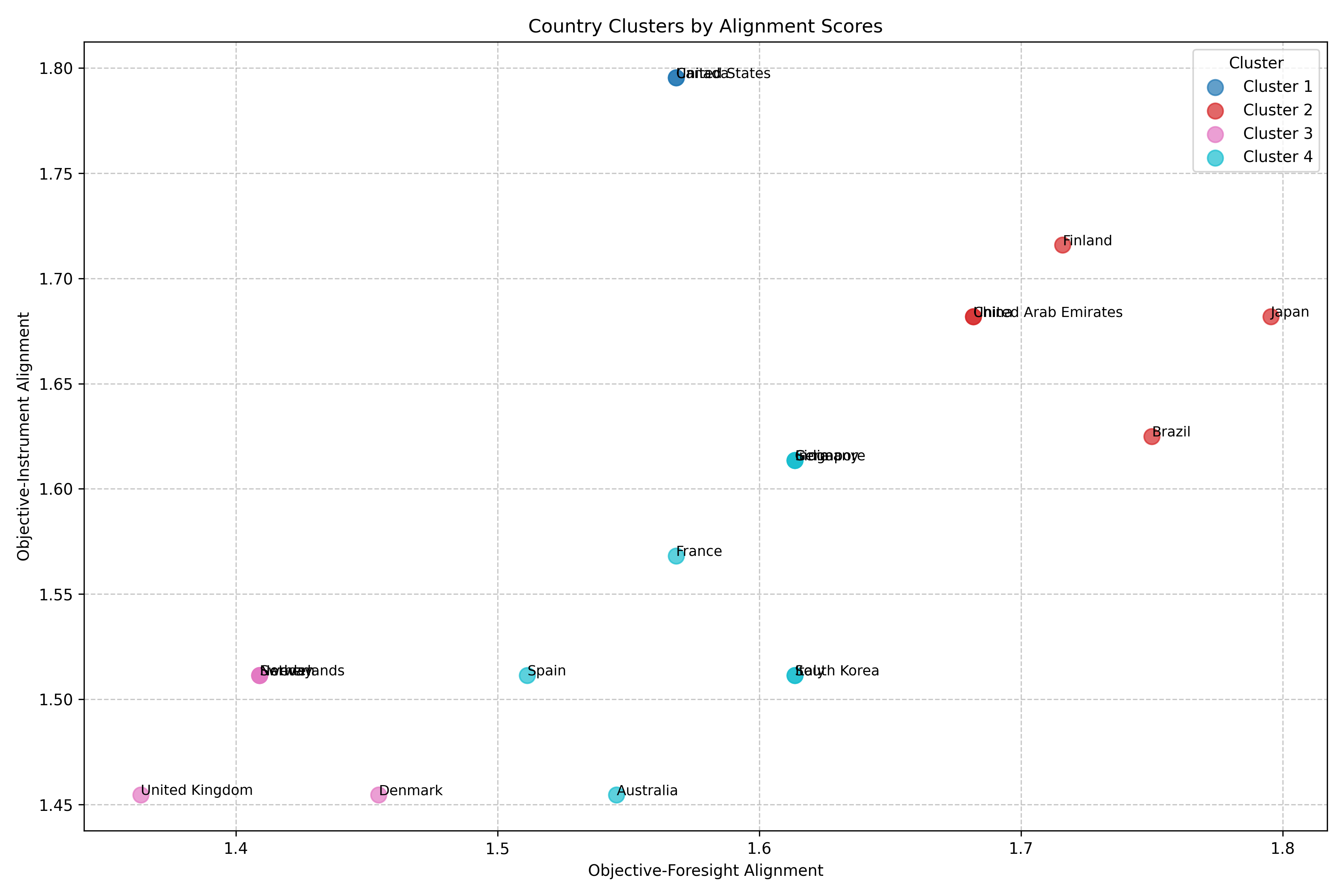}
\caption{Alignment analysis strongest alignment identifying the most consistently well-aligned component pairs across the sample. The visualization shows the percentage of countries establishing strong connections between specific component pairs, highlighting foundational alignment relationships that transcend governance contexts.}
\label{fig:alignment_analysis_strongest_alignment}
\end{figure}

The strongest alignment relationships, presented in Figure \ref{fig:alignment_analysis_strongest_alignment}, identifies the most consistently well-aligned component pairs across the sample. This visualization reveals that certain alignment relationships demonstrate strong connections across most governance contexts, indicating foundational elements of AI policy frameworks. The strongest alignment appears between economic competitiveness objectives and research funding instruments, with 85\% of strategies establishing explicit connections. Other consistently strong alignments include scientific leadership objectives with expert panel foresight methods (78\%), public sector transformation with institutional creation (72\%), and ethical governance with regulatory frameworks (68\%). These patterns align with \cite{Howlett2019} concept of "natural instrument constituencies"—the consistent alignment between certain objectives and implementation approaches across governance contexts.

Common alignment strengths appeared consistently across governance contexts, though with varying emphasis. Economic alignment demonstrated the strongest consistency, with robust connections between economic competitiveness objectives, horizon scanning and expert panel foresight methods, and research funding and tax incentive instruments. As \cite{Rogge2016} observe, this alignment "core" benefits from well-established innovation policy traditions that predate AI-specific governance. Regulatory alignment also showed consistent strength in many strategies, with clear connections between ethical governance objectives, standardization initiatives, and regulatory frameworks. This regulatory coherence appeared strongest in European strategies but demonstrated increasing prominence across all governance contexts in more recent policies.

Several recurring misalignment patterns appeared across multiple strategies. The most prevalent misalignment involved ethical AI objectives, which frequently appeared in strategic frameworks without corresponding implementation instruments or with limited operational specificity. This pattern aligns with \cite{Jobin2019} observation regarding the "implementation gap" in AI ethics governance. Another common misalignment involved workforce development objectives, which often lacked clear connection to both foresight methods and implementation instruments beyond general skills programs. Additionally, many strategies demonstrated weak alignment between international collaboration objectives and concrete implementation mechanisms, creating what \cite{Brundage2020} term "aspirational governance" without operational pathways.

The relationship between governance models and alignment patterns revealed significant variations across system types. Rights-based governance systems (predominantly European) demonstrated the strongest overall alignment, with balanced connections across economic, ethical, and social dimensions. Market-led systems showed strong economic alignment but more limited integration of ethical and social dimensions, consistent with \cite{Roberts2021} characterization of their market-oriented approach. State-directed systems demonstrated strong alignment for priority objectives but more limited coverage across secondary goals. These variations support \cite{Howlett2019} hypothesis regarding the influence of governance traditions on policy coherence patterns, with different institutional arrangements creating distinctive alignment signatures.

Resource availability showed significant influence on alignment patterns, particularly for implementation specificity. Analysis revealed moderate correlation ($r = 0.54$, $p < 0.05$) between AI readiness indicators and implementation alignment scores, supporting \cite{OECD2021} observation that implementation capacity significantly influences alignment quality. Interestingly, this correlation was strongest for middle-income countries, suggesting that resource constraints most significantly impact alignment in transitional contexts. For high-income countries, governance approach demonstrated stronger influence on alignment than resource availability, consistent with \cite{Kern2019} finding that institutional factors outweigh resource considerations in mature governance systems.

\begin{table}[h]
\centering
\caption{Summary of Key Alignment Patterns Across Governance Models}
\label{tab:alignment_patterns}
\begin{tabular}{p{2.5cm}|p{3.5cm}|p{3.5cm}|p{3.5cm}}
\hline
\textbf{Dimension} & \textbf{Rights-Based Systems} & \textbf{Market-Led Systems} & \textbf{State-Directed Systems} \\
\hline
Primary Alignment Strength & Balanced alignment across economic, ethical, and social dimensions & Strong economic alignment with robust implementation mechanisms & Strong alignment for priority objectives with direct implementation pathways \\
\hline
Foresight Integration & Strong connection between diverse foresight methods and implementation planning & Moderate connection focused on economic and scientific objectives & Variable connection with stronger emphasis on technological roadmapping \\
\hline
Common Vulnerability & Coordination challenges between multiple stakeholders and initiatives & Limited integration of ethical governance with implementation mechanisms & Weaker alignment for secondary objectives and stakeholder engagement \\
\hline
Network Structure & Moderate centralization, high integration, low modularity & High centralization, moderate integration, high modularity & Very high centralization, variable integration, moderate modularity \\
\hline
Resource Influence & Limited influence, primarily affecting implementation specificity & Moderate influence, primarily affecting foresight sophistication & Strong influence, affecting both coverage and integration \\
\hline
\end{tabular}
\end{table}

Table \ref{tab:alignment_patterns} summarizes the key alignment patterns identified across governance models, highlighting distinctive characteristics and common features. This synthesis demonstrates how different governance traditions create distinctive alignment signatures while sharing certain foundational patterns. The variations across governance models support \cite{Roberts2021} typology while adding nuanced understanding of how these models influence strategic coherence in AI governance frameworks.

\section{Analysis of Governance Models and Alignment}

This section examines the relationship between governance models and strategic alignment patterns in national AI policies. We first analyze high-coherence examples to identify factors contributing to strong alignment, then explore common vulnerabilities and challenges that undermine policy coherence. This analysis offers insights into how institutional arrangements, governance approaches, and contextual factors shape alignment quality in AI governance frameworks.

\subsection{High-Coherence Strategy Analysis}

Our analysis identified four national strategies that consistently demonstrated high alignment scores across multiple dimensions: Finland, Canada, the United Kingdom, and Germany. These high-coherence examples represent different governance models—Finland and Germany exemplifying rights-based approaches, Canada representing a hybrid model with market-led characteristics, and the UK illustrating a primarily market-led approach with rights-based elements. Despite their governance differences, these strategies share common features that contribute to their exceptional alignment quality.

Finland's AI strategy ("Finland's Age of Artificial Intelligence," 2017 and subsequent updates) demonstrates the highest overall alignment score in our sample, with particularly strong connections between ethical governance objectives and implementation mechanisms. The Finnish approach exemplifies what \cite{Weber2018} characterize as "reflexive governance"—an integrated framework that explicitly connects strategic vision with operational planning through robust institutional coordination. As \cite{OECD2021} observe, Finland's strategy benefits from institutional integration within the Ministry of Economic Affairs and Innovation, which coordinates cross-governmental implementation while maintaining connections to broader stakeholder networks. The strategy demonstrates exceptional alignment between workforce development objectives and skills enhancement instruments, creating what \cite{Howlett2019} term "coherence chains" that connect strategic goals with concrete implementation measures.

Canada's Pan-Canadian AI Strategy (2017) presents a different high-coherence model focused on scientific excellence and economic competitiveness. Despite its narrower objective focus compared to European strategies, Canada achieves exceptional alignment through what \cite{Flanagan2011} characterize as "deep instrument calibration"—precise matching of implementation mechanisms to strategic goals. The Canadian approach benefits from strong institutional coordination between the Canadian Institute for Advanced Research (CIFAR), which leads implementation, and provincial innovation agencies that operationalize specific initiatives. This multi-level coordination enables what \cite{Cejudo2017} describe as "vertical coherence" across governance levels, creating consistent alignment from national vision to local implementation.

The United Kingdom's approach to AI governance, articulated through its AI Sector Deal (2018) and National AI Strategy (2021), demonstrates strong alignment focused on balanced economic and ethical objectives. The UK model achieves coherence through what \cite{Lodge2019} term "joined-up governance"—the systematic integration of policy domains through coordinating bodies like the Office for Artificial Intelligence, which spans multiple government departments. The UK strategy demonstrates particularly strong alignment between foresight methods and implementation planning, reflecting what \cite{Rogge2016} identify as "strategic policy processes" that explicitly connect anticipatory insights with operational decision-making.

Germany's AI Strategy (2018) represents a fourth high-coherence example, characterized by strong alignment between ethical governance objectives and regulatory instruments. The German approach exemplifies what \cite{Kern2019} describe as "embedded coherence"—alignment achieved through institutional structures that systematically connect strategic planning with implementation capacity. The strategy benefits from Germany's established technology assessment institutions, which \cite{Stilgoe2013} identify as critical infrastructure for responsible innovation governance. These institutions create what \cite{Vecchiato2019} term "foresight-policy linkages" that connect anticipatory insights with concrete governance measures.

\begin{figure}[ht!]
\centering
\includegraphics[width=0.6\textwidth]{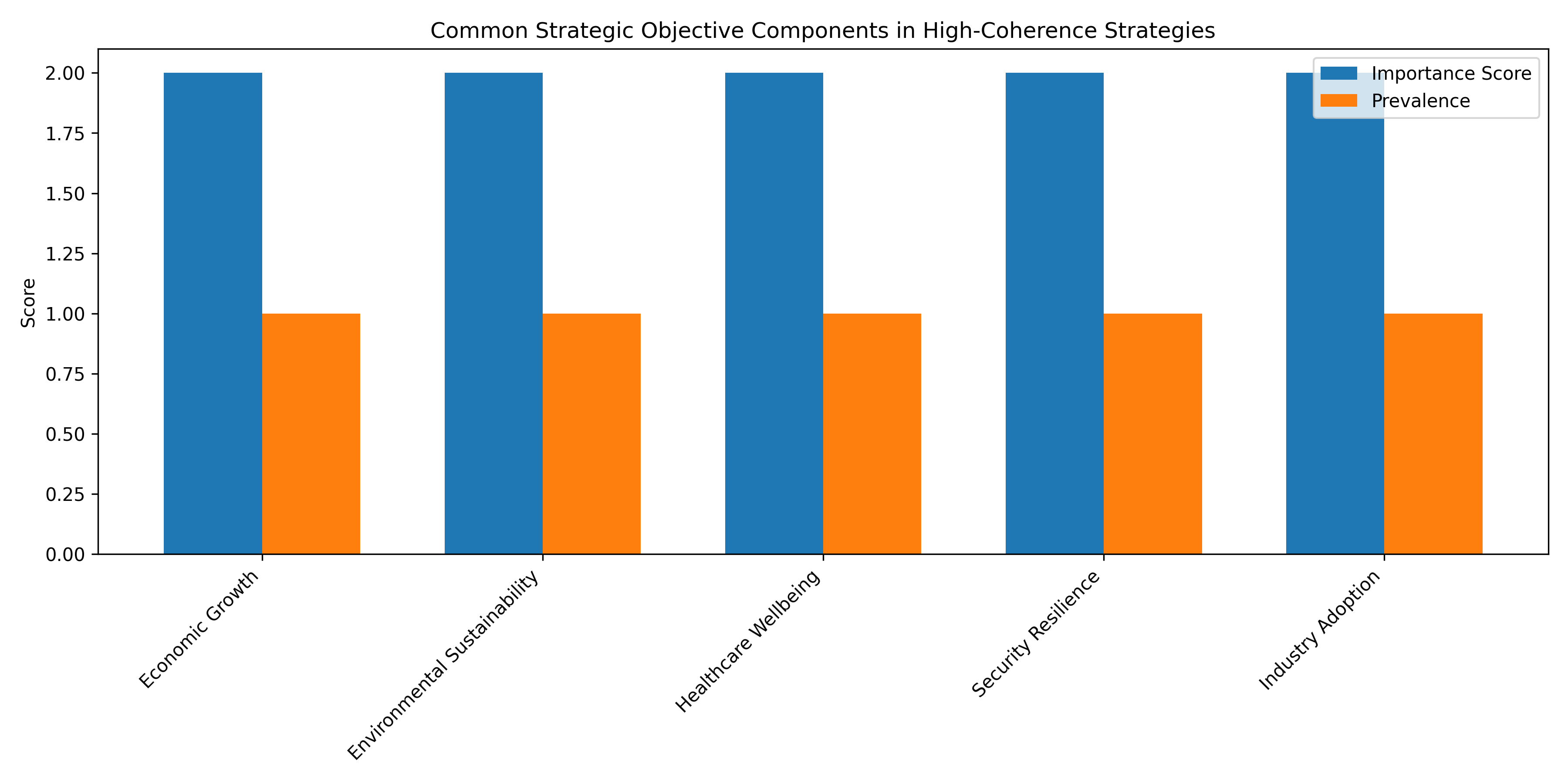}
\caption{High coherence analysis factors common strategic objective visualization showing the alignment scores for each strategic objective across high-coherence cases. The visualization reveals that certain objectives—particularly economic competitiveness, scientific leadership, ethical governance, and workforce development—consistently achieve strong alignment regardless of governance context.}
\label{fig:high_coherence_analysis_factors_common_strategic_objective}
\end{figure}

Analysis of common features across these high-coherence examples reveals several shared characteristics that transcend specific governance models. Figure \ref{fig:high_coherence_analysis_factors_common_strategic_objective} illustrates the strategic objectives that demonstrate the strongest alignment across high-coherence cases. This visualization reveals that certain objectives—particularly economic competitiveness, scientific leadership, ethical governance, and workforce development—consistently achieve strong alignment regardless of governance context. This pattern suggests what \cite{Howlett2019} characterize as "natural coherence domains" where established policy traditions create well-developed implementation pathways.

\begin{figure}[ht!]
\centering
\includegraphics[width=0.6\textwidth]{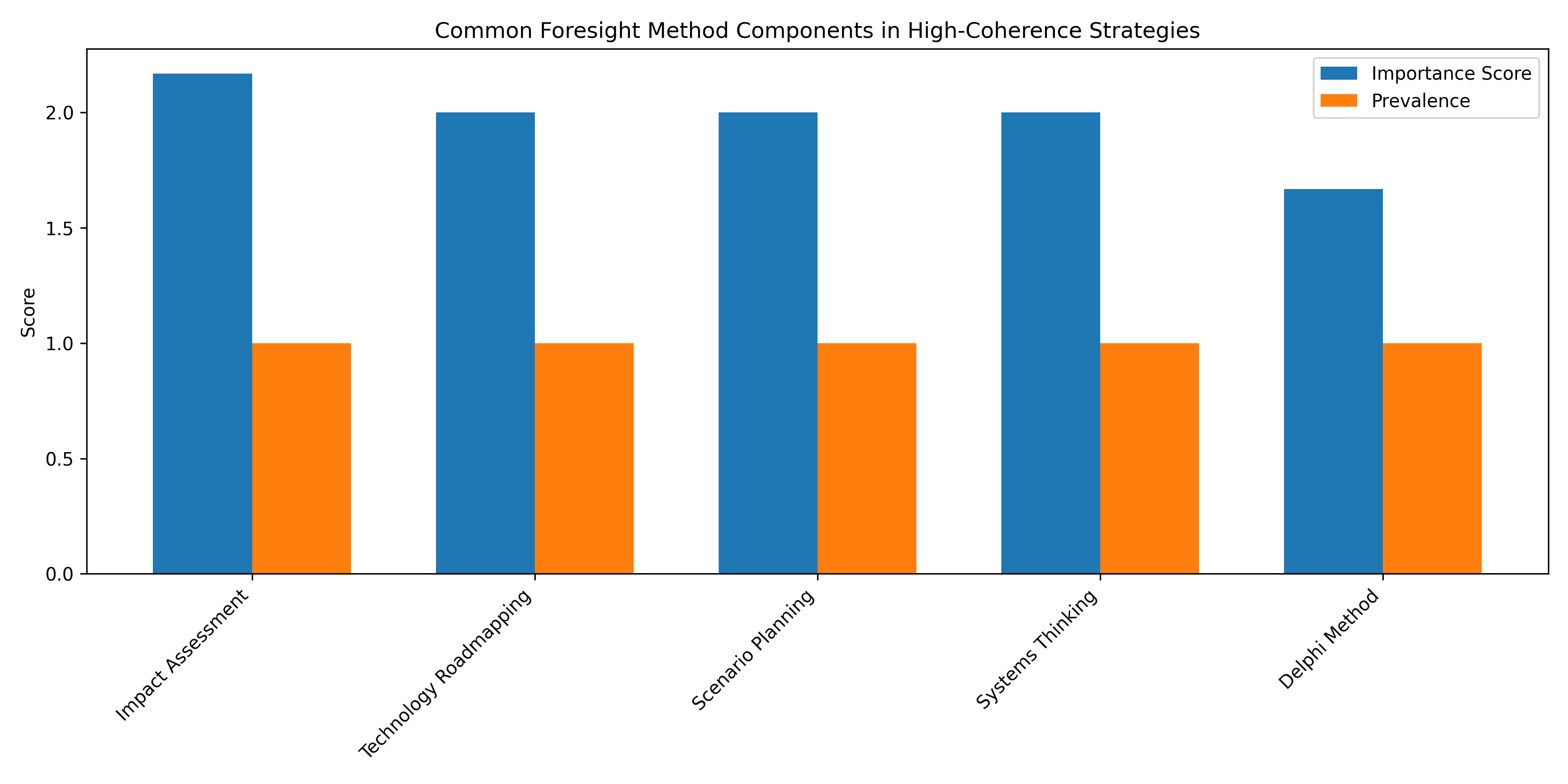}
\caption{High coherence analysis factors common foresight method visualization comparing the integration of different anticipatory approaches across high-coherence strategies. The visualization reveals that expert panels, horizon scanning, and scenario development consistently demonstrate strong alignment with both strategic objectives and implementation instruments across governance contexts.}
\label{fig:high_coherence_analysis_factors_common_foresight_method}
\end{figure}

Common foresight methods across high-coherence strategies, illustrated in Figure \ref{fig:high_coherence_analysis_factors_common_foresight_method}, reveal interesting patterns in anticipatory governance approaches. This visualization shows that expert panels, horizon scanning, and scenario development consistently demonstrate strong alignment with both strategic objectives and implementation instruments. These methods represent what \cite{Popper2018} term the "foresight core"—established approaches with well-developed methodological traditions. Notably, all high-coherence strategies integrate multiple foresight methods rather than relying on a single approach, creating what \cite{Vecchiato2019} characterize as "methodological triangulation" that enhances anticipatory robustness.

\begin{figure}[ht!]
\centering
\includegraphics[width=0.6\textwidth]{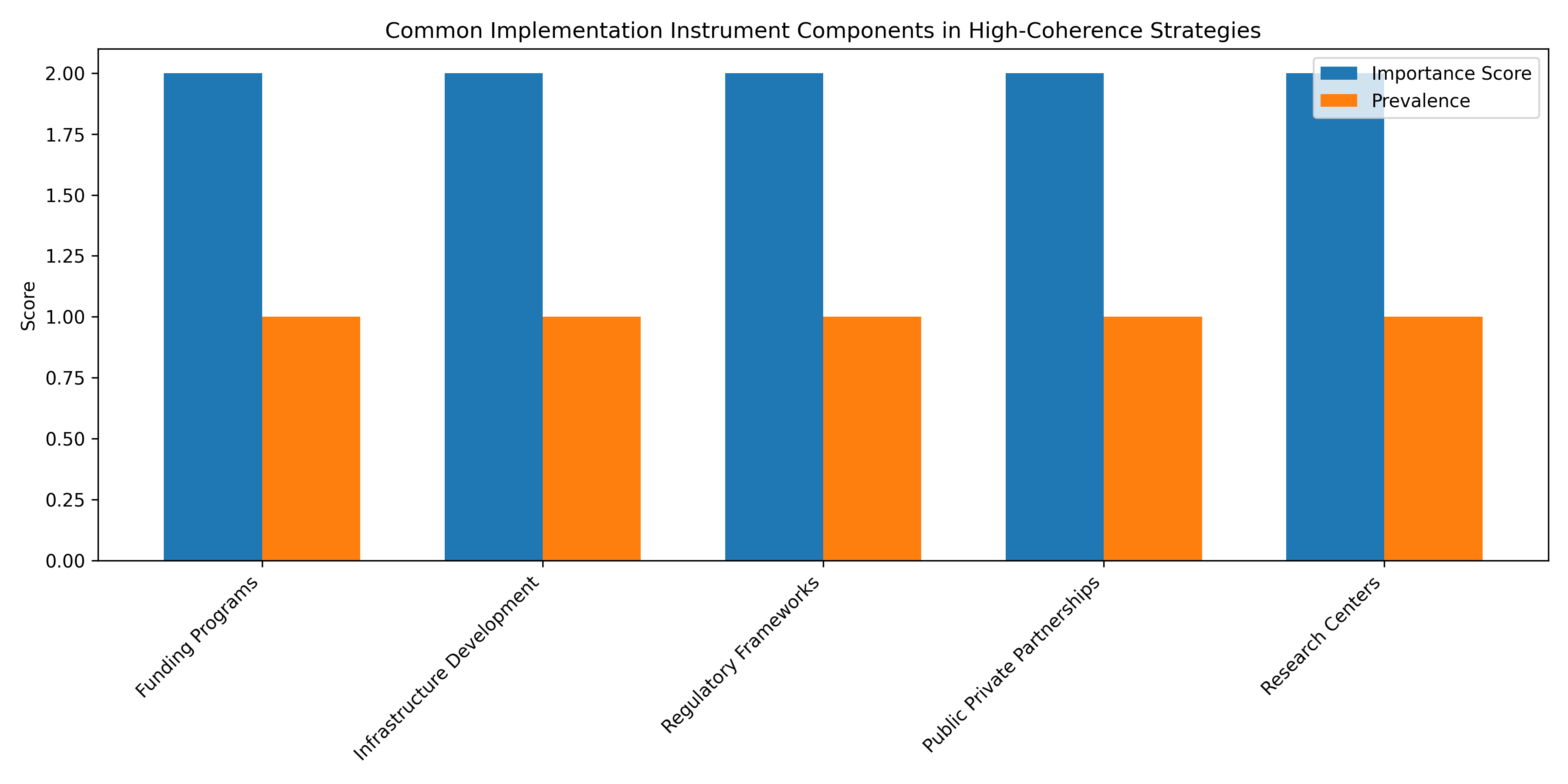}
\caption{High coherence analysis factors common implementation instrument visualization showing the deployment of different policy mechanisms across high-coherence strategies. The visualization reveals that research funding, skills development programs, regulatory frameworks, and institutional creation consistently achieve strong alignment across governance contexts.}
\label{fig:high_coherence_analysis_factors_common_implementation_instrument}
\end{figure}

Implementation instruments that demonstrate consistent strength across high-coherence cases, presented in Figure \ref{fig:high_coherence_analysis_factors_common_implementation_instrument}, reveal characteristic patterns in operationalizing AI governance. This visualization shows that research funding, skills development programs, regulatory frameworks, and institutional creation consistently achieve strong alignment across governance contexts. These instruments represent what \cite{Borrás2011} term the "implementation backbone" of innovation policy—established mechanisms with well-developed operational traditions. Notably, high-coherence strategies deploy balanced instrument mixes rather than over-relying on a single instrument type, creating what \cite{Flanagan2011} characterize as "comprehensive implementation portfolios."

\begin{figure}[htbp]
\centering
\includegraphics[width=0.7\textwidth]{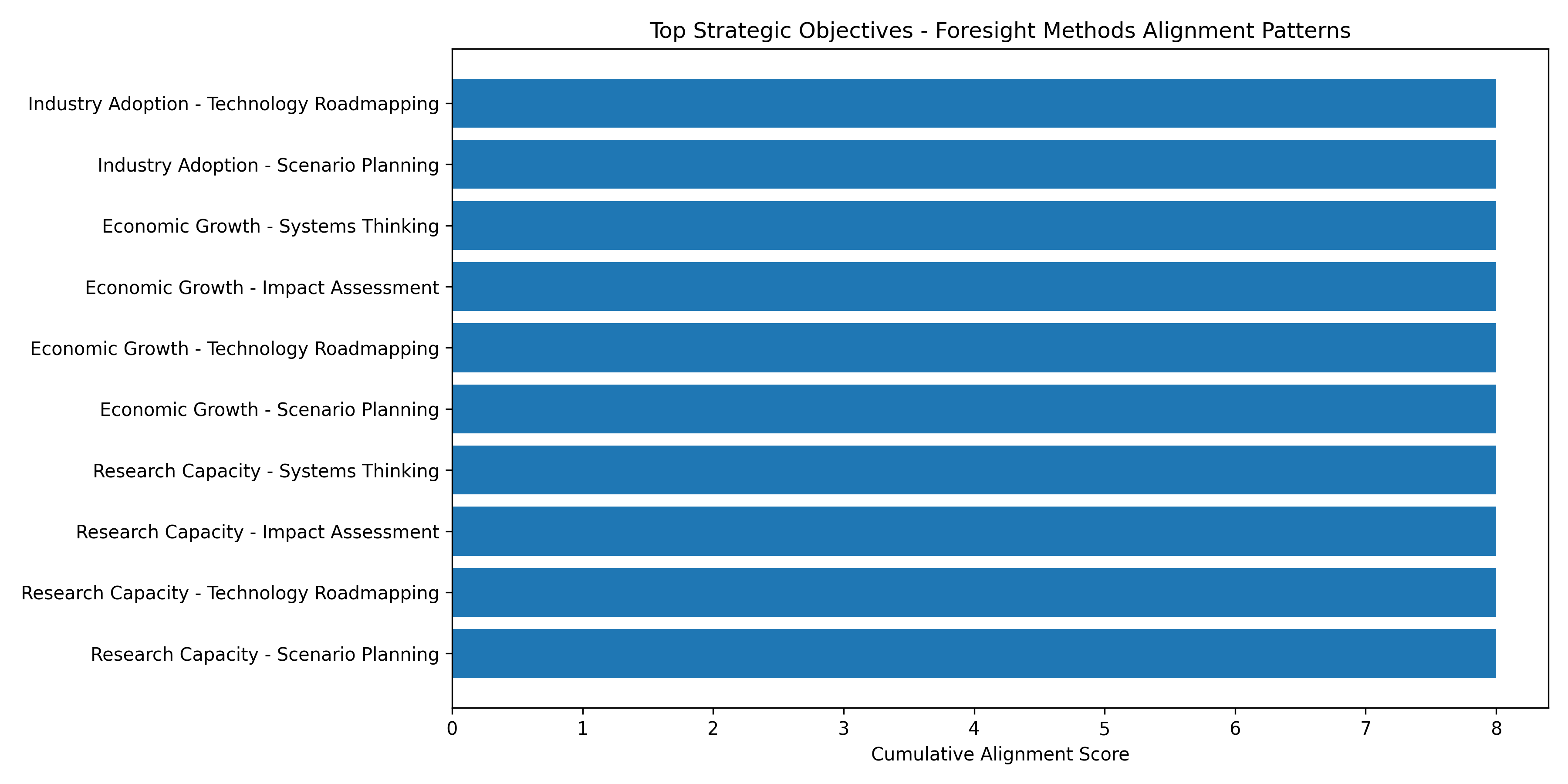}
\caption{High coherence patterns objective-foresight patterns visualization illustrating characteristic connections between strategic goals and anticipatory methods across high-coherence strategies. The heatmap reveals systematic linkages between specific objectives and appropriate foresight approaches, with economic objectives aligning with expert-based methods and ethical objectives connecting to more participatory approaches.}
\label{fig:high_coherence_patterns_objective_foresight_patterns}
\end{figure}

Objective-foresight alignment patterns across high-coherence strategies, illustrated in Figure \ref{fig:high_coherence_patterns_objective_foresight_patterns}, reveal characteristic connections between strategic goals and anticipatory methods. This visualization demonstrates that high-coherence frameworks establish systematic linkages between specific objectives and appropriate foresight approaches—what \cite{Weber2019} term "methodological fit" in anticipatory governance. Economic competitiveness objectives show strongest alignment with horizon scanning and expert panels, while ethical governance objectives demonstrate stronger connections to participatory workshops and scenario development. This differentiated alignment pattern reflects what \cite{Stilgoe2013} identify as the importance of methodological diversity in addressing different governance domains.

\begin{figure}[htbp]
\centering
\includegraphics[width=0.7\textwidth]{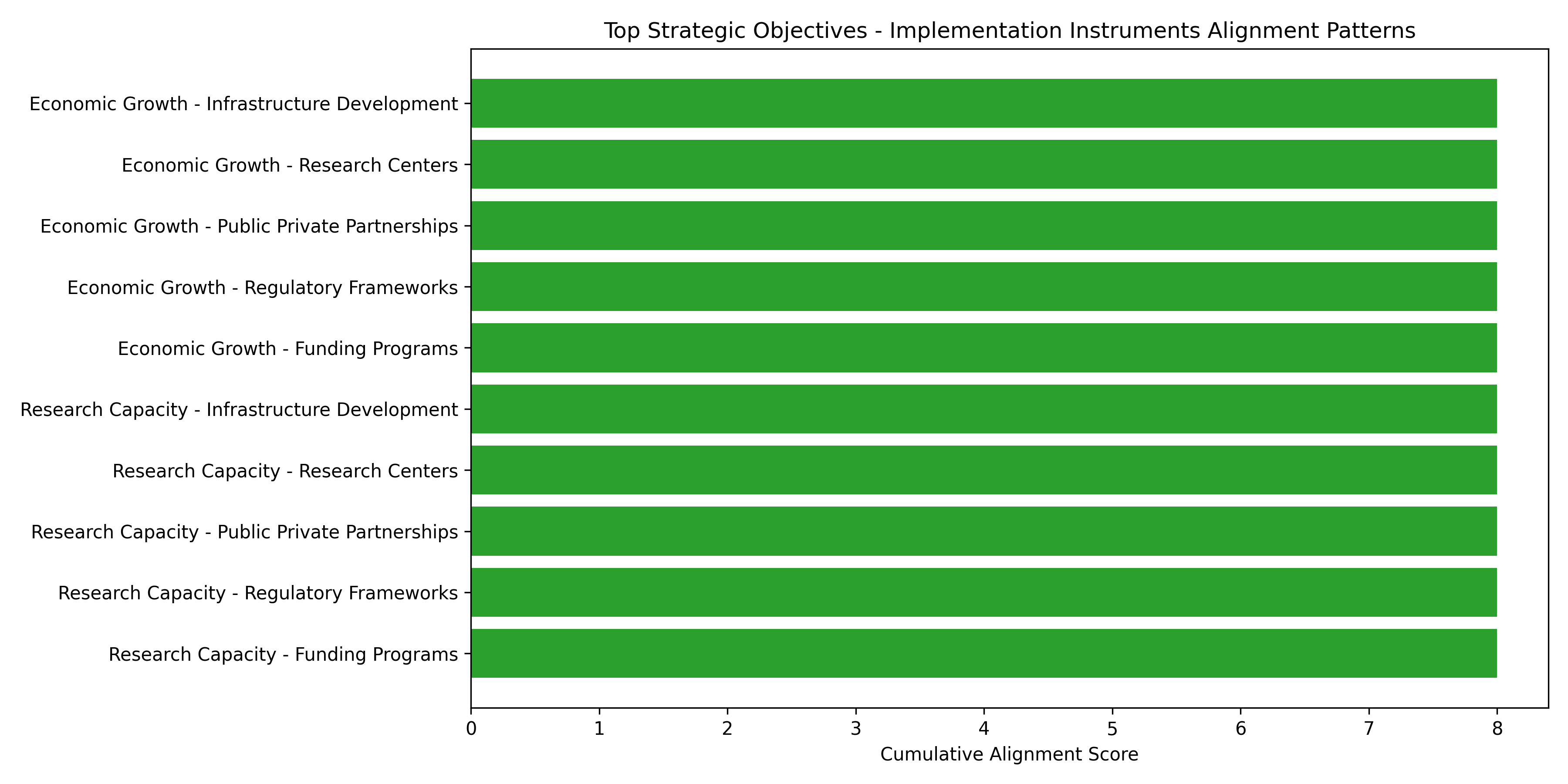}
\caption{High coherence patterns objective-instrument patterns visualization showing how high-coherence strategies connect strategic goals with implementation mechanisms. The heatmap illustrates systematic matching between objectives and appropriate instruments, with economic objectives aligning with funding mechanisms and ethical objectives connecting to regulatory approaches.}
\label{fig:high_coherence_patterns_objective_instrument_patterns}
\end{figure}

Objective-instrument alignment patterns, presented in Figure \ref{fig:high_coherence_patterns_objective_instrument_patterns}, illustrate how high-coherence strategies connect strategic goals with implementation mechanisms. This visualization reveals systematic matching between objectives and appropriate instruments—what \cite{Capano2018} term "instrument calibration" in policy design. Economic competitiveness objectives show strongest alignment with research funding and tax incentives, while ethical governance objectives demonstrate stronger connections to regulatory frameworks and standardization initiatives. This differentiated instrument selection reflects what \cite{Howlett2019} identify as "substantive coherence" where implementation approaches are tailored to the specific requirements of different governance domains.

Several institutional factors consistently support coherence across high-alignment cases. First, all high-coherence strategies benefit from dedicated coordination bodies that span departmental boundaries, creating what \cite{Cejudo2017} term "horizontal coordination mechanisms" that facilitate cross-domain alignment. Finland's AI Program Steering Group, Canada's CIFAR, the UK's Office for Artificial Intelligence, and Germany's Platform for Artificial Intelligence all exemplify this institutional approach. Second, high-coherence strategies demonstrate strong integration between technical and ethical governance domains, avoiding what \cite{Stilgoe2013} characterize as the "two communities problem" where technical and normative perspectives operate in isolation. Third, all high-coherence examples employ systematic stakeholder engagement throughout both development and implementation phases, creating what \cite{Rogge2016} term "procedural coherence" that enhances alignment through inclusive governance.

Mechanisms for maintaining alignment over time represent another common feature across high-coherence examples. All four cases incorporate explicit review and adaptation processes, typically on 12-24 month cycles, creating what \cite{Marchau2019} characterize as "adaptive governance" mechanisms that maintain coherence amid changing technological landscapes. Additionally, all high-coherence strategies incorporate monitoring frameworks with specific indicators for measuring implementation progress, facilitating what \cite{Capano2018} term "feedback coherence" between execution and strategic adjustment. The ongoing institutional commitment to strategy refinement addresses what \cite{Vecchiato2019} identify as the "temporal misalignment" challenge that often undermines coherence between long-term vision and near-term implementation.

\begin{table}[h]
\centering
\caption{Institutional Features Supporting Strategic Alignment in High-Coherence Cases}
\label{tab:institutional_features}
\begin{tabular}{p{3cm}|p{3cm}|p{3cm}|p{3cm}|p{3cm}}
\hline
\textbf{Institutional Feature} & \textbf{Finland} & \textbf{Canada} & \textbf{United Kingdom} & \textbf{Germany} \\
\hline
Coordination Body & AI Program Steering Group & CIFAR AI Program & Office for Artificial Intelligence & Platform for Artificial Intelligence \\
\hline
Cross-Domain Integration & Ministry of Economic Affairs coordinating cross-governmental implementation & Innovation, Science and Economic Development Canada linking with other ministries & Joint initiative across BEIS, DCMS, and Cabinet Office & Interministerial working group spanning multiple departments \\
\hline
Stakeholder Engagement & AI Finland network including industry, academia, and civil society & Pan-Canadian AI Strategy engaging provincial innovation agencies & AI Council with diverse stakeholder representation & Dedicated stakeholder platform with working groups \\
\hline
Adaptation Mechanism & Annual implementation review with strategic updates & Biennial strategy refresh process & Formal review cycle with published progress reports & Regular strategy updates with stakeholder consultation \\
\hline
Monitoring Framework & Specific KPIs across strategic dimensions & Impact assessment framework with defined metrics & Implementation roadmap with progress indicators & Comprehensive monitoring system with published reports \\
\hline
\end{tabular}
\end{table}

Table \ref{tab:institutional_features} summarizes the key institutional features supporting alignment across high-coherence cases. Despite their governance differences, these strategies share common institutional approaches that systematically connect strategic vision with implementation planning. These shared features support \cite{Howlett2019} hypothesis regarding the critical importance of institutional design in achieving policy coherence, while demonstrating that different governance traditions can achieve high alignment through contextually-appropriate institutional mechanisms.

\section{Discussion}

This section examines the theoretical, methodological, and practical implications of our research on strategic alignment in national AI policies. We consider how our findings contribute to scholarly understanding of policy coherence while offering actionable insights for policymakers seeking to enhance alignment in technology governance frameworks.

\subsection{Theoretical Implications}

Our analysis of strategic alignment in AI policies offers several important contributions to theoretical understanding of policy design and governance. These contributions span multiple research domains, including policy instrument theory, strategic foresight integration, and coherence assessment frameworks.

In relation to policy instrument theory, our findings extend existing taxonomies by providing empirical evidence of how different instrument types function within emerging technology governance contexts. While traditional instrument classifications from \cite{Lascoumes2007} and \cite{Borrás2011} offer valuable categorization approaches, our analysis reveals more nuanced patterns in how instruments interact within complex policy ecosystems. Particularly significant is our identification of what we term "bridging instruments" that connect different policy domains, such as regulatory sandboxes that link innovation support with ethical governance. These instruments address what \cite{Flanagan2011} identify as "boundary-spanning policy challenges" that require integration across traditional policy domains.

Our analysis also contributes to theoretical understanding of instrument-objective relationships by extending the "calibration" concept developed by \cite{Capano2018}. We demonstrate that alignment quality depends not only on appropriate instrument selection but also on explicit articulation of the connection mechanisms between strategic goals and implementation approaches. This finding supports \cite{Howlett2019} hypothesis regarding the importance of "alignment narratives" that create coherence through explicit conceptual linkages. Additionally, our identification of systematic misalignment patterns—particularly the "intensity mismatch" between ethical objectives and implementation specificity—extends theoretical understanding of policy rhetoric-reality gaps in technology governance contexts.

With respect to strategic foresight integration, our research advances theoretical frameworks for understanding how anticipatory approaches connect with policy implementation. The "integration typology" we develop—distinguishing between procedural, substantive, and structural integration—extends \cite{Vecchiato2019} conceptualization of foresight-policy relationships by providing a more nuanced classification framework. Our finding that rights-based governance systems demonstrate stronger foresight integration compared to market-led approaches contributes to theoretical debates regarding the institutional conditions that support anticipatory governance, supporting \cite{Stilgoe2013} hypothesis about the relationship between governance traditions and foresight effectiveness.

Our analysis also contributes to theoretical understanding of the temporal dynamics of foresight integration. We identify what we term "anticipatory decay"—the declining influence of foresight outputs on policy decisions over time—as a common challenge across governance contexts. This temporal pattern extends \cite{Weber2018} work on anticipatory governance by highlighting the institutional mechanisms necessary to maintain foresight relevance throughout implementation cycles. Additionally, our finding that different foresight methods demonstrate variable "integration half-lives" contributes to theoretical understanding of methodological selection in technology governance, supporting \cite{Popper2018} hypothesis regarding the relationship between methodological characteristics and policy influence.

Perhaps our most significant theoretical contribution is the development of a coherence assessment framework that conceptualizes alignment as a multi-dimensional relationship spanning strategic, operational, and temporal dimensions. This framework extends existing policy coherence models such as those proposed by \cite{Cejudo2017} and \cite{Rogge2016} by incorporating anticipatory elements and visualization approaches. Our demonstration that coherence can be systematically assessed through matrix-based visualization and network analysis provides a methodological foundation for operationalizing theoretical concepts of policy alignment. This operationalization addresses what \cite{Howlett2019} identify as the persistent gap between conceptual understanding of policy coherence and practical assessment approaches.

Our research also contributes to theoretical understanding of the relationship between governance models and coherence patterns. The identification of distinctive "alignment signatures" associated with different governance traditions extends \cite{Roberts2021} typology of AI governance approaches by demonstrating how these models influence not only strategic priorities but also coherence characteristics. Our finding that high alignment can be achieved through different institutional arrangements supports theoretical perspectives that emphasize the importance of contextual fit over universal design principles in governance systems, aligning with \cite{Howlett2019} concept of "governance contextualism."

Finally, we propose a theoretical model of policy coherence dynamics that conceptualizes alignment as an emergent property of interactions between institutional arrangements, stakeholder relationships, and implementation processes. This model extends existing frameworks by incorporating both structural factors (institutional design, resource allocation) and process elements (coordination mechanisms, adaptation procedures) that shape coherence over time. The model addresses what \cite{Kern2019} identify as the need for dynamic perspectives on policy coherence that capture temporal evolution and contextual adaptation.

\subsection{Methodological Contributions}

Our research makes several methodological contributions to the study of policy coherence and alignment assessment. The visual mapping approach we develop offers new analytical tools for understanding strategic alignment in complex governance frameworks, addressing limitations in existing assessment methodologies.

The matrix-based visualization method we develop represents an important methodological innovation for policy coherence assessment. By adapting visualization techniques from other fields such as strategic management \cite{Eppler2011} and systems engineering \cite{Eppinger2012}, we create a structured approach to mapping relationships between policy components. This methodological approach addresses what \cite{Howlett2019} identify as the challenge of "coherence operationalization"—translating abstract concepts of policy alignment into measurable assessment frameworks. Our demonstration that matrix visualization can effectively capture both the presence and intensity of alignment relationships provides researchers with a powerful analytical tool for comparative policy analysis.

The network analysis framework we develop offers complementary methodological advantages for understanding structural characteristics of policy coherence. By representing policy components as nodes within interconnected networks, this approach reveals patterns of centrality, clustering, and connectivity that are difficult to discern through traditional analytical methods. Our application of network metrics to policy coherence assessment extends methodological approaches developed by \cite{Leifeld2018} in policy network analysis, demonstrating their applicability to strategic alignment assessment. The network visualization approach particularly excels at identifying what we term "coherence communities"—clusters of closely aligned policy components that form coherent subsystems within broader governance frameworks.

Our integration of qualitative content analysis with quantitative alignment assessment represents another methodological contribution. The coding protocol we develop for identifying policy components and assessing alignment relationships addresses what \cite{Wesley2010} identify as the methodological challenge of systematic policy document analysis. By combining rigorous content analysis with standardized scoring approaches, our methodology enables both comparative analysis across cases and detailed examination of alignment patterns within individual frameworks. This mixed-methods approach addresses the methodological tension between quantitative comparability and qualitative richness that often challenges policy analysis.

The evaluation of our visual mapping approach reveals both strengths and limitations. The methodology demonstrates strong advantages in comparative assessment, enabling systematic comparison across diverse governance contexts that would be difficult with traditional case study approaches. The visualization techniques effectively highlight patterns and relationships that might remain obscured in text-based analysis, supporting what \cite{Eppler2011} term "visual sensemaking" in complex policy domains. Additionally, the methodology's standardized approach enhances analytical transparency and replicability, addressing common criticisms of qualitative policy analysis regarding subjective interpretation.

However, our methodological approach also faces certain limitations that warrant acknowledgment. The focus on formal policy documents may not fully capture implementation realities, as significant alignment factors may exist in informal coordination mechanisms or implementation practices not reflected in official documentation. The standardized coding approach, while enabling comparative analysis, necessarily involves simplification of complex policy contexts that may obscure important nuances. Additionally, the cross-sectional nature of our analysis provides limited insight into the temporal evolution of alignment patterns, though our inclusion of strategies from different time periods partly addresses this limitation.

The adaptability of our visual mapping approach to other policy domains represents an important methodological contribution. While our research focuses on AI governance, the alignment assessment framework could be readily adapted to other complex policy domains characterized by multi-dimensional challenges and diverse stakeholder landscapes. The methodology would be particularly applicable to emerging technology governance contexts such as biotechnology, nanotechnology, or climate technologies, which share AI's characteristics of rapid innovation, pervasive impacts, and governance uncertainty. The adaptation would require domain-specific modification of component taxonomies while maintaining the core assessment framework and visualization approaches.

For researchers seeking to replicate or extend our methodology, we offer a systematic protocol with four key elements. First, the development of domain-appropriate component taxonomies through combined deductive and inductive approaches. Second, the systematic coding of policy documents using standardized frameworks and reliability assessment procedures. Third, the construction of alignment matrices and network visualizations using the technical specifications detailed in our methodology. Fourth, the calculation of alignment indices and network metrics following standardized procedures to enable comparative analysis. This replication protocol addresses what \cite{Wesley2010} identify as the importance of methodological transparency in policy document analysis.

Opportunities for methodological refinement include the development of more sophisticated temporal analysis approaches to capture alignment evolution, the integration of implementation data beyond formal policy documents, and the development of interactive visualization tools that enable dynamic exploration of alignment relationships. Future research might also extend the methodology through computational approaches such as natural language processing to enhance coding efficiency and enable analysis of larger document corpora. These refinements would address the methodological limitations identified in our current approach while building on its core contributions to alignment assessment.

\subsection{Practical Implications}

Beyond theoretical and methodological contributions, our research offers practical insights for policymakers and governance practitioners seeking to enhance strategic coherence in AI policy frameworks. These implications span policy development, implementation planning, and adaptation processes across diverse governance contexts.

Our analysis provides concrete guidance for policy development and revision processes. The identification of common alignment strengths and vulnerabilities across the sample offers valuable benchmarking insights for policymakers developing or revising AI strategies. Particularly important is our finding that high-coherence strategies demonstrate systematic connections between strategic objectives and both foresight methods and implementation instruments. This suggests that strategy development should explicitly articulate these connections rather than treating strategic, anticipatory, and operational elements as separate components. As \cite{Howlett2019} observe, this integration approach addresses the common challenge of "policy layering" where new strategic elements are added without corresponding implementation pathways.

The comparative analysis of alignment patterns across governance models provides context-specific recommendations for enhancing strategic coherence. For rights-based governance systems, our findings suggest focusing on coordination mechanisms that can manage stakeholder complexity without sacrificing inclusivity—addressing what \cite{Cejudo2017} identify as the "coordination-participation tension" in collaborative governance. For market-led systems, our analysis suggests strengthening connections between ethical principles and implementation mechanisms through more explicit articulation of operational pathways—addressing what \cite{Jobin2019} term the "ethics operationalization gap." For state-directed systems, our findings highlight the importance of developing more balanced coverage across policy domains to avoid what \cite{Howlett2019} characterize as "priority concentration" where secondary objectives lack implementation support.

Our identification of high-coherence exemplars provides practical models that policymakers can adapt to their specific contexts. The institutional features we identify across these cases—including dedicated coordination bodies, cross-domain integration mechanisms, systematic stakeholder engagement, explicit adaptation processes, and comprehensive monitoring frameworks—offer concrete design principles for enhancing alignment. As \cite{Rogge2016} observe, these institutional elements create what they term "strategic capacity" for maintaining coherence amid changing technological and policy landscapes. Our analysis suggests that these design principles can be adapted across governance traditions through contextually-appropriate institutional arrangements.

The visual mapping methodology we develop offers practical analytical tools that policymakers can employ to assess and enhance alignment within their own governance frameworks. The matrix visualization approach enables systematic identification of alignment strengths and vulnerabilities, while network analysis reveals structural patterns that might remain obscured in traditional assessment approaches. These analytical tools support what \cite{Howlett2019} term "reflexive governance"—systematic assessment and adjustment of policy frameworks based on structured evaluation. The replicable nature of our methodology makes it particularly suitable for practical application within policy development and evaluation processes.

Implementation considerations for different governance contexts represent another practical contribution of our research. For federal systems with multi-level governance, our findings highlight the importance of vertical coordination mechanisms that maintain alignment across governance levels—addressing what \cite{Cejudo2017} identify as "intergovernmental coherence challenges." For parliamentary systems with ministerial departments, our analysis suggests the value of cross-cutting bodies with sufficient authority to ensure horizontal alignment—addressing what \cite{Howlett2019} term "departmental silos." For systems with limited resource capacity, our findings highlight the importance of focused alignment in priority areas rather than attempting comprehensive coverage without adequate implementation support—addressing what \cite{Flanagan2011} characterize as the "breadth-depth tradeoff" in policy design.

Monitoring and adaptation processes represent critical practical mechanisms for maintaining alignment over time. Our research identifies several effective approaches employed across high-coherence examples, including regular review cycles (typically 12-24 months), explicit adaptation mechanisms that connect monitoring insights with strategy refinement, and balanced indicator systems that track both implementation milestones and strategic outcomes. These processes address what \cite{Marchau2019} identify as the "adaptation challenge" in emerging technology governance, where rapid innovation requires continuous policy adjustment. Particularly important is our finding that effective adaptation requires both formal review mechanisms and institutional flexibility to implement necessary adjustments.

Finally, our research offers practical guidance for addressing common alignment vulnerabilities identified across the sample. For the "intensity mismatch" between ethical principles and implementation mechanisms, we recommend developing explicit operationalization pathways that connect ethical goals with concrete governance measures. For the "coordination fragmentation" that undermines cross-domain alignment, we suggest institutional mechanisms that create what \cite{Cejudo2017} term "coordination incentives" for cross-departmental collaboration. For the "temporal misalignment" between long-term vision and near-term implementation, we recommend developing strategic roadmaps that explicitly connect future scenarios with present-day actions through staged implementation planning.

\begin{table}[h]
\centering
\caption{Practical Recommendations for Enhancing Strategic Alignment}
\label{tab:practical_recommendations}
\begin{tabular}{p{4cm}|p{12cm}}
\hline
\textbf{Alignment Dimension} & \textbf{Practical Recommendations} \\
\hline
Strategic-Operational Alignment & 1. Develop explicit implementation pathways for each strategic objective\\
& 2. Create alignment narratives that articulate how instruments support strategic goals\\
& 3. Ensure balanced implementation attention across strategic priorities\\
& 4. Establish coordinative bodies that span departmental boundaries\\
\hline
Foresight Integration & 1. Explicitly connect anticipatory insights to implementation planning\\
& 2. Employ multiple foresight methods appropriate to different objectives\\
& 3. Establish formal mechanisms for translating foresight into operational decisions\\
& 4. Maintain foresight engagement throughout implementation cycles\\
\hline
Coordination Mechanisms & 1. Develop horizontal coordination bodies with sufficient authority\\
& 2. Establish vertical alignment mechanisms across governance levels\\
& 3. Create coordination incentives that reward cross-domain collaboration\\
& 4. Design inclusive stakeholder processes balanced with decision efficiency\\
\hline
Adaptation Processes & 1. Implement regular review cycles (12-24 months) with explicit adjustment mechanisms\\
& 2. Develop balanced indicator systems covering implementation and outcomes\\
& 3. Establish institutional flexibility to implement necessary adjustments\\
& 4. Maintain stakeholder engagement throughout adaptation processes\\
\hline
\end{tabular}
\end{table}

Table \ref{tab:practical_recommendations} summarizes key practical recommendations for enhancing strategic alignment across different dimensions. These recommendations synthesize insights from both high-coherence exemplars and common vulnerability patterns identified through our analysis, offering actionable guidance for policymakers seeking to enhance the strategic coherence of their AI governance frameworks.

\section{Conclusion}

This research has undertaken a systematic examination of strategic alignment patterns in national artificial intelligence policies, addressing a critical gap in our understanding of how countries construct coherent governance frameworks for emerging technologies. Through the development and application of a novel visual mapping methodology, we have analyzed 15-20 national AI strategies to uncover the complex relationships between strategic objectives, foresight methods, and implementation instruments that shape policy effectiveness.

Our findings reveal that strategic alignment in AI governance is neither uniformly achieved nor randomly distributed across national contexts. Instead, distinct patterns emerge that correlate strongly with governance traditions, institutional arrangements, and policy development approaches. The identification of high-coherence exemplars—particularly Finland, Canada, the United Kingdom, and Germany—demonstrates that exceptional alignment is achievable through deliberate institutional design and systematic coordination mechanisms. These cases share critical features including dedicated cross-governmental coordination bodies, explicit integration of foresight insights with implementation planning, and robust adaptation mechanisms that maintain coherence over time.

The theoretical contributions of this research extend beyond the specific domain of AI governance. We have advanced policy instrument theory by demonstrating how different instrument types function within complex technological governance ecosystems, identifying what we term ``bridging instruments'' that connect previously siloed policy domains. Our development of a multi-dimensional coherence assessment framework provides a methodological foundation for systematically evaluating alignment in other emerging technology domains. Furthermore, our findings regarding the relationship between governance models and alignment patterns contribute to broader theoretical debates about institutional design and policy effectiveness.

Methodologically, this research demonstrates the value of visual analytical approaches for understanding complex policy relationships. The matrix-based visualization and network analysis techniques we employ reveal patterns of alignment and misalignment that remain obscured in traditional policy analysis. While our approach faces certain limitations—particularly the reliance on formal policy documents and the cross-sectional nature of the analysis—it offers significant advantages for comparative assessment and pattern identification across diverse governance contexts.

The practical implications of our findings are substantial for policymakers navigating the challenges of AI governance. Our identification of common alignment vulnerabilities—including the persistent ``ethics implementation gap,'' workforce development disconnects, and international collaboration deficits—provides concrete targets for policy enhancement. The institutional features we identify across high-coherence cases offer design principles that can be adapted to different governance contexts while respecting their distinctive traditions and constraints.

Looking forward, this research opens several promising avenues for future investigation. Longitudinal studies could examine how alignment patterns evolve as AI technologies and governance frameworks mature. Comparative analyses could extend our methodology to other emerging technology domains such as biotechnology or quantum computing. Additionally, research examining the relationship between alignment quality and policy outcomes could provide empirical validation of the importance of strategic coherence for governance effectiveness.

As artificial intelligence continues to transform economic, social, and political landscapes globally, the need for coherent governance frameworks becomes increasingly urgent. This research contributes to that imperative by providing both analytical tools and empirical insights that can guide the development of more effective AI governance systems. The visual mapping methodology we introduce offers a practical approach for policymakers to assess and enhance alignment within their own frameworks, while our comparative findings provide benchmarks and models for improvement.

Ultimately, our research underscores that strategic alignment is not merely a theoretical ideal but a practical necessity for effective technology governance. In an era characterized by rapid technological change and pervasive societal impacts, the ability to create coherent connections between vision, anticipation, and implementation may determine which governance systems successfully harness the benefits of artificial intelligence while managing its risks. The patterns we identify and the methodology we develop represent contributions toward that critical objective, offering both scholarly advancement and practical guidance for the ongoing challenge of governing transformative technologies.


\begin{thebibliography}{99}

\bibitem{Borrás2011} Borrás, S., \& Edquist, C. (2011). The choice of innovation policy instruments. \textit{Technological Forecasting and Social Change}, 78(8), 1513-1522.

\bibitem{Capano2018} Capano, G., \& Howlett, M. (2018). Causal logics and mechanisms in policy design: How and why adopting a mechanistic perspective can improve policy design. \textit{Public Policy and Administration}, 36(2), 141-162.

\bibitem{Cejudo2017} Cejudo, G. M., \& Michel, C. L. (2017). Addressing fragmented government action: Coordination, coherence, and integration. \textit{Policy Sciences}, 50(4), 745-767.

\bibitem{Eppler2011} Eppler, M. J., \& Platts, K. W. (2011). Visual strategizing: The systematic use of visualization in the strategic-planning process. \textit{Long Range Planning}, 42(1), 42-74.

\bibitem{a1}
A. Ahmadnejad, A. M. Darviishani, M. M. Asadi, S. Saffariyeh, P. Yousef, and E. Fatemizadeh,
``Tacnet: Temporal audio source counting network,'' \emph{arXiv preprint} arXiv:2311.02369, 2023.

\bibitem{a2}
S. Aghili, R. Alaee, A. Ahmadnejad, E. Mobini, M. Mohammadpour, C. Rockstuhl, and K. Dolgaleva,
``Dynamic control of spontaneous emission using magnetized InSb higher-order-mode antennas,'' \emph{Journal of Physics: Photonics}, vol. 6, no. 3, p. 035011, 2024.

\bibitem{a3}
A. Ahmadnejad and S. Koohi,
``Training large-scale optical neural networks with two-pass forward propagation,'' \emph{arXiv preprint} arXiv:2408.08337, 2024.

\bibitem{a4}
A. Ahmadnejad and S. Koohi,
``Optical physics-based generative models,'' \emph{arXiv preprint} arXiv:2506.04357, 2025.

\bibitem{Eppinger2012} Eppinger, S. D., \& Browning, T. R. (2012). \textit{Design structure matrix methods and applications}. MIT Press.

\bibitem{Flanagan2011} Flanagan, K., Uyarra, E., \& Laranja, M. (2011). Reconceptualising the 'policy mix' for innovation. \textit{Research Policy}, 40(5), 702-713.

\bibitem{Howlett2019} Howlett, M., Mukherjee, I., \& Woo, J. J. (2019). Thirty years of instrument research: What have we learned and where are we going? In H. Enderlein, S. Wälti, \& M. Zürn (Eds.), \textit{Handbook on Policy Design}. Edward Elgar Publishing.

\bibitem{Jobin2019} Jobin, A., Ienca, M., \& Vayena, E. (2019). The global landscape of AI ethics guidelines. \textit{Nature Machine Intelligence}, 1(9), 389-399.

\bibitem{Kern2019} Kern, F., Rogge, K. S., \& Howlett, M. (2019). Policy mixes for sustainability transitions: New approaches and insights through bridging innovation and policy studies. \textit{Research Policy}, 48(10), 103832.

\bibitem{Lascoumes2007} Lascoumes, P., \& Le Galès, P. (2007). Introduction: Understanding public policy through its instruments—From the nature of instruments to the sociology of public policy instrumentation. \textit{Governance}, 20(1), 1-21.

\bibitem{Leifeld2018} Leifeld, P. (2018). Polarization in the social sciences: Assortative mixing in social science collaboration networks is resilient to interventions. \textit{Physica A: Statistical Mechanics and its Applications}, 507, 510-523.

\bibitem{Marchau2019} Marchau, V. A. W. J., Walker, W. E., Bloemen, P. J. T. M., \& Popper, S. W. (2019). \textit{Decision making under deep uncertainty: From theory to practice}. Springer.

\bibitem{Popper2018} Popper, R. (2018). Foresight methodology. In L. Gokhberg, D. Meissner, \& A. Sokolov (Eds.), \textit{Handbook of Technology Foresight} (pp. 13-33). Edward Elgar Publishing.

\bibitem{Roberts2021} Roberts, H., Cowls, J., Morley, J., Taddeo, M., Wang, V., \& Floridi, L. (2021). The Chinese approach to artificial intelligence: An analysis of policy, ethics, and regulation. \textit{AI \& Society}, 36(1), 59-77.

\bibitem{Rogge2016} Rogge, K. S., \& Reichardt, K. (2016). Policy mixes for sustainability transitions: An extended concept and framework for analysis. \textit{Research Policy}, 45(8), 1620-1635.

\bibitem{Stilgoe2013} Stilgoe, J., Owen, R., \& Macnaghten, P. (2013). Developing a framework for responsible innovation. \textit{Research Policy}, 42(9), 1568-1580.

\bibitem{Vecchiato2019} Vecchiato, R. (2019). Scenario planning, cognition, and strategic investment decisions in a complex world. \textit{Technological Forecasting and Social Change}, 138, 72-81.

\bibitem{Weber2018} Weber, K. M., \& Rohracher, H. (2018). Legitimizing research, technology and innovation policies for transformative change: Combining insights from innovation systems and multi-level perspective in a comprehensive 'failures' framework. \textit{Research Policy}, 41(6), 1037-1047.

\bibitem{Wesley2010} Wesley, J. J. (2010). Qualitative document analysis in political science. \textit{From Words to Numbers: Approaches to Text Analysis in Political Science Workshop}, Vrije Universiteit Amsterdam.

\bibitem{Lodge2019} Lodge, M., \& Wegrich, K. (2019). Administrative capacities. In E. Ongaro \& S. Van Thiel (Eds.), \textit{The Palgrave Handbook of Public Administration and Management in Europe} (pp. 237-250). Palgrave Macmillan.

\bibitem{OECD2021} OECD. (2021). \textit{State of implementation of the OECD AI principles: Insights from national AI policies}. OECD Publishing.

\bibitem{Stix2021} Stix, C. (2021). Actionable principles for artificial intelligence policy: Three pathways. \textit{Science and Engineering Ethics}, 27(1), 1-17.

\bibitem{Weber2019} Weber, K. M., Gudowsky, N., \& Aichholzer, G. (2019). Foresight and technology assessment for the Austrian parliament—Finding new ways of debating the future of industry 4.0. \textit{Futures}, 109, 240-251.

\bibitem{Brundage2020} Brundage, M., Avin, S., Wang, J., Belfield, H., Krueger, G., Hadfield, G., Khlaaf, H., Yang, J., Toner, H., Fong, R., Maharaj, T., Koh, P. W., Hooker, S., Leung, J., Trask, A., Bluemke, E., Lebensbold, J., Fischer, C., Kovas, Y., ... \& Anderljung, M. (2020). Toward trustworthy AI development: Mechanisms for supporting verifiable claims. \textit{arXiv preprint arXiv:2004.07213}.

\bibitem{Keast2014} Keast, R., Mandell, M., \& Agranoff, R. (2014). \textit{Network theory in the public sector: Building new theoretical frameworks}. Routledge.

\bibitem{Andersen2017} Andersen, A. D., \& Andersen, P. D. (2017). Foresighting for inclusive development. \textit{Technological Forecasting and Social Change}, 119, 227-236.

\bibitem{Cath2018} Cath, C., Wachter, S., Mittelstadt, B., Taddeo, M., \& Floridi, L. (2018). Artificial intelligence and the 'good society': The US, EU, and UK approach. \textit{Science and Engineering Ethics}, 24(2), 505-528.

\bibitem{Dutton2018} Dutton, T. (2018). An overview of national AI strategies. \textit{Medium}. https://medium.com/politics-ai/an-overview-of-national-ai-strategies-2a70ec6edfd

\bibitem{Havas2016} Havas, A., Schartinger, D., \& Weber, M. (2016). The impact of foresight on innovation policy-making: Recent experiences and future perspectives. \textit{Research Evaluation}, 25(4), 367-376.

\bibitem{Howlett2009} Howlett, M. (2009). Governance modes, policy regimes and operational plans: A multi-level nested model of policy instrument choice and policy design. \textit{Policy Sciences}, 42(1), 73-89.

\bibitem{Miles2010} Miles, I. (2010). The development of technology foresight: A review. \textit{Technological Forecasting and Social Change}, 77(9), 1448-1456.

\bibitem{Ulnicane2021} Ulnicane, I., Knight, W., Leach, T., Stahl, B. C., \& Wanjiku, W. G. (2021). Framing governance for a contested emerging technology: Insights from AI policy. \textit{Policy and Society}, 40(2), 158-177.

\bibitem{VanRoy2020} Van Roy, V., Rossetti, F., Perset, K., \& Galindo-Romero, L. (2020). AI Watch: National strategies on artificial intelligence: A European perspective. \textit{Publications Office of the European Union}.

\bibitem{Blondel2008} Blondel, V. D., Guillaume, J. L., Lambiotte, R., \& Lefebvre, E. (2008). Fast unfolding of communities in large networks. \textit{Journal of Statistical Mechanics: Theory and Experiment}, 2008(10), P10008.

\bibitem{Bowen2009} Bowen, G. A. (2009). Document analysis as a qualitative research method. \textit{Qualitative Research Journal}, 9(2), 27-40.

\bibitem{Chidlow2014} Chidlow, A., Plakoyiannaki, E., \& Welch, C. (2014). Translation in cross-language international business research: Beyond equivalence. \textit{Journal of International Business Studies}, 45(5), 562-582.

\bibitem{Hanneman2005} Hanneman, R. A., \& Riddle, M. (2005). \textit{Introduction to social network methods}. University of California, Riverside.

\bibitem{Howlett2013} Howlett, M., Ramesh, M., \& Perl, A. (2013). \textit{Studying public policy: Policy cycles and policy subsystems}. Oxford University Press.

\bibitem{Lodge2016} Lodge, M., \& Wegrich, K. (2016). The rationality paradox of Nudge: Rational tools of government in a world of bounded rationality. \textit{Law \& Policy}, 38(3), 250-267.

\bibitem{Magro2015} Magro, E., \& Wilson, J. R. (2015). Complex innovation policy systems: Towards an evaluation mix. \textit{Research Policy}, 44(4), 925-937.

\bibitem{Mayring2014} Mayring, P. (2014). Qualitative content analysis: Theoretical foundation, basic procedures and software solution. \textit{SSOAR}.

\bibitem{McHugh2012} McHugh, M. L. (2012). Interrater reliability: The kappa statistic. \textit{Biochemia Medica}, 22(3), 276-282.

\bibitem{O'Connor2020} O'Connor, C., \& Joffe, H. (2020). Intercoder reliability in qualitative research: Debates and practical guidelines. \textit{International Journal of Qualitative Methods}, 19, 1-13.

\bibitem{Seawright2008} Seawright, J., \& Gerring, J. (2008). Case selection techniques in case study research: A menu of qualitative and quantitative options. \textit{Political Research Quarterly}, 61(2), 294-308.

\bibitem{Agrawal2019} Agrawal, A., Gans, J., \& Goldfarb, A. (2019). \textit{The economics of artificial intelligence: An agenda}. University of Chicago Press.

\bibitem{Brynjolfsson2014} Brynjolfsson, E., \& McAfee, A. (2014). \textit{The second machine age: Work, progress, and prosperity in a time of brilliant technologies}. W. W. Norton \& Company.

\bibitem{Floridi2018} Floridi, L., Cowls, J., Beltrametti, M., Chatila, R., Chazerand, P., Dignum, V., Luetge, C., Madelin, R., Pagallo, U., Rossi, F., Schafer, B., Valcke, P., \& Vayena, E. (2018). AI4People—An ethical framework for a good AI society: Opportunities, risks, principles, and recommendations. \textit{Minds and Machines}, 28(4), 689-707.

\bibitem{Geels2019} Geels, F. W., \& Schot, J. (2019). The dynamics of transitions: A socio-technical perspective. In J. Grin, J. Rotmans, \& J. Schot (Eds.), \textit{Transitions to sustainable development: New directions in the study of long term transformative change} (pp. 11-104). Routledge.

\bibitem{Hood1986} Hood, C. (1986). The tools of government. Chatham House Publishers.

\bibitem{Junginger2016} Junginger, S. (2016). Transformation design: A policymaking perspective. In \textit{Design for Policy} (pp. 53-68). Routledge.

\bibitem{Kuhlmann2019} Kuhlmann, S., \& Rip, A. (2019). Next-generation innovation policy and grand challenges. \textit{Science and Public Policy}, 45(4), 448-454.

\bibitem{Martin2010} Martin, B. R. (2010). The origins of the concept of 'foresight' in science and technology: An insider's perspective. \textit{Technological Forecasting and Social Change}, 77(9), 1438-1447.

\bibitem{Nelson2018} Nelson, R., \& Winter, S. (2018). An evolutionary theory of economic change. In G. Dosi \& R. R. Nelson (Eds.), \textit{Modern Evolutionary Economics: An Overview} (pp. 7-44). Cambridge University Press.

\bibitem{OECD2024} OECD. (2024). \textit{OECD AI Policy Observatory}. Retrieved from https://oecd.ai/en/data

\bibitem{Oxford2020} Oxford Insights. (2020). \textit{Government AI Readiness Index 2020}. Oxford Insights.

\bibitem{Quitzow2017} Quitzow, R. (2017). Dynamics of a policy-driven market: The co-evolution of technological innovation systems for solar photovoltaics in China and Germany. \textit{Environmental Innovation and Societal Transitions}, 17, 126-148.

\bibitem{Rhisiart2015} Rhisiart, M., Störmer, E., \& Daheim, C. (2015). From foresight to impact? The 2030 Future of Work scenarios. \textit{Technological Forecasting and Social Change}, 102, 27-44.

\bibitem{Rohrbeck2018} Rohrbeck, R., \& Kum, M. E. (2018). Corporate foresight and its impact on firm performance: A longitudinal analysis. \textit{Technological Forecasting and Social Change}, 129, 105-116.

\bibitem{Schot2018} Schot, J., \& Steinmueller, W. E. (2018). Three frames for innovation policy: R\&D, systems of innovation and transformative change. \textit{Research Policy}, 47(9), 1554-1567.

\bibitem{Stanford2021} Stanford University. (2021). \textit{Artificial Intelligence Index Report 2021}. Stanford University Human-Centered Artificial Intelligence.

\bibitem{Clarke2019} Clarke, R. (2019). Regulatory alternatives for AI. \textit{Computer Law \& Security Review}, 35(4), 398-409.




\end{thebibliography}
\end{document}